\documentclass[a4paper,onecolumn,showpacs,notitlepage,floatfix,nofootinbib,superscriptaddress,prd]{revtex4-2}

\pdfoutput=1

\usepackage[utf8]{inputenc}

\usepackage{graphicx}
\usepackage{dcolumn}
\usepackage{bm}
\usepackage{mathtools}
\usepackage{amsfonts,amsmath,amssymb}
\usepackage{graphicx, caption, subcaption}
\usepackage{ulem}
\usepackage{enumitem}

\usepackage{xcolor}

\newcommand{\be}{\begin{eqnarray}}
\newcommand{\ee}{\end{eqnarray}}
\newcommand{\nn}{\nonumber}
\newcommand{\nl}{\nonumber \\}
\newcommand{\pd}{\partial}

\newcommand{\bul}{\overset{\underset{\bullet}{}}}

\def\approxprop{%
  \def\p{%
    \setbox0=\vbox{\hbox{$\propto$}}%
    \ht0=0.6ex \box0 }%
  \def\s{%
    \vbox{\hbox{$\sim$}}%
  }%
  \mathrel{\raisebox{0.7ex}{%
      \mbox{$\underset{\s}{\p}$}%
    }}%
}

\DeclareFontFamily{U}{mathx}{}
\DeclareFontShape{U}{mathx}{m}{n}{<-> mathx10}{}
\DeclareSymbolFont{mathx}{U}{mathx}{m}{n}
\DeclareMathAccent{\widehat}{0}{mathx}{"70}
\DeclareMathAccent{\widecheck}{0}{mathx}{"71}

\allowdisplaybreaks

\begin{document} 

\author{Syo Kamata}
\email{skamata11phys@gmail.com}
\affiliation{National Centre for Nuclear Research, 02-093 Warsaw, Poland}
\affiliation{Department of Physics, The University of Tokyo, Tokyo 113-0033, Japan}

\title{Resurgence for the non-conformal Bjorken flow \\ with Fermi-Dirac and Bose-Einstein statistics}
\begin{abstract}    
We consider resurgence for the non-conformal Bjorken flow with Fermi-Dirac and Bose-Einstein statistics in the extended relaxation-time approximation. Firstly, we examine the full formal transseries expanded around the equilibrium and then construct the resurgent relation by looking at the structure of Borel-transformed ODEs. We form a conjecture of the resurgent relation based on the consideration that the Stokes constants constituting the resurgent relation originate only from singularities of dissipative variables on the Borel plane, and that the other variables such as temperature and chemical potential become Borel non-summable through nonlinear terms with the dissipative variables. We numerically check the conjecture for fundamental variables by explicitly evaluating the values of the dominant Stokes constant depending on the initial conditions and the particle mass. We also comment on some issues related to transseries structure and resurgence, such as the case of broken ${\rm U(1)}$ symmetry, the massless case, generalized relaxation-time, and attractor solutions.
\end{abstract}

\maketitle

\tableofcontents

\section{Introduction}

Relativistic hydrodynamics and kinetic theoretical approaches are powerful methods for describing high-energy nuclear collisions theoretically and understanding their non-equilibrium physics based on QCD \cite{Florchinger:2011qf, Akamatsu:2016llw, Romatschke:2017vte, Romatschke:2017ejr}. The Chapman-Enskog (CE) expansion, based on the gradient expansion, encodes the Boltzmann equation into hydrodynamics and has successfully described the non-equilibrium physics of high-energy nuclear collisions in a certain setup called the conformal Bjorken flow, as an IR effective theory \cite{Boyd1999}. The CE expansion can be regarded as a sort of asymptotic expansion in terms of the Knudsen number around the continuous flow limit determined by ${\rm Kn} \ll 1$. The conformal Bjorken flow, with certain approximations in the collision kernel and its kinetic theoretical approach, can be well-formulated by introducing appropriate orthogonal polynomials and a flow time \cite{Florchinger:2011qf, Akamatsu:2016llw, Romatschke:2017vte, Romatschke:2017ejr}.

When considering an asymptotic expansion around a near-equilibrium by using a certain flow time, dissipative effects, such as shear viscosity, are generally divergent series, i.e., the radius of convergence is zero \cite{Boyd1999}. In such a case, these asymptotic solutions can be approximations, but there exists a limitation for reducing the error from their exact solution depending on the value of the flow time \cite{Boyd1999}. However, instead of this disadvantage, this situation implies that a nontrivial relationship between hydro (perturbative) and nonhydro (nonperturbative) modes might be constructable. This relationship can be formulated by employing a mathematical tool called resurgence theory \cite{Mitschi2016, LodayRichaud2016}. Resurgence theory is a method to extract nonperturbative information from a perturbative expansion (or vice versa) through Borel resummation. These perturbative and nonperturbative sectors can be expressed as transseries, which are a sort of extension of a standard asymptotic expansion by introducing ingredients called transmonomials such as exponential decay and logarithmic convergence, and by using Borel resummation, coefficients in one sector can be expressed by those in the other sectors using singularities on the Borel plane. This relationship is called the \textit{resurgent relation} \cite{Marino:2012zq, Aniceto:2013fka, Dorigoni:2014hea, Aniceto:2018bis, Costin2006TopologicalCO, sauzin2014introduction, Mitschi2016, LodayRichaud2016}. In the context of high-energy nuclear collisions, the transseries and resurgence analysis of the conformal Bjorken flow were investigated in Refs. \cite{Heller:2015dha, Basar:2015ava, Aniceto:2015mto, Behtash:2019txb, Behtash:2018moe, Behtash:2020vqk}. These analyses help us clarify nonperturbative physics out of equilibrium and fundamental questions in hydrodynamics and kinetic theoretical approaches, such as the existence of nonhydro modes and a sort of universality of attractors.

The relevance of the symmetry of a Boltzmann distribution to transseries and its resurgent relation is an interesting question from both physical and mathematical perspectives. Since QCD is not a conformal theory, its nonconformal effects become significant when the temperature is lower relative to particle masses during the time evolution \cite{Moore:2008ws, Noronha-Hostler:2014dqa}. In addition, the nonconformal effects due to particle mass and ${\rm U(1)}$ (baryon number) become non-negligible at lower energies in the non-equilibrium process of QGP \cite{Li:2018fow, Du:2019obx, Denicol:2018wdp}. Therefore, the assumption of conformal symmetry is broken in the late-time regime of non-equilibrium QCD, and the conformal Bjorken flow needs to be modified to describe such non-equilibrium physics. Transport coefficients and hydrodynamics with/without ${\rm U(1)}$ symmetry for the nonconformal case have been investigated in Refs. \cite{Denicol:2014vaa, Jaiswal:2014isa, Li:2018fow, Denicol:2018wdp, Du:2019obx, Florkowski:2015lra, Dash:2021ibx, Ambrus:2022vif}, and the transseries structure of the nonconformal Bjorken flow without ${\rm U(1)}$ symmetry changes to extremely nontrivial forms due to the particle mass effect \cite{Kamata:2022jrc}.

In this paper, we study the transseries structure and resurgence of the nonconformal Bjorken flow with FD and BE statistics by imposing conservation laws on both the energy-momentum tensor and the current density. We employ the \textit{extended relaxation-time approximation} to be compatible with microscopic and macroscopic conservation laws \cite{Teaney:2013gca, Hoult:2021gnb, Banerjee:2012iz, Jensen:2012jh, Kovtun:2019hdm, Dash:2021ibx}. Firstly, we derive the full formal transseries expanded around the equilibrium. We particularly address the effects of conserved ${\rm U(1)}$ symmetry and conformal symmetry breaking on the transseries structure. After deriving the formal transseries, we construct the resurgent relation. The main nontrivial points for the construction in our problem are as follows:
\begin{itemize}
\item Our dynamical system is a multi-variable nonlinear ODE consisting of two types: $\frac{d Y}{d w} = -S Y + \frac{1}{w} \left[ a +b Y \right] + O(Y^2, w^{-2})$ and $\frac{d Y}{d w} = \frac{Y}{w} \left[ a + b Y \right] + O(Y^3 w^{-1},Y w^{-2})$, and a variable determined by either type of ODE appears in the other type as (non)linear terms.
\item The resurgent relation has a nontrivial dependence on parameters in the theory, such as initial conditions (integration constants) and particle mass.
\end{itemize}
In order to overcome the above problems, we form a conjecture of the resurgent relation by making observations of the large order behavior of the perturbative sector and then by focusing on the Borel transformed ODEs.

This paper is organized as follows: In Sec.~\ref{sec:setup}, we briefly review the nonconformal Bjorken flow with FD and BE statistics. In Sec.~\ref{sec:trans_IRdomain}, we derive the full formal IR transseries from ODEs. In Sec.~\ref{sec:Resurgence_analysis}, we consider the resurgence of the IR transseries. In Sec.~\ref{sec:add_comments}, we make some comments on issues related to transseries and resurgence, such as the case of broken ${\rm U(1)}$ symmetry, the massless case, generalized relaxation-time, and the attractor solution. Sec.~\ref{sec:summary} is devoted to the summary and discussion. Technical details used in Secs. \ref{sec:setup}--\ref{sec:add_comments} are summarized in the appendices. We take $g_{\mu \nu} = {\rm diag}(1,0,0,-\tau^2)$ and $g^{\mu \nu} = {\rm diag}(1,0,0,-1/\tau^2)$ to define the Milne coordinate. We perform the transseries analysis by imposing the Landau frame condition as $u_{\nu} T^{\nu \mu} = {\cal E} u^{\mu}$ and taking the local rest frame as $u^{\mu}=(1,0,0,0)$ in the Milne coordinate. For convenience, we refer to the early time and late time regimes as the UV and IR regimes, respectively.

\section{Review of nonconformal Bjorken flow with FD and BE statistics} \label{sec:setup}
In this section, we briefly review our setup.
In order to avoid complexity, we summarize definitions of symbols used in the below analysis, such as $\Xi_{k}$, $\Xi_{k}^{(n)}$, $\aleph^{(n,s)}_{k}$, and explicit forms of transport coefficients in App.\ref{sec:prep}.
{ Refs.~\cite{Florkowski:2015lra,Dash:2021ibx,Ambrus:2022vif,Denicol:2010xn} are helpful for the derivation.
}

We would start with the Boltzmann equation with the relaxation-time approximation.
In this work, we employ \textit{extended relaxation-time approximation} (ERTA) taking the following form\cite{Teaney:2013gca,Hoult:2021gnb,Banerjee:2012iz,Jensen:2012jh,Kovtun:2019hdm,Dash:2021ibx}:
\be
p^{\mu} \pd_\mu f(x,{\bf p}) = C[f], \qquad
C[f]:=- \frac{E_{\bf p}}{\tau_R(x)} \left[ f(x,{\bf p}) - f^{*}_{{\rm eq}}(x,{\bf p}) \right], \label{eq:ERTA}
\ee
where $f(x,{\bf p})$ is the distribution function,  ${\bf p}$ is the spacial component of the particle momentum $p^{\mu}$ satisfying the on-shell condition given by $p^2 = m^2$, $\tau_R$ is the relaxation-time, and $E_{\bf p}:= u \cdot p$ is energy on a frame (fluid velocity) $u^{\mu}$.
In the ERTA, we introduce $f^*_{\rm eq}(x,{\bf p})$ which is the local equilibrium with \textit{thermodynamic frame} to be compatible with  hydrodynamics and kinetic theory in the collision term.
It is defined as
\be
f^*_{\rm eq}(x,{\bf p}) = \frac{1}{\exp \left[ E^{*}_{\bf p} \beta_*(x) + \alpha_*(x) \right] + a},
\ee
where $\beta$ is the inverse temperature, $\alpha$ is defined as $\alpha:= \mu \beta$ using the chemical potential denoted by $\mu$.
Additionally, $a$ is a parameter of particle statistics, and taking $a=0, +1$, and $-1$ corresponds to Maxwell-Boltzmann (MB), Fermi-Dirac (FD), and Bose-Einstein (BE) statistics, respectively.
Here, the asterisk, ``$*$'', denotes variables in the thermodynamic frame.
These variables, $(u_*^\mu, \beta_*, \alpha_*)$, include corrections from the gradient expansion and can be decomposed into the zero-th order part and the correction as ${\cal O}_* = {\cal O} + \delta {\cal O}$, where  $\delta {\cal O}=O(\pd)$.
In this paper, we consider only NS hydro, so that we assume that the second or higher order corrections in $\delta {\cal O}$ is negligible and that $(\delta {\cal O})^2 \approx 0$.
We should emphasize that in the ERTA the thermodynamic frame, $u^{\mu}_*$, does not need to be the same as $u^{\mu}$, and hydro variables are defined by using $u^{\mu}$, as we would explain later.
The latter fluid velocity, $u^{\mu}$, is sometimes called as \textit{hydrodynamic frame} to distinguish from the thermodynamic frame.
Furthermore, on the one hand $(\beta_*, \alpha_*)$ are thermodynamic variables defined in the local equilibrium denoted by $f_{\rm eq}^*$ on the kinetic theory side, but on the other hand $(\beta, \alpha)$ are auxiliary variables constituting hydro variables.
Hydro variables can be formally defined by imposing a frame condition (Landau frame in our case) to $u^{\mu}$, and the perfect fluid is defined from the zero-th order gradient part in $f^*_{\rm eq}$, which we denote it as $f_{\rm eq}$.
See also Refs.~\cite{Teaney:2013gca, Dash:2021ibx} in detail.
From now on, when we say ``equilibrium'', we assume that it means hydrodynamic equilibrium, $f_{\rm eq}$.

Then, we define hydro variables using a hydrodynamic frame $u^{\mu}$.
The energy-momentum (EM) tensor, $T^{\mu \nu}$, and the current density, $N^{\mu}$, are defined from the distribution as 
\be
&& T^{\mu \nu} := \int_p p^{\mu} p^{\nu} f, \qquad N^{\mu} := \int_p p^{\mu} f,  \\ 
&& \int_p := \int \frac{d^4 p}{(2 \pi)^3 \sqrt{-\det g}} \delta(p^2) \cdot 2 \theta(p^0), \qquad p^2 = m^2, \nn
\ee
with the delta function $\delta(x)$ and the step function $\theta(x)$.
We impose the conservation laws to them as
\be
{\cal D}_{\mu} T^{\mu \nu} = 0, \qquad {\cal D}_{\mu} N^{\mu} = 0,
\ee
where ${\cal D}_\mu$ is the covariant derivative, and these determine dynamics of $\beta$ and $\alpha$.
The Landau matching condition and the kernel condition that
\be
\int_p  C[f] = \int_p p^\mu C[f] = 0
\ee
give identification of hydro variables in the EM tensor and the current density as
\be
&& T^{\mu \nu} = {\cal E}_0 u^{\mu} u^{\nu} - (P + \Pi) \Delta^{\mu \nu} + \pi^{\mu \nu}, \\
&& N^{\mu} = n_0 u^{\mu} + n^{\mu},
\ee
where ${\cal O}_0$ denotes the variables evaluated by the local equilibrium, $f_{\rm eq}$.
One can define these hydro variables from the Boltzmann distribution by performing the Chapman-Enskog (CE)-like expansion as introducing a book-keeping parameter $\epsilon$ corresponding to the Knudsen number.
By expanding the distribution as $f \approx f_{\rm eq} + \epsilon f^{[1]}$ and substituting it into Eq.(\ref{eq:ERTA}), one can determine $f^{[1]}$ as $f^{[1]} = f^{[1]}_{*} + f^{[1]}_{\epsilon}$, where $f^{[1]}_{*}$ and $f^{[1]}_{\epsilon}$ is the correction from the thermodynamic equilibrium and the gradient part from $f_{\rm eq}$, respectively, given by
\be
&& f^{[1]}_* := f^{*}_{\rm eq} - f_{\rm eq} \approx \sum_{{\cal O} \in \{ u^{\mu}, \beta, \alpha \}} \frac{\pd f_{\rm eq}}{\pd {\cal O}} \delta {\cal O}, \qquad 
f^{[1]}_\epsilon = - \frac{\tau_R}{E_{\bf p}} p^{\mu} \pd_\mu f_{\rm eq}.
\label{eq:f_1st}
\ee
The contributions from the thermodynamic equilibrium, $(\delta u^{\mu}, \delta \beta, \delta \alpha)$, can be determined by a frame condition.
Finally, we take $\epsilon = 1$ by assuming that $|f_{\rm eq}| \ll |f^{[1]}|$.
From the distribution, the hydro variables are defined as
\be
&& {\cal E}_0 := \int_p E_{\bf p}^2 f_{\rm eq}, \qquad P := - \frac{\Delta_{\alpha \beta}}{3} \int_p p^\alpha p^\beta f_{\rm eq}, \qquad n_0 := \int_p E_{\bf p} f_{\rm eq}, \\
&& \Pi := - \frac{\Delta_{\alpha \beta}}{3} \int_p p^{\alpha} p^{\beta} f^{[1]}, \qquad \pi^{\mu \nu} := \Delta^{\mu \nu}_{\alpha \beta}\int_p p^{\alpha} p^{\beta} f^{[1]}, \qquad n^{\mu} := \Delta^{\mu}_{\ \alpha} \int_p p^{\alpha} f^{[1]},
\ee
where $\Delta^{\mu \nu} := g^{\mu \nu} - u^\mu u^\nu$, and $\Delta^{\mu \nu}_{\alpha \beta} := \frac{1}{2} ( \Delta^{\mu}_{\ \alpha} \Delta^{\nu}_{\ \beta} + \Delta^{\nu}_{\ \alpha} \Delta^{\mu}_{\ \beta} ) - \frac{1}{3} \Delta^{\mu \nu}\Delta_{\alpha \beta}$.
The perfect fluid part is written as 
\be
{\cal E}_0 = \Xi_{2}, \qquad P = \frac{\Xi_{2} - m^2 \Xi_{0}}{3}, \qquad n_0 = \Xi_{1}. \label{eq:E0Pn0_1}
\ee
Notice that the traceless condition of the EM tensor is broken by the particle mass:
\be
g_{\mu \nu} T^{\mu \nu}_{{\rm eq}} = m^2 \Xi_0.
\ee
The boost invariant hydro is given under the condition that any variables are invariant under the Bjorken symmetry $ISO(2) \times SO(1,1) \times {\mathbb Z}_2$, i.e. ${\cal L}_{\xi} {\cal O} = 0$ and $R_{\zeta }{\cal O} = {\cal O}$, where the ${\cal L}_\xi$ is the Lie derivative with the Killing vector of Bjorken symmetry, $\xi$, and $R_\zeta$ is the reflection of rapidity, $\zeta$\cite{Bjorken:1982qr,Gubser:2010ze}.
By employing the Milne coordinate, the spacetime-dependence in the distribution reduces to only $\tau$, where $\tau$ is the Milne time given by $\tau := \sqrt{t^2 -z^2}$ using the Minkowski coordinate.
By using the Boltzmann equation (\ref{eq:ERTA}), one can derive ODEs taking the following forms: 
\be
\frac{d \beta}{d \tau} &=&  \frac{\beta}{\tau} \left[\chi_\beta + \gamma_\beta  ( \Pi - \widehat{\pi}) \right], \\
\frac{d \alpha}{d \tau} &=&  \frac{1}{\tau} \left[ \chi_\alpha - \gamma_\alpha  ( \Pi - \widehat{\pi} ) \right], \\
\frac{d \Pi}{d \tau} &=& - \frac{\Pi}{\tau_R}  - \frac{1}{\tau} \left[ \widetilde{\beta}_\Pi + \delta_{\Pi \Pi} \Pi - \lambda_{\Pi \pi} \widehat{\pi} \right], \\
\frac{d \widehat{\pi}}{d \tau} &=& - \frac{\widehat{\pi}}{\tau_R} + \frac{1}{\tau} \left[ \frac{4}{3} \beta_\pi  - \left( \frac{1}{3} \tau_{\pi \pi}  + \delta_{\pi \pi} \right) \widehat{\pi} + \frac{2}{3} \lambda_{\pi \Pi} \Pi \right],
\ee
where $\widehat{\pi}:= \pi_{\zeta}^{\ \zeta}$ is the shear viscosity, and ${\Pi}$ is the bulk viscosity.
The explicit form of the transport coefficients and the derivation of ODEs are summarized in Apps.\ref{sec:prep} and \ref{sec:derive_ODEs}, respectively.
The contribution from $f_{\rm eq}^*$, i.e. $(\delta u^\mu, \delta \beta, \delta \alpha)$, can be also determined by the Landau matching condition.
If $\tau_R$ does not have the momentum dependence, it is given by
\be
\delta u^{\mu} = 0, \qquad \delta \beta = 0, \qquad  \delta \alpha = \frac{\tau_R}{\tau} C_\alpha, \label{eq:deltabau2}
\ee
where
\be
C_\alpha := \chi_\alpha + \frac{m^2 \beta}{3} \cdot \frac{\Xi^{(1)}_2 \Xi^{(1)}_1 - \Xi^{(1)}_3 \Xi^{(1)}_0}{(\Xi^{(1)}_2)^2 - \Xi^{(1)}_3 \Xi^{(1)}_1}. \label{eq:Caa}
\ee

From now on, we take the mass unit ($m=1$) for the simplified notation\footnote{
In our notation, we use $\beta$ for $z=\beta m$ when taking $m=1$.
}, and we choose the relaxation-time as $\tau_R = \beta \theta_0$.
For the technical convenience, we redefine the viscous variables as 
$\bar{\Pi}:= \gamma_\beta \Pi$, $\bar{\pi}:= \gamma_\beta \widehat{\pi}$, where $\gamma_\beta := - \frac{\beta^{-1} \Xi^{(1)}_{1} }{(\Xi^{(1)}_{2})^2 - \Xi^{(1)}_{3} \Xi^{(1)}_{1}}$ and introduce ${\cal R}:=e^{-(\beta+\alpha)}$ instead of dealing with $\alpha$.
Furthermore, we also introduce a flow time,{
\be
w:= \frac{\tau \theta_0}{\tau_R} = \frac{\tau}{\beta},
\ee
}which is useful for the construction of resurgent relation that we would discuss later.
After all,  the modified ODEs are defined as
\be
\frac{d \beta}{d w} &=& \frac{\beta}{w} \cdot \frac{\chi_\beta +  \bar{\Pi} - \bar{\pi}}{1 - \left( \chi_\beta +  \bar{\Pi} - \bar{\pi} \right)} \quad =: {\cal F}_\beta, \label{eq:ODE_b_w} \\
\frac{d {\cal R}}{d w} &=&
-\frac{{\cal R}}{w} \cdot \frac{\beta \chi_\beta + \chi_\alpha + \beta ( 1 -  \frac{\Xi^{(1)}_{2}}{\Xi^{(1)}_{1}}) (\bar{\Pi}-\bar{\pi})}{1 - \left( \chi_\beta +  \bar{\Pi} - \bar{\pi} \right)} \quad =: {\cal F}_{\cal R}, \label{eq:ODE_R_w} \\
\frac{d \bar{\Pi}}{d w} &=& -\frac{\frac{\bar{\Pi}}{\theta_0} + \frac{1}{w} \left[ \bar{\beta}_\Pi + C_\Pi \bar{\Pi} - \lambda_{\Pi \pi} \bar{\pi} - D  \left( \bar{\Pi} - \bar{\pi} \right) \bar{\Pi} \right]}{1 - \left( \chi_\beta +  \bar{\Pi} - \bar{\pi} \right)} \quad =: {\cal F}_{\bar{\Pi}}, \label{eq:ODE_bP_w} \\
\frac{d \bar{\pi}}{d w}  &=& - \frac{\frac{\bar{\pi}}{\theta_0} + \frac{1}{w} \left[ - \frac{4}{3} \bar{\beta}_\pi  + C_\pi \bar{\pi} - \frac{2}{3} \lambda_{\pi \Pi} \bar{\Pi} - D \left(\bar{\Pi} - \bar{\pi} \right) \bar{\pi} \right]}{1 - \left( \chi_\beta +  \bar{\Pi} - \bar{\pi} \right)} \quad =: {\cal F}_{\bar{\pi}}, \label{eq:ODE_bp_w} 
\ee
where
\be
&& \bar{\beta}_\Pi := \gamma_\beta \widetilde{\beta}_\Pi, \qquad \bar{\beta}_\pi := \gamma_\beta \beta_\pi, \nl
&& C_{\Pi} := \delta_{\Pi \Pi} - \left( \beta  \chi_\beta  \pd_\beta + \chi_\alpha \pd_\alpha \right) \log \gamma_\beta, \nl
&& C_{\pi} := \frac{1}{3} \tau_{\pi \pi} + \delta_{\pi \pi} - \left( \beta  \chi_\beta  \pd_\beta + \chi_\alpha \pd_\alpha \right) \log \gamma_\beta, \nl
&& D := \beta \left( \pd_\beta   - \frac{\Xi^{(1)}_{2}}{\Xi^{(1)}_{1}} \pd_\alpha \right) \log \gamma_\beta.
\ee
Notice that the ODE of $\alpha$ is given by
\be
\frac{d \alpha}{d w} &=& \frac{1}{w} \cdot \frac{\chi_\alpha - \beta \frac{\Xi^{(1)}_{2}}{\Xi^{(1)}_{1}} ( \bar{\Pi} - \bar{\pi} )}{1 - \left( \chi_\beta +  \bar{\Pi} - \bar{\pi} \right)} \quad =: {\cal F}_\alpha. \label{eq:ODE_a_w} 
\ee
If one wants to obtain translation between $\tau$ and $w$, it can be obtained by solving
\be
\frac{d w}{d \tau} 
&=& \frac{1 - \left( \chi_\beta +  \bar{\Pi} - \bar{\pi} \right)}{\beta}. \label{eq:tautow}
\ee

The conformally broken effect can be directly observed from the leading order of the temperature.
Since $\chi_\beta \rightarrow \frac{2}{3}$ as $\beta \rightarrow + \infty$, the leading order expanded around $\tau = +\infty$ is obtained as
\be
&& \frac{d \beta}{d w} \sim \frac{2 \beta}{w}, \qquad \frac{d w}{d \tau} \sim \frac{1}{3 \beta} \quad \Rightarrow \quad \beta \sim \sigma_\beta w^2, \qquad  w  \sim \left( \frac{\tau}{\sigma_\beta} \right)^{1/3}, \qquad (\tau \gg 1)
\ee
with the integration constant, $\sigma_\beta \in {\mathbb R}_+$.
One can immediately see that the inverse temperature has a power law with the exponent, $2/3$, which is different from the conformal case, $\beta \approxprop \tau^{1/3}$.
This fact is true even for the MB case ($a=0$) such that $\alpha$ (or ${\cal R}$) is decoupled in the other ODEs.
It is because the conserved ${\rm U(1)}$ symmetry imposed by ${\cal D}_\mu N^{\mu}=0$
changes the functional form of speed of sound, $c_s^2 = \chi_\beta$.

\section{Transseries in the IR regime} \label{sec:trans_IRdomain}
{
In this section, we derive transseries of the variables in the IR regime, $w \gg 1$ by beginning with the ODEs (\ref{eq:ODE_b_w})-(\ref{eq:ODE_a_w}).
This part is one of the main results in this paper, and  in Sec.~\ref{sec:Resurgence_analysis} we will seek resurgent relations of these variables by using the result. 
}
As we will see in detail later, the transseries are given as the following forms:
\be
&& \beta \sim \sigma_\beta w^2 \sum_{{\bf n} \in {\mathbb N}_0^2} \sum_{k \in {\mathbb N}_0} a_{\beta}^{[{\bf n},k]} \bm{\zeta}^{\bf n} w^{-k}, \nl
&& {\cal R} \sim \sigma_{\cal R} \sum_{{\bf n} \in {\mathbb N}_0^2} \sum_{k \in {\mathbb N}_0} a_{\cal R}^{[{\bf n},k]} \bm{\zeta}^{\bf n} w^{-k}, \label{eq:full_trans_massive} \\
&& \bar{X} \sim \sum_{{\bf n} \in {\mathbb N}_0^2} \sum_{k \in {\mathbb N}_0} a_{\bar{X}}^{[{\bf n},k]} \bm{\zeta}^{\bf n} w^{-k}, \qquad (\bar{X} \in \{ \bar{\Pi},\bar{\pi} \})
\nn
\ee
where $a^{[{\bf n},k]}_{\cal O} \in {\mathbb R}$ is a real coefficient which is a function of integration constants, $(\sigma_{\beta}, \sigma_{\cal R})$, and the relaxation scale, $\theta_0$.
Additionally, $\bm{\zeta}^{\bf n}$ is the product of higher transmonomials, $\zeta_\pm$, containing exponential decay defined as
\be
 \bm{\zeta}^{\bf n} &:=& \zeta_+^{n_+} \zeta_-^{n_-}, \qquad \ \zeta_\pm :=  \sigma_{\pm} \frac{e^{-S_\pm w}}{w^{\rho_{\pm}}}, \nl
 S_\pm &=& S := \frac{3}{\theta_0}, \qquad \rho_{\pm} := 5 - 12 C^{\bar{\beta}_{\pi}}_0(\sigma_{{\cal R}, a}) \mp \frac{\sqrt{1285}}{5}, \label{eq:lam_rho0} \\
 C^{\bar{\beta}_{\pi}}_0(\sigma_{{\cal R}, a}) &=& \frac{4 {\rm Li}_{\frac{1}{2}}\left(\sigma_{{\cal R}, a}\right) {\rm Li}_{\frac{5}{2}}\left(\sigma_{{\cal R}, a}\right)}{15 {\rm Li}_{\frac{1}{2}}\left(\sigma_{{\cal R}, a}\right) {\rm Li}_{\frac{5}{2}}\left(\sigma_{{\cal R}, a}\right)-9 {\rm Li}_{\frac{3}{2}}\left(\sigma_{{\cal R}, a} \right)^2} = \frac{2}{3} - \frac{\sigma_{{\cal R}, a}}{4 \sqrt{2}} + O(\sigma_{{\cal R}, a}^2), \label{eq:C_bpi0}
\ee
where $\sigma_\pm$ is the integration constant, $ \sigma_{{\cal R}, a} := - a \sigma_{\cal R}$, and ${\bf n}=(n_+,n_-) \in {\mathbb N}_0^2$

Below, we would consider the transseries structure in the following definition labeled by sectors\cite{Kamata:2022jrc}:
\be
\mbox{Perfect fluid (PF) sector} &:& \mbox{No-dependence of $\theta_0$}, \nl
\mbox{Perturbative (PT) sector} &:& \mbox{Power expansion of $\theta_0$}, \nl
\mbox{Nonperturbative (NP) sector(s)} &:& \mbox{Exponentially damping term of $\theta_0^{-1}$} \nl
&& \mbox{$\times$ power expansion of $\theta_0$} \nn
\ee
The relationship among each of the sectors can be schematically expressed by
\be
&& \mbox{Full transseries} = \mbox{PT sector} \oplus \mbox{NP sectors}, \nl
&& 
\mbox{PF sector} \subset \mbox{PT sector}, \qquad \mbox{NP sectors} = \bigoplus_{\substack{{\bf n} \in {\mathbb N}_0^2 \\ |{\bf n}|>0}} \mbox{${\bf n}$-th NP sector}, \nn
\ee
where the ${\bf n}$-th NP sector is labeled by the exponential factor, $e^{- {\bf n} \cdot  {\bf S} w/\theta_0}$, and the PF sector can be extracted from the PT sector by taking the small $\theta_0$ limit as\footnote{
  In this paper, we take the slightly different definition from Ref.~\cite{Kamata:2022jrc} for the PF sector.
  In Ref.~\cite{Kamata:2022jrc}, the PF sector is defined to be excluded from the PT sector.
}
\be
\lim_{\theta_0 \rightarrow 0_+} \mbox{PT sector} = \mbox{PF sector}. \nn
\ee
{
We derive the transseries of each the sector in the below subsections, Sec.~\ref{sec:trans_PF}-\ref{sec:trans_NP}.
We also mention transseries of other variables, such as the energy density and the bulk pressure, in Sec.~\ref{sec:trans_others}.
}

{We emphasize that, in the below analysis based on the ODEs (\ref{eq:ODE_b_w})-(\ref{eq:ODE_a_w}), the approximation which we will use is only $w \gg 1$. Notice that, in the special cases such as the massless limit, such a limit  has to be taken \textit{before} taking asymptotic limit, i.e. $w \rightarrow + \infty$ because these limits are in general non-commutative with each other.
  Such a non-commutativity usually arises when symmetry changes by a certain limit, and as a result it causes some change of the transseries structure.
See Secs.~\ref{sec:U1broken} and \ref{sec:massless}.}

\subsection{PF sector} \label{sec:trans_PF}
\begin{figure}[tp]
  \begin{center}
    \begin{tabular}{cc}
      \begin{minipage}{1.\hsize}
        \begin{center} 
          \includegraphics[clip, width=100mm]{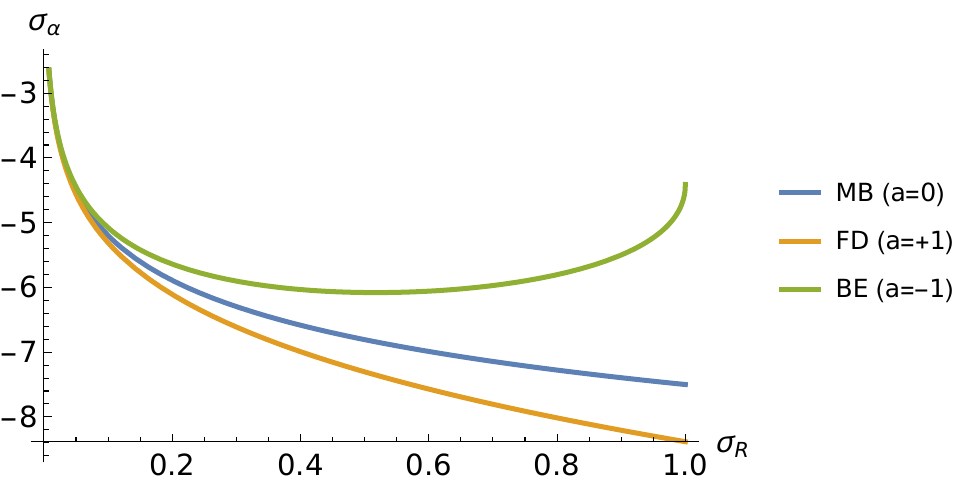}
        \end{center}
      \end{minipage}      
    \end{tabular} 
    \caption{The plot of $\sigma_\alpha$ vs. $\sigma_{\cal R}$ in Eq.(\ref{eq:sigsig}).}
    \label{fig:sigsig}
  \end{center}
\end{figure}

Firstly, we consider the PF sector by starting with the ODEs of $(\beta, \alpha)$.
The PF sector is defined by eliminating the viscous variables, $(\bar{\Pi}, \bar{\pi})$, from the ODEs {(\ref{eq:ODE_b_w})(\ref{eq:ODE_a_w})}, and taking $(\bar{\Pi}, \bar{\pi}) \rightarrow (0,0)$ reduces them to
\be
\frac{d \beta_{\rm pf} }{d w} = \frac{\beta_{\rm pf}}{w} \cdot \frac{\chi_\beta}{1-\chi_\beta}, \qquad \frac{d \alpha_{\rm pf}}{d w} &=& \frac{1}{w} \cdot \frac{\chi_\alpha}{1 - \chi_\beta} \label{eq:ODE_pf}.
\ee
According to Eqs.(\ref{eq:chi_b_asym_zinf})(\ref{eq:chi_a_asym_zinf}), the asymptotic forms of $\chi_\beta$ and $\chi_\alpha$ expanded around $\beta = + \infty$ are given by
\be
\chi_\beta &\sim& \frac{2}{3} - C^{\chi}_0({\cal R}_{a}) \beta^{-1} + O(\beta^{-2}), \label{eq:chi_b_lead} \\
\chi_\alpha &\sim& -\frac{2}{3} \beta + C^{\chi}_0({\cal R}_{a}) + O(\beta^{-1}), \label{eq:chi_a_lead} \\
 C^{\chi}_0({\cal R}_{a}) &:=& \frac{15 {\rm Li}_{\frac{3}{2}}({\cal R}_a) {\rm Li}_{\frac{5}{2}}({\cal R}_a) - 35 {\rm Li}_{\frac{1}{2}}({\cal R}_a) {\rm Li}_{\frac{7}{2}}({\cal R}_a)}{18 {\rm Li}_{\frac{3}{2}}({\cal R}_a)^2- 30 {\rm Li}_{\frac{1}{2}}({\cal R}_a) {\rm Li}_{\frac{5}{2}}({\cal R}_a)} \label{eq:C^chi} \\
&=& \frac{5}{3} - \frac{35}{96 \sqrt{2}} {\cal R}_a + O({\cal R}_a^2), \nn
\ee
where ${\cal R}_a := -a {\cal R}$, and ${\rm Li}_n(x)$ is the logarithmic integral function, so that the first some leading orders are obtained as
\be
\beta_{\rm pf} \sim \sigma_\beta w^2 + O(w^0), \qquad \alpha_{\rm pf} \sim - \sigma_\beta w^2 
+ \sigma_\alpha + O(w^{-2}), \label{eq:ba_pf}
\ee
where $\sigma_\beta \in {\mathbb R}_+$ and $\sigma_\alpha \in {\mathbb R}$ are integration constants.
This result implies that the chemical potential in the IR limit gives
\be
\lim_{w \rightarrow +\infty} \mu_{\rm pf} =\lim_{w \rightarrow +\infty} \alpha_{\rm pf}/\beta_{\rm pf} = -1 \quad (= - m).
\ee
In addition, one can obtain the PF sector of ${\cal R}_a$ (or ${\cal R}$) in the similar way.
Since
\be
\frac{d (\beta_{\rm pf} + \alpha_{\rm pf})}{d w} = \frac{1}{w} \cdot \frac{\beta_{\rm pf} \chi_\beta + \chi_\alpha}{1 - \chi_\beta} = \frac{1}{w} \cdot O(\beta_{\rm pf}^{-1}),
\ee
the leading order of $\beta_{\rm pf} + \alpha_{\rm pf}$ is a real constant, and thus,
\be
{\cal R}_a \sim  \sigma_{{\cal R}, a} + O(w^{-2}), \qquad \sigma_{{\cal R}, a}:= - a \sigma_{\cal R}. \qquad (\sigma_{\cal R} \in {\mathbb R}_+)
\ee
Because of that, the leading order of $C^{\chi}_0({\cal R}_{a})$ in Eq.(\ref{eq:C^chi}) is a constant, i.e., $C^{\chi}_0(\sigma_{{\cal R}, a})$.
Now, either $\sigma_\alpha$ or $\sigma_{\cal R}$ is a function of the other one.
The relationship between $\sigma_\alpha$ and $\sigma_{\cal R}$ is given by\footnote{
  By taking into account of $O(w^0)$ for the temperature, one can find $\beta_{\rm pf} \sim \sigma_\beta w^2 + \frac{9}{2} C^{\chi}_0(\sigma_{{\cal R}, a}) + O(w^{-2})$.
  Thus,
  \be
  && {\cal R}_{\rm pf} = e^{-\beta_{\rm pf} - \alpha_{\rm pf}} \sim e^{-\frac{9}{2} C^{\chi}_0(\sigma_{{\cal R}, a}) - \sigma_{\alpha}} + O(w^{-2}) \quad \Rightarrow \quad  e^{-\frac{9}{2} C^{\chi}_0(\sigma_{{\cal R}, a}) - \sigma_{\alpha}} = \sigma_{\cal R}.
  \ee  
}
\be
\sigma_\alpha &=& - \log \sigma_{\cal R} - \frac{9}{2} C^{\chi}_0(\sigma_{{\cal R}, a}) \label{eq:sigsig} \\
&=&  - \log \sigma_{\cal R} -  \frac{15}{2} +  \frac{105}{64  \sqrt{2}} \sigma_{{\cal R},a} + O(\sigma_{{\cal R},a}^2). \nn
\ee
This equality gives a constraint for the domain of $\sigma_\alpha$ and $\sigma_{\cal R}$.
Fig.~\ref{fig:sigsig} shows the plot of Eq.(\ref{eq:sigsig}).
In order to make $\sigma_\alpha$ a real number, $\sigma_{\cal R}$ has to be real positive, and $\sigma_{\cal R} < 1$ when $a=-1$.
In contrast, $\sigma_\alpha$ is negative except regions near divergent points.
By using the asymptotic forms of $\chi_{\beta,\alpha}$ obtained in Eqs.(\ref{eq:chi_b_asym_zinf})(\ref{eq:chi_a_asym_zinf}), one can obtain the PF sector as
\be
\beta_{\rm pf} &\sim& \sigma_\beta w^2 \sum_{k \in {\mathbb N}_0} a_{\beta_{\rm pf}}^{[k]} w^{-2k} \nl
&=& \sigma_\beta w^{2} \left[ 1.0000 + \sigma^{-1}_\beta \left( 7.5000 - 1.1601 \sigma_{{\cal R},a}  \right) w^{-2} \right. \nl
  && \left. + \sigma^{-2}_\beta \left( - 33.7500 + 9.0581 \sigma_{{\cal R},a} \right) w^{-4} + O(w^{-6},\sigma_{{\cal R},a}^2) \right], \label{eq:beta_pf_w} \\ \nl
     {\cal R}_{\rm pf} &\sim& {\sigma_{\cal R}} \sum_{k \in {\mathbb N}_0} a_{{\cal R}_{\rm pf}}^{[k]} w^{-2k} \nl
     &=& \sigma_{\cal R} \left[ 1.0000 + \sigma^{-1}_\beta \left( 1.8750 - 0.9115 \sigma_{{\cal R},a}  \right) w^{-2} \right. \nl
       && \left. + \sigma^{-2}_\beta \left( - 14.1797 + 7.7400 \sigma_{{\cal R},a} \right) w^{-4} + O(w^{-6},\sigma_{{\cal R},a}^2) \right], \label{eq:R_pf_w}
\ee
Here, the coefficients are functions of the integration constants, $a_{{\cal O}_{\rm pf}}^{[k]} = a_{{\cal O}_{\rm pf}}^{[k]}[\sigma_\beta, \sigma_{{\cal R},a}]$, and we took $a^{[k=0]}_{\beta_{\rm pf}} = a^{[k=0]}_{{\cal R}_{\rm pf}} = 1$ for the normalization.
{
  Eqs.(\ref{eq:beta_pf_w})(\ref{eq:R_pf_w}) are expanded forms around $\sigma_{{\cal R},a} = 0$, which is originated from the fact that $C^\chi_0({\cal R}_a)$ in Eq.(\ref{eq:C^chi}) is the formal series expansion of ${\cal R}_a$.
  In contrast, $\sigma_\beta$ is an overall factor determined by the order of $w$.
}

Let us make sure if the contribution from $f_{\rm eq}^*$, i.e. $\delta \alpha$, is small enough relative to $\alpha_{\rm pf}$ in the IR regime.
Since $C_\alpha$ defined in Eq.(\ref{eq:Caa}) can be expanded as
\be
C_\alpha &\sim& - \frac{2}{3} \beta + \frac{7}{3} - \frac{35 {\cal R}_a}{96 \sqrt{2}} + O(\beta^{-1},{\cal R}_a^2), \qquad (\beta \gg 1) \label{eq:C_a_sim}
\ee
substituting the asymptotic solution in Eq.(\ref{eq:ba_pf}) into Eq.(\ref{eq:C_a_sim}) yields $C_\alpha \sim - \frac{2}{3} \sigma_\beta w^{2}$.
Hence, the behavior of $\delta \alpha$ in the IR regime is obtained from Eq.(\ref{eq:deltabau2}) as
\be
\delta \alpha = \frac{\tau_R}{\tau} C_\alpha \sim -  \frac{2}{3} \sigma_\beta \theta_0 w,
\ee
and thus, $|\alpha_{\rm pf}| \gg |\delta \alpha|$ as $w \rightarrow +\infty$.
Therefore, it is consistent with our assumption that $|f_{\rm eq}| \ll |f^{[1]}_*|$ in the IR regime.

We would comments on some remarkable facts;
\begin{itemize}
\item The leading order of $\beta_{\rm pf}$ is determined by $\chi_\beta$, which is the speed of sound, $c_s^2 = \chi_\beta$, and then propagate it to that of $\alpha_{\rm pf}$ (or ${\cal R}_{\rm pf}$).  
  The leading orders do not have the $a$-dependence, meaning that they are irrelevant to the statistics such as MB, FD, and BE.
  As we can see in analysis of the PT and NP sectors,  transseries in the higher sectors is determined by the lower sectors, so that the speed of sound is a seed to determine transmonomials in all sectors.
  See also Ref.~\cite{Kamata:2022jrc}.
\item $\sigma_\beta$ can be regarded to be the similar role to the relaxation scale for $w$, i.e. $a^{[k]}_{{\cal O}_{\rm pf}} \propto \sigma_\beta^{-k}$, and taking larger $\sigma_\beta$ drops the contribution from higher orders of $w^{-1}$.
As one can see by explicitly turning on the mass such that $\sigma_\beta \rightarrow m \sigma_\beta$, taking large $\sigma_\beta$ essentially corresponds to the heavy mass limit.
\end{itemize}

\subsection{PT sector} \label{sec:trans_PT}
Then, we consider the PT sector.
The PT sector is obtained by taking into account the contribution of viscosities, $(\bar{\Pi},\bar{\pi})$, in the ODEs.
From $(\beta_{\rm pf}, {\cal R}_{\rm pf})$ and Eqs.(\ref{eq:ODE_bP_w})(\ref{eq:ODE_bp_w}), the leading order of $(\bar{\Pi}, \bar{\pi})$ can be obtained as
\be
&& - \frac{\bar{\Pi}}{\theta_0} + \frac{7}{3 w} \bar{\pi} \sim 0, \qquad - \frac{\bar{\pi}}{\theta_0} + \frac{4}{3 w} C^{\bar{\beta}_{\pi}}_0(\sigma_{{\cal R}, a}) \sim 0 \quad \mbox{as} \quad w \rightarrow +\infty \nl
&\Rightarrow \quad& \bar{\Pi} \sim \frac{28}{9} C^{\bar{\beta}_{\pi}}_0(\sigma_{{\cal R}, a}) \left( \frac{\theta_0}{w} \right)^2, \qquad \bar{\pi} \sim \frac{4}{3} C^{\bar{\beta}_{\pi}}_0(\sigma_{{\cal R}, a})\frac{\theta_0}{w},
\ee
where $C^{\bar{\beta}_\pi}_0$ is defined in Eq.(\ref{eq:C_bpi0}).
One can easily compute higher orders recursively from the ODEs and obtain the formal solutions of the PT sector as
\be
&& \beta_{\rm pt} \sim \sigma_\beta w^2 \sum_{k \in {\mathbb N}_0} a_{\beta}^{[{\bf 0},k]}  w^{-k}, \qquad {\cal R}_{\rm pt} \sim \sigma_{\cal R}  \sum_{k \in {\mathbb N}_0} a_{\cal R}^{[{\bf 0},k]}  w^{-k}, \nl
&& \bar{\Pi}_{\rm pt} \sim \sum_{k \in {\mathbb N}_0+2} a_{\bar{\Pi}}^{[{\bf 0},k]}  w^{-k}, \qquad \quad  \ \bar{\pi}_{\rm pt} \sim \sum_{k \in {\mathbb N}_0+1} a_{\bar{\pi}}^{[{\bf 0},k]}  w^{-k}.
\ee
One can take $a_{\beta}^{[{\bf 0},k=0]}=a_{\cal R}^{[{\bf 0},k=0]}=1$ for the normalization of $(\sigma_{\beta},\sigma_{\cal R})$ without the loss of generality.
More explicitly, these can be written down as
\be
\beta_{\rm pt} &\sim& \sigma_\beta w^{2} \left[ 1.0000 + \left( 8.0000 -  2.1213 \sigma_{{\cal R},a}\right) \theta_0 w^{-1} \right. \nl
  && \left. + \left\{ 5.3333 -11.3137 \sigma_{{\cal R},a}
  + \left( 7.5000 - 1.1601 \sigma_{{\cal R},a} \right) (\sigma_\beta \theta_0^2)^{-1} \right\} \theta_{0}^2 w^{-2}  \right. \nl
  && \left. + O(w^{-3},\sigma_{{\cal R},a}^2) \right] , \\ \nl
{\cal R}_{\rm pt} &\sim& \sigma_{\cal R} \left[ 1.0000 + \left( 4.0000 -  2.4749 \sigma_{{\cal R},a}\right) \theta_0 w^{-1} \right. \nl
  && \left. + \left\{ -5.3333 - 5.1855 \sigma_{{\cal R},a} + \left( 1.8750 - 0.9115 \sigma_{{\cal R},a} \right) (\sigma_\beta \theta_0^2)^{-1} \right\} \theta_{0}^2 w^{-2}  \right. \nl
  && \left. + O(w^{-3},\sigma_{{\cal R},a}^2) \right] , \\ \nl
\bar{\Pi}_{\rm pt} &\sim&  \left( 2.0741 -  0.5500 \sigma_{{\cal R},a}\right) \theta_0^{2} w^{-2} + \left( -4.8395 - 0.9166 \sigma_{{\cal R},a} \right) \theta_{0}^3 w^{-3} \nl
&& +\left\{ 19.6807 + 1.3810 \sigma_{{\cal R},a} + \left(- 9.3333 + 2.0968 \sigma_{{\cal R},a} \right) (\sigma_\beta \theta_0^2)^{-1}\right\} \theta_0^{4} w^{-4} \nl
  && + O(w^{-5},\sigma_{{\cal R},a}^2), \\ \nl
\bar{\pi}_{\rm pt} &\sim&  \left( 0.8889 - 0.2357 \sigma_{{\cal R},a}\right) \theta_0 w^{-1} + \left( -1.4815 - 0.5500 \sigma_{{\cal R},a} \right) \theta_{0}^2 w^{-2} \nl
&& +\left\{ 6.9531 + 0.7752 \sigma_{{\cal R},a} + \left(- 3.1111 + 0.5843 \sigma_{{\cal R},a} \right) (\sigma_\beta \theta_0^2)^{-1}\right\} \theta_0^{3} w^{-3} \nl
  && + O(w^{-4},\sigma_{{\cal R},a}^2). 
\ee
The coefficients depend on $(\sigma_\beta,\sigma_{{\cal R},a},\theta_0)$, and their expanded form with respect to $(\sigma_\beta \theta_0^2)^{-1}$ is a polynomials with a degree determined by $k$ as
\be
&& a^{[{\bf 0},k]}_{\cal O}[\sigma_\beta,\sigma_{{\cal R}, a},\theta_0] = \sum_{s=0}^{s_{\max}} a^{[{\bf 0},k]}_{{\cal O}(s)} [\sigma_{{\cal R}, a},\theta_0] (\sigma_\beta \theta_0^{2})^{-s} \qquad (a^{[{\bf 0},k]}_{{\cal O}(s)}[\sigma_{{\cal R}, a}, \theta_0] \in {\mathbb R}) \nl
&& \mbox{with} \quad s_{\max} =
\begin{cases}
    \lfloor k/2 \rfloor & \mbox{for ${\cal O} \in \{ \beta, {\cal R} \}$} \\
  \lfloor (k-1)/2 \rfloor & \mbox{for ${\cal O} \in \{ \bar{\Pi}, \bar{\pi} \}$}
\end{cases}, \label{eq:a_para_deps0} 
\ee
where $\lfloor x \rfloor$ is the floor function, and we defined $a^{[{\bf 0},k]}_{{\cal O}(s)}[\sigma_{{\cal R},a},\theta_0]$ such that $a^{[{\bf 0},k]}_{{\cal O}(s)}[\sigma_{{\cal R},a},\theta_0]\propto \theta_0^k$.
{
  Similar to the PF sector, $a^{[{\bf 0},k]}_{{\cal O}(s)}[\sigma_{{\cal R},a},\theta_0]$ is a function and can be expanded around $\sigma_{{\cal R},a}=0$, which yield a formal series expansion.
}
The small $\theta_0$ limit extracts the PF sector from the PT sector as
\be
&& \lim_{\theta_0 \rightarrow 0_+} a^{[{\bf 0},k]}_{\beta, {\cal R}}[\sigma_\beta,\sigma_{{\cal R}, a},\theta_0] =
\begin{cases}
  0 & \quad \mbox{for odd $k$} \\
  a^{[k/2]}_{\beta_{\rm pf},{\cal R}_{\rm pf}}[\sigma_\beta, \sigma_{{\cal R},a}] & \quad \mbox{for even $k$} 
\end{cases}, \label{eq:theta0_lim} \\
&& \lim_{\theta_0 \rightarrow 0_+} a^{[{\bf 0},k]}_{\bar{\Pi},\bar{\pi}}[\sigma_\beta,\sigma_{{\cal R}, a},\theta_0] = 0.
\ee
If one takes the low temperature (or heavy mass) limit as $\sigma_{\beta} \rightarrow +\infty$, then the $\sigma_\beta$-dependence in $a^{[0,k]}_{{\cal O}}[\sigma_\beta, \sigma_{\cal R},\theta_0]$
goes away.
Thus, the polynomial becomes a monomial proportional to $\theta_0^k$, i.e.,
\be
&& \lim_{\sigma_\beta \rightarrow +\infty} a^{[{\bf 0},k]}_{\cal O}[\sigma_\beta,\sigma_{{\cal R}, a},\theta_0] = a^{[{\bf 0},k]}_{{\cal O}(s=0)}[\sigma_{{\cal R}, a},\theta_0] \propto \theta_{0}^k. \label{eq:m_inf_pt}
\ee
Notice that the order of $\theta_0$ is the same as that of $w^{-1}$.

\subsection{NP sectors} \label{sec:trans_NP}

\begin{figure}[tp]
  \begin{center}
    \begin{tabular}{cc}
      \begin{minipage}{1.\hsize}
        \begin{center} 
          \includegraphics[clip, width=80mm]{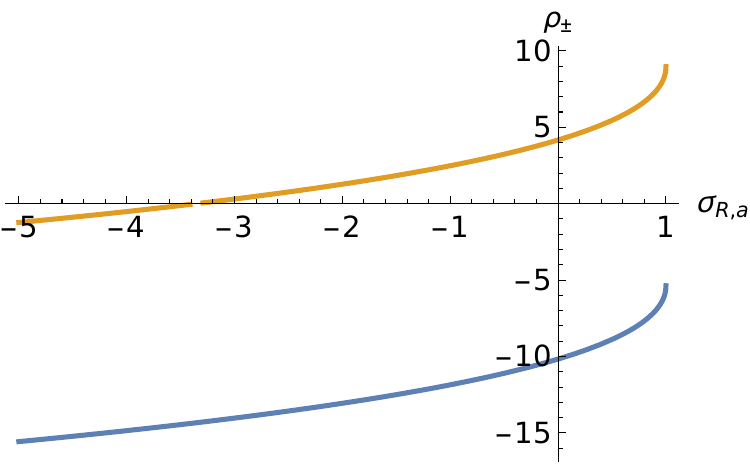}
        \end{center}
      \end{minipage}      
    \end{tabular} 
    \caption{The plot of $\rho_\pm$ vs. $\sigma_{{\cal R},a} (= -a \sigma_{\cal R})$.
      The colored lines denote $\rho_+$ (Blue) and $\rho_-$ (Orange).}
    \label{fig:rho_sigma}
  \end{center}
\end{figure}

Finally, we consider the NP sectors.
We summarize the derivation of the NP transmonomials in App.\ref{sec:transseries}.
By identification of transmonomials in the NP sector, one can easily obtain the formal transseries from the ODEs in the similar way to computation of the PT sector.
Those can be expressed by the following forms:
\be
&& \beta_{\rm np} \sim \sigma_{\beta} w^{2} \sum_{\substack{{\bf n} \in {\mathbb N}_0^2 \\ |{\bf n}|>0}} \sum_{k \in {\mathbb N}} a_{\beta}^{[{\bf n},k]} \bm{\zeta}^{\bf n} w^{-k},  \qquad  {\cal R}_{\rm np} \sim \sigma_{\cal R} \sum_{\substack{{\bf n} \in {\mathbb N}_0^2 \\ |{\bf n}|>0}} \sum_{k \in {\mathbb N}} a_{\cal R}^{[{\bf n},k]} \bm{\zeta}^{\bf n} w^{-k}, \nl
&& \bar{\Pi}_{\rm np} \sim  \sum_{\substack{{\bf n} \in {\mathbb N}_0^2 \\ |{\bf n}|>0}} \sum_{k \in {\mathbb N}_0} a_{\bar{\Pi}}^{[{\bf n},k]} \bm{\zeta}^{\bf n} w^{-k}, \qquad \qquad  \bar{\pi}_{\rm np} \sim  \sum_{\substack{{\bf n} \in {\mathbb N}_0^2 \\ |{\bf n}|>0}} \sum_{k \in {\mathbb N}_0} a_{\bar{\pi}}^{[{\bf n},k]} \bm{\zeta}^{\bf n} w^{-k}, 
\ee
where ${\bm \zeta}^{\bf n}$ is defined in Eq.(\ref{eq:lam_rho0}).
The NP transmonomial, $\zeta_{+} (\zeta_-)$, contains an integration constant denoted by $\sigma_+ (\sigma_-)$, and $\sigma_\pm$ appears only as coupling to the exponential decay, $e^{-S_\pm w}$, in the full transseries.
As we can see from Eq.(\ref{eq:lam_rho0}), $\rho_\pm$ is independent on $\theta_0$ but a nontrivial function of $\sigma_{{\cal R},a}$.
Fig.~\ref{fig:rho_sigma} shows the plot of $\rho_\pm$.
Notice that this plot is consistent with the observation from Fig.~\ref{fig:sigsig}, i.e. $\sigma_{\cal R} \in {\mathbb R}_+$ for $a=0$ and $+1$, and $\sigma_{\cal R} \in (-\infty,1)$ for $a=-1$.
For $|{\bf n}|=1$, for example, the explicit values of the first some leading orders are obtained as\footnote{
  These expanded forms do not include $\sigma_{{\cal R},a}$ originated from the expanded form of $\rho_\pm$ in the NP transmonomials.
  If one expands $\zeta_\pm$ with respect to $\sigma_{{\cal R},a}$, then a $\log$-type transmonomial appears from $\zeta_\pm$.
  This point is somewhat relevant to construction of the resurgent relation, which we would discuss in Sec.\ref{sec:resurgence}.
}
\be
\beta_{\rm np} &\sim& \sigma_\beta w^{2} \zeta_+ \left[ 5.5706 \theta_0 w^{-1} \right. \nl
  && \left. +  \left\{ 136.4337 -24.7230 \sigma_{{\cal R},a} + \left( -83.5586 + 12.9248 \sigma_{{\cal R},a} \right) (\sigma_\beta \theta_0^2)^{-1} \right\} \theta_0^2 w^{-2} \right. \nl
  && \left. +  \left\{ 1311.2652 - 503.7184 \sigma_{{\cal R},a} + \left( -1690.3866 + 496.7208 \sigma_{{\cal R},a} \right) (\sigma_\beta \theta_0^2)^{-1} \right. \right. \nl
  && \left. \left. + \left( 626.6897 - 193.8722 \sigma_{{\cal R},a} \right) (\sigma_\beta \theta_0^2)^{-2} \right\} \theta_0^3 w^{-3} + O(w^{-4},\sigma_{{\cal R},a}^2) \right] \nl
&& + \sigma_\beta w^{2} \zeta_- \left[ -0.4039 \theta_0 w^{-1} \right. \nl
  && \left. +  \left\{ - 27.2671 + 6.2281 \sigma_{{\cal R},a} + \left( 6.0586 - 0.9371 \sigma_{{\cal R},a} \right) (\sigma_\beta \theta_0^2)^{-1} \right\} \theta_0^2 w^{-2} \right. \nl
  && \left. +  \left\{ - 835.8880 + 401.0524 \sigma_{{\cal R},a} + \left( 345.3330 - 138.4836 \sigma_{{\cal R},a} \right) (\sigma_\beta \theta_0^2)^{-1} \right. \right. \nl
  && \left. \left. + \left( - 45.4397 + 14.0572 \sigma_{{\cal R},a} \right) (\sigma_\beta \theta_0^2)^{-2} \right\} \theta_0^3 w^{-3} + O(w^{-4},\sigma_{{\cal R},a}^2) \right] + O(\zeta_{\pm}^2), \label{eq:b_np_ex} \\ \nl
 {\cal R}_{\rm np} &\sim& \sigma_{\cal R} \zeta_+ \left[ \left( 2.7853 -0.9847 \sigma_{{\cal R},a} \right) \theta_0 w^{-1} \right. \nl
  && \left. +  \left\{ 57.0757 - 37.4645 \sigma_{{\cal R},a} + \left( -41.7793 + 21.2336 \sigma_{{\cal R},a} \right) (\sigma_\beta \theta_0^2)^{-1} \right\} \theta_0^2 w^{-2} \right. \nl
  && \left. +  \left\{ 442.1846 - 457.0197 \sigma_{{\cal R},a} + \left( -683.2985 + 535.8722 \sigma_{{\cal R},a} \right) (\sigma_\beta \theta_0^2)^{-1} \right. \right. \nl
  && \left. \left. + \left( 313.3448 - 207.7202 \sigma_{{\cal R},a} \right) (\sigma_\beta \theta_0^2)^{-2} \right\} \theta_0^3 w^{-3} + O(w^{-4},\sigma_{{\cal R},a}^2) \right] \nl
&& + \sigma_{\cal R} \zeta_- \left[ \left( -0.2020 + 0.0714 \sigma_{{\cal R},a} \right) \theta_0 w^{-1} \right. \nl
  && \left. +  \left\{ - 12.8257 + 8.0056 \sigma_{{\cal R},a} + \left( 3.0293 - 1.5396 \sigma_{{\cal R},a} \right) (\sigma_\beta \theta_0^2)^{-1} \right\} \theta_0^2 w^{-2} \right. \nl
  && \left. +  \left\{ - 367.7182 + 340.5605 \sigma_{{\cal R},a} + \left( 160.9279 - 129.4686 \sigma_{{\cal R},a} \right) (\sigma_\beta \theta_0^2)^{-1} \right. \right. \nl
   && \left. \left. + \left( - 22.7198 + 15.0613 \sigma_{{\cal R},a} \right) (\sigma_\beta \theta_0^2)^{-2} \right\} \theta_0^3 w^{-3} + O(w^{-4},\sigma_{{\cal R},a}^2) \right] + O(\zeta_{\pm}^2), \label{eq:R_np_ex} \\ \nl
 \bar{\Pi}_{\rm np} &\sim&  \zeta_+ \left[ -0.8569  \right. \nl
  && \left. +  \left\{ -20.1776 + 2.6306 \sigma_{{\cal R},a} + \left( 12.8529 - 1.9881 \sigma_{{\cal R},a} \right) (\sigma_\beta \theta_0^2)^{-1} \right\} \theta_0 w^{-1} \right. \nl
  && \left. +  \left\{ -184.3400 + 49.5267 \sigma_{{\cal R},a} + \left( 250.3210 - 57.3032 \sigma_{{\cal R},a} \right) (\sigma_\beta \theta_0^2)^{-1} \right. \right. \nl
  && \left. \left. + \left( -96.3966 + 29.8212 \sigma_{{\cal R},a} \right) (\sigma_\beta \theta_0^2)^{-2} \right\} \theta_0^2 w^{-2} + O(w^{-3},\sigma_{{\cal R},a}^2) \right] \nl
&& +  \zeta_- \left[ 1.1346  \right. \nl
  && \left. +  \left\{  70.1035 - 15.8039 \sigma_{{\cal R},a} + \left( - 17.0195 + 2.6326 \sigma_{{\cal R},a} \right) (\sigma_\beta \theta_0^2)^{-1} \right\} \theta_0 w^{-1} \right. \nl
  && \left. +  \left\{  1944.4395 - 928.1535 \sigma_{{\cal R},a} + \left( - 918.5201 + 353.3727 \sigma_{{\cal R},a} \right) (\sigma_\beta \theta_0^2)^{-1} \right. \right. \nl
   && \left. \left. + \left( 127.6466 - 39.4886 \sigma_{{\cal R},a} \right) (\sigma_\beta \theta_0^2)^{-2} \right\} \theta_0^2 w^{-2} + O(w^{-3},\sigma_{{\cal R},a}^2) \right] + O(\zeta_{\pm}^2), \\ \nl
 \bar{\pi}_{\rm np} &\sim&  \zeta_+ \left[ 1.0000  \right. \nl
  && \left. +  \left\{ 23.5483 - 2.9844 \sigma_{{\cal R},a} + \left( - 15.0000 + 2.3202 \sigma_{{\cal R},a} \right) (\sigma_\beta \theta_0^2)^{-1} \right\} \theta_0 w^{-1} \right. \nl
  && \left. +  \left\{ 214.8146 - 55.9638 \sigma_{{\cal R},a} + \left( - 291.5030 + 65.5335 \sigma_{{\cal R},a} \right) (\sigma_\beta \theta_0^2)^{-1} \right. \right. \nl
  && \left. \left. + \left( 112.5000 - 34.8029 \sigma_{{\cal R},a} \right) (\sigma_\beta \theta_0^2)^{-2} \right\} \theta_0^2 w^{-2} + O(w^{-3},\sigma_{{\cal R},a}^2) \right] \nl
&& +  \zeta_- \left[ 1.0000  \right. \nl
  && \left. +  \left\{  61.7850 - 13.9452 \sigma_{{\cal R},a} + \left( - 15.0000 + 2.3202 \sigma_{{\cal R},a} \right) (\sigma_\beta \theta_0^2)^{-1} \right\} \theta_0 w^{-1} \right. \nl
  && \left. +  \left\{  1713.8204 - 817.5552 \sigma_{{\cal R},a} + \left( - 814.6306 + 311.8033 \sigma_{{\cal R},a} \right) (\sigma_\beta \theta_0^2)^{-1} \right. \right. \nl
  && \left. \left. + \left( 112.5000 - 34.8029 \sigma_{{\cal R},a} \right) (\sigma_\beta \theta_0^2)^{-2} \right\} \theta_0^2 w^{-2} + O(w^{-3},\sigma_{{\cal R},a}^2) \right] + O(\zeta_{\pm}^2). \label{eq:bp_np_ex}
\ee
Here, we took $a_{\bar{\pi}}^{[(1,0),k=0]} = a_{\bar{\pi}}^{[(0,1),k=0]} = 1$ for the normalization of $\sigma_\pm$.
The expanded form of the coefficients with respect to $(\sigma_\beta \theta_0^2)^{-1}$ is found as
\be
&& a^{[{\bf n} \ne {\bf 0} ,k]}_{\cal O}[\sigma_\beta,\sigma_{{\cal R}, a},\theta_0]
= \sum_{s=0}^{s_{\max}} a^{[{\bf n},k]}_{{\cal O}(s)} [\sigma_{{\cal R}, a},\theta_0] (\sigma_\beta \theta_0^2 )^{-s} \qquad (a^{[{\bf n},k]}_{{\cal O}(s)}[\sigma_{{\cal R}, a}, \theta_0] \in {\mathbb R}) \nl
&& \mbox{with} \quad s_{\max} =
\begin{cases}
  k-1 & \mbox{for ${\cal O} \in \{ \beta, {\cal R} \}$} \\
  k  & \mbox{for ${\cal O} \in \{ \bar{\Pi}, \bar{\pi} \}$} 
\end{cases}. \label{eq:a_para_deps_np0}
\ee
In addition, same as the PT sector, $a^{[{\bf n},k]}_{{\cal O}(s)}[\sigma_{{\cal R},a},\theta_0]$ was defined such that $a^{[{\bf n},k]}_{{\cal O}(s)}[\sigma_{{\cal R},a},\theta_0] \propto \theta_0^k$,
{and expanding it around $\sigma_{{\cal R},a}=0$ gives a formal series expansion.}
When taking $\theta_0 \rightarrow 0_+$, coefficients in the NP sector generally becomes divergent, but ${\cal O}_{\rm np}$ vanishes because of the exponential decay in $\zeta_\pm$, i.e. $e^{-S_\pm w/\theta_0}$ with $S_\pm=3/\theta_0$.
Taking the low temperature (or heavy mass) limit gives the similar form to the case of the PT sector in Eq.(\ref{eq:m_inf_pt}), as
\be
&& \lim_{\sigma_\beta \rightarrow +\infty} a^{[{\bf n} \ne {\bf 0},k]}_{\cal O}[\sigma_\beta,\sigma_{{\cal R}, a},\theta_0] = a^{[{\bf n} \ne {\bf 0},k]}_{{\cal O}(s=0)}[\sigma_{{\cal R}, a},\theta_0] \propto \theta_{0}^k. \label{eq:m_inf_np}
\ee

Before closing the part of transseries construction, we would make comments on the low temperature (heavy mass) limit, $\sigma_\beta \rightarrow +\infty$.
As we can see in App.\ref{sec:asy_exp}, when taking this limit, the transport coefficients, $\chi_{\beta}$, and $ z \chi_{\beta} + \chi_{\alpha}$ reduce to functions of only ${\cal R}_a$.
Since the heavy mass limit does not change symmetry imposed to the equilibrium, this limit to the formal transseries can be taken without discontinuity happening.
However, there exists symmetry emerged by this limit, that is scale invariance in the dynamical system.
This heavy mass limit makes the ODEs given by Eqs.(\ref{eq:ODE_b0})-(\ref{eq:ODE_bp0}) using $\tau_R=\theta_0 \beta$ to be invariant under the scale transformation defined as $(\beta, \bar{\Pi},\bar{\pi},{\cal R},\tau) \rightarrow (\lambda^{-\Delta_\beta }\beta, \lambda^{-\Delta_{\bar{\Pi}}} \bar{\Pi},\lambda^{-\Delta_{\bar{\pi}}} \bar{\pi},\lambda^{-\Delta_{\cal R}} {\cal R}, \lambda \tau)$ with $\lambda \in {\mathbb R}$, $\Delta_{\beta} = -1$, and $\Delta_{\bar{\Pi}} = \Delta_{\bar{\pi}} = \Delta_{\cal R} = 0$\footnote{
  If one uses $\alpha$ instead of ${\cal R}$, D-symmetry can not be seen apparently.
}.
This symmetry is called as ``D-scale symmetry'' in Ref.~\cite{Kamata:2022jrc}.
The interesting fact in our case is that D-symmetry appears in both the massless and heavy mass limits in contrast to the case of ${\cal D}_\mu N^\mu \ne 0$ studied in Ref.~\cite{Kamata:2022jrc}.
When ODEs have D-symmetry, one can drop or decouple one degree of freedom from the other ODEs by using an appropriate scale invariant quantity as a flow time.
Moreover, since $w = \theta_0 \tau/\tau_R = \tau/\beta$ is D-scale invariant, $1/w$-expansion is essentially the same as the CE expansion in the similar to the conformal Bjorken flow using the same flow time.
Indeed, this situation is realized in Eqs.(\ref{eq:m_inf_pt})(\ref{eq:m_inf_np}).
Oppositely speaking, it can be also said that, for IR transseries of the nonconformal cases, taking care of orders deviating from $(\theta_0/w)^k$, i.e. $a^{[{\bf n},k]}_{{\cal O}(s>0)}$, is significant to appropriately count contributions from particle kinetic energy.

\subsection{Transseries of other variables} \label{sec:trans_others}
Transseries of other variables which are functions of $(\beta,{\cal R},\bar{\Pi},\bar{\pi})$ can be obtained from the transseries solutions of those.
Such a variable can be decomposed into PF, PT, and NP sectors in the similar way to the case of $(\beta,{\cal R},\bar{\Pi},\bar{\pi})$, e.g. by taking $\theta_0 \rightarrow 0_+$ for the identification of the PF sector.
We briefly describe formal transseries of  $({\cal E}_0, P, n_0)$ given by Eq.(\ref{eq:E0Pn0_1}).
Additionally, we also look at transverse and longitudinal pressures, $(P_{\perp}, P_{\parallel})$, defined as
\be
&& P_{\perp} :=  P +  \Pi + \frac{\widehat{\pi}}{2} = P + \left( \bar{\Pi} + \frac{\bar{\pi}}{2} \right)/\gamma_\beta, \\
&& P_{\parallel} :=  P +  \Pi -  \widehat{\pi} = P +  \left( \bar{\Pi} -  \bar{\pi} \right)/\gamma_\beta.
\ee

The formal transseries expanded around $w = +\infty$ is given by
\be
&& {\cal E}_0 \sim \frac{\sigma_{\cal R}}{\sigma_\beta^{3/2}} w^{-3} \sum_{{\bf n} \in {\mathbb N}_0^2} \sum_{k \in {\mathbb N}_0} a_{{\cal E}_0}^{[{\bf n},k]} \bm{\zeta}^{\bf n} w^{-k}, \qquad  P \sim \frac{\sigma_{\cal R}}{\sigma_\beta^{5/2}} w^{-5} \sum_{{\bf n} \in {\mathbb N}_0^2} \sum_{k \in {\mathbb N}_0} a_{P}^{[{\bf n},k]} \bm{\zeta}^{\bf n} w^{-k}, \nl
&& n_0 \sim \frac{\sigma_{\cal R}}{\sigma_\beta^{3/2}} w^{-3} \sum_{{\bf n} \in {\mathbb N}_0^2} \sum_{k \in {\mathbb N}_0} a_{n_0}^{[{\bf n},k]} \bm{\zeta}^{\bf n} w^{-k}, \\
&& P_{\perp} \sim \frac{\sigma_{\cal R}}{\sigma_\beta^{5/2}} w^{-5} \sum_{{\bf n} \in {\mathbb N}_0^2} \sum_{k \in {\mathbb N}_0} a_{P_\perp}^{[{\bf n},k]} \bm{\zeta}^{\bf n} w^{-k}, \qquad  P_{\parallel} \sim \frac{\sigma_{\cal R}}{\sigma_\beta^{5/2}} w^{-5} \sum_{{\bf n} \in {\mathbb N}_0^2} \sum_{k \in {\mathbb N}_0} a_{P_{\parallel}}^{[{\bf n},k]} \bm{\zeta}^{\bf n} w^{-k}. \nn
\ee
We summarize the explicit values of coefficients in the first some leading orders 
in App.\ref{sec:trans_obs}.
All of these coefficients depend on $(\sigma_\beta, \sigma_{{\cal R},a}, \theta_0)$ and are polynomials of $(\sigma_\beta \theta_0^2)^{-1}$ with the similar form to $(\beta, {\cal R})$ in Eqs.(\ref{eq:a_para_deps0})(\ref{eq:a_para_deps_np0}).
For example, the PT sector of $({\cal E}_0, P, n_0)$ is given by
\be
   {\cal E}_0 &\sim& \frac{\sigma_{\cal R}}{\sigma_\beta^{3/2}} w^{-3}  \left[ 0.0635 + 0.0224 \sigma_{{\cal R},a}  \right. \nl
  &&   \left. + \left( - 0.5079 - 0.0449 \sigma_{{\cal R},a} \right) \theta_0 w^{-1} + O(w^{-2}, \sigma_{{\cal R},a}^2) \right], \\ \nl
   P &\sim& \frac{\sigma_{\cal R}}{\sigma_\beta^{5/2}} w^{-5} \left[ 0.0635 + 0.0112 \sigma_{{\cal R},a} \right. \nl
   && \left. + \left(-1.0159 + 0.0449 \sigma_{{\cal R},a} \right) \theta_0 w^{-1} + O(w^{-2}, \sigma_{{\cal R},a}^2) \right], \\ \nl
   n_0 &\sim& \frac{\sigma_{\cal R}}{\sigma_\beta^{3/2}} w^{-3} \left[ 0.0635 + 0.0224 \sigma_{{\cal R},a} \right. \nl
   && \left. + \left( -0.5079 - 0.0449 \sigma_{{\cal R},a} \right) \theta_0 w^{-1} + O(w^{-2}, \sigma_{{\cal R},a}^2) \right], 
\ee
and the $\theta_0$-dependence enters into the coefficients from the next leading order.
As opposed to $\beta_{\rm pf}$ and ${\cal R}_{\rm pf}$, the leading order contains $\sigma_{{\cal R},a}$, so that the large order behavior around the equilibrium depends on the particle statistics, MB, FD, or BE.
In addition, the particle mass dependence appears as their overall factor by turning on it as $\sigma_\beta \rightarrow m \sigma_\beta $.
In the similar way, $(P_\perp,P_\parallel)$ are obtained as
\be
   P_{\perp} &\sim& \frac{\sigma_{\cal R}}{\sigma_\beta^{5/2}} w^{-5} \left[ 0.0635 + 0.0112 \sigma_{{\cal R},a} \right. \nl
     && \left. + \left(-0.9736 + 0.0524 \sigma_{{\cal R},a} \right) \theta_0 w^{-1} + O(w^{-2}, \sigma_{{\cal R},a}^2) \right], \\ \nl
   P_{\parallel} &\sim& \frac{\sigma_{\cal R}}{\sigma_\beta^{5/2}} w^{-5} \left[ 0.0635 + 0.0112 \sigma_{{\cal R},a} \right. \nl
   && \left. + \left(-1.1006 + 0.0299 \sigma_{{\cal R},a} \right) \theta_0 w^{-1} + O(w^{-2}, \sigma_{{\cal R},a}^2) \right].
\ee
Notice that, around the equilibrium, the anisotropy correctly vanishes as $P_\perp \sim P_\parallel \, \propto \, w^{-5} = (\sigma_\beta^{-1} \tau)^{-5/3}$.

\section{Resurgence analysis} \label{sec:Resurgence_analysis}
We consider resurgence for the nonconformal Bjorken flow.
In Sec.\ref{sec:asymptotic}, we firstly consider the large order behavior and see Borel summability of fundamental variables in our model, which is helpful to construct the resurgent relation\footnote{
{In our paper, the resurgent relation means a relation among coefficients of the PT and NP sectors, that is also called \textit{the large order relation}.}
}.
Then, in Sec.\ref{sec:resurgence}, we construct the resurgent relation.
We summarize the review of Borel resummation and the technical details for construction of the resurgent relation in Apps.\ref{app:borel_resum} and \ref{app:resurgence}, respectively.
Definitions of symbols related to Borel resummation is also summarized in App.\ref{app:borel_resum}.
For technical simplicity, we use $T=1/\beta$ defined as
\be
&& T \sim \sigma_\beta^{-1} w^{-2} \sum_{{\bf n} \in {\mathbb N}_0^2} \sum_{k \in {\mathbb N}_0} a_{T}^{[{\bf n},k]} \bm{\zeta}^{\bf n} w^{-k}, \label{eq:T_trans}
\ee
in place of $\beta$\footnote{
  One can perform the same analysis described below by using $\beta$.
}.

\subsection{Large order behavior} \label{sec:asymptotic}
Before considering resurgence, we firstly see the large order behavior of the PT sector and remind how it relates to the first NP sector.
In our analysis, instead of dealing with $(\bar{\Pi},\bar{\pi})$, we consider $\widetilde{X}_\pm$ defined as $(\widetilde{X}_+,\widetilde{X}_-)^{\top} = U (\bar{\Pi},\bar{\pi})^{\top}$, where $U$ is a constant matrix given by Eq.(\ref{eq:U_Uinv}).
It is because the linearized equation of $\widetilde{X}_\pm$ gives the NP transmonomials, $\zeta_\pm$, as two independent solutions, so that it can be expected that a relation of the large order behavior of the PT sector to the NP sector can be more clearly seen than keeping $(\bar{\Pi},\bar{\pi})$.
Additionally, as we saw in Sec.\ref{sec:trans_IRdomain}, the coefficients are polynomials of $(\sigma_\beta \theta_0^2)^{-1}$.
For now, we expand the coefficients with respect to $(\sigma_\beta \theta_0^2)^{-1}$ such as Eq.(\ref{eq:a_para_deps0}) and consider the spanned formal series expansion labeled by a fixed order of $(\sigma_\beta \theta_0^2)^{-1}$ defined as
\be
\widetilde{X}_{\pm(s)}:= \sum_{k \in {\mathbb N0}} a^{[{\bf 0},k]}_{\widetilde{X}_\pm (s)}[\sigma_{{\cal R}, a}, \theta_0]w^{-k}. \label{eq:X_s}
\ee
Thus, $\widetilde{X}_{\pm} = \sum_{s \in {\mathbb N}_0} \widetilde{X}_{\pm(s)} (\sigma_{\beta} \theta_0^2)^{-s}$.
It is notable that, since $\widetilde{X}_\pm$ is a dimensionless variable, the particle mass appear as coupling only to $\sigma_\beta$ in the transseries when explicitly turning on $m$.
In this sense, it would be worth to take the form (\ref{eq:X_s}) to consider the mass dependence in the resurgent relation.
We write down the form of transseries of $\widetilde{X}_\pm$ in App.\ref{sec:trans_XX}.

Let us firstly remind the relation between the large order behavior of the PT sector and the first NP sector by taking the following ansatz to the large $k$-th order of $a^{[{\bf 0},k]}_{\widetilde{X}_{\pm}(s)}$:
\be
\frac{ a^{[{\bf 0},k+1]}_{\widetilde{X}_\pm(s)}}{a^{[{\bf 0},k]}_{\widetilde{X}_\pm(s)}}
\simeq  b_{\pm(s)} k + c_{\pm(s)},  \qquad (k \gg 1) \label{eq:a_rad}
\ee
where $b_{\pm(s)} \in {\mathbb R}_+$ and $c_{\pm(s)} \in {\mathbb R}$.
The positive $b_{\pm(s)}$ guarantees a positive or negative series, that means that $\widetilde{X}_{\pm(s)}$ is Borel nonsummable.
If this ansatz really works, then a coefficient of the higher order can be approximated as
\be
a^{[{\bf 0},k]}_{\widetilde{X}_\pm(s)} &\simeq& M_{\pm(s)} b_{\pm(s)}^{k-1} (c_{\pm(s)}/b_{\pm(s)} + 1)_{k-1} \qquad (k \gg 1) \nl
&=& M_{\pm(s)} b_{\pm(s)}^{k-1} \frac{\Gamma(k + c_{\pm(s)}/b_{\pm(s)})}{\Gamma(1 + c_{\pm(s)}/b_{\pm(s)})} \label{eq:a_approx}
\ee
with  $M_{\pm(s)} \in {\mathbb R}$, where $(a)_n$ is the Pochhammer symbol.
Since acting the Borel transform to $\widetilde{X}_{\pm(s)}$ yields
\be
&& {\cal B}[\widetilde{X}_{\pm(s)}]  = 
\sum_{k \in {\mathbb N}} \frac{a^{[{\bf 0},k]}_{\widetilde{X}_\pm(s)}}{\Gamma(k)} \xi^{k-1} \simeq \frac{M_{\pm(s)}}{\left( 1 - b_{\pm(s)} \xi \right)^{c_{\pm(s)}/b_{\pm(s)}+1}},  \qquad (k \gg 1)
\ee
the leading order of discontinuity on the real positive axis on the Borel plane is found as
\be
\left( {\cal S}_{0_+} - {\cal S}_{0_-} \right)[\widetilde{X}_{\pm(s)}]
&\approxprop&
\frac{e^{-w/b_{\pm(s)}}}{w^{-c_{\pm(s)}/b_{\pm(s)}}}. \qquad (k \gg 1) \label{eq:S0p_S0m_X}
\ee
Eq.(\ref{eq:S0p_S0m_X}) means that the large order behavior (\ref{eq:a_rad}) directly relates to the transmonomial in the NP sector and suggests that the $b_{\pm (s)}$ and $c_{\pm (s)}$ can be expressed by using $S$ and $\rho_\pm$ given by Eq.(\ref{eq:lam_rho0}).

\begin{figure}[tp]
  \begin{center}
    \begin{tabular}{cc}
      \begin{minipage}{0.5\hsize}
        \begin{center} 
          \includegraphics[clip, width=67.5mm]{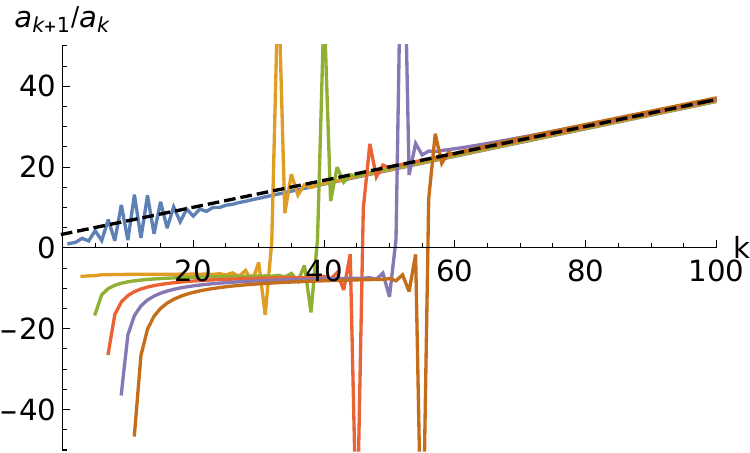}
        \end{center}
        \subcaption {$\widetilde{X}_+$}
      \end{minipage} 
      \begin{minipage}{0.5\hsize}
        \begin{center} 
          \includegraphics[clip, width=80mm]{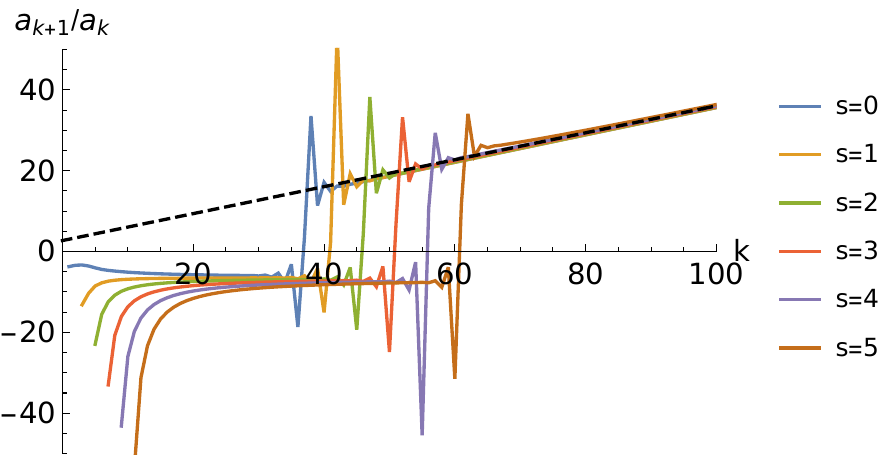}
        \end{center}
        \subcaption {$\widetilde{X}_-$}
      \end{minipage} \\
      \begin{minipage}{0.5\hsize}
        \begin{center} 
          \includegraphics[clip, width=67.5mm]{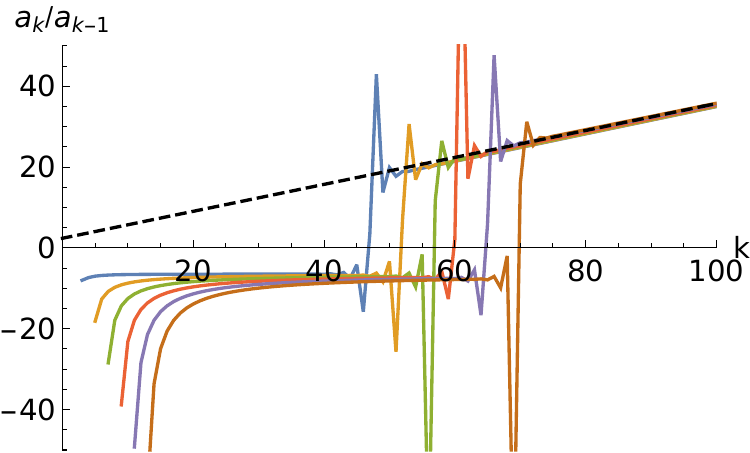}
        \end{center}
        \subcaption {$T$}
      \end{minipage} 
      \begin{minipage}{0.5\hsize}
        \begin{center} 
          \includegraphics[clip, width=80mm]{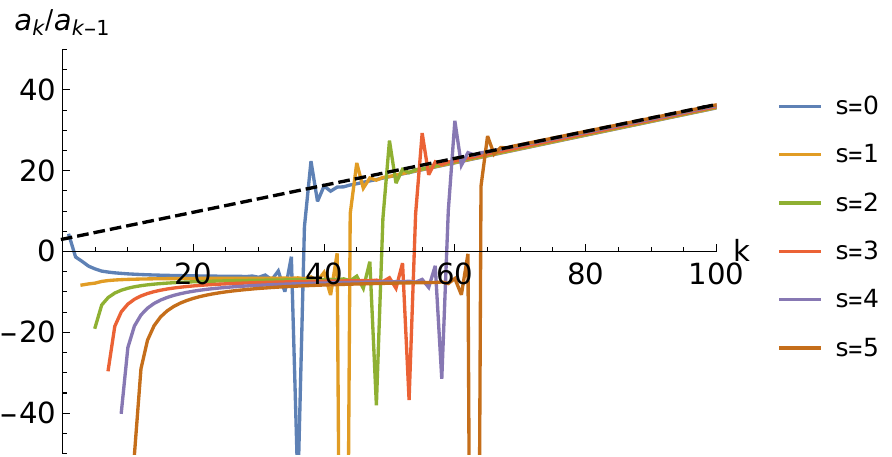}
        \end{center}
        \subcaption {${\cal R}$}
      \end{minipage}      
    \end{tabular} 
    \caption{The ratio, $a^{[{\bf 0},k+1]}_{{\cal O}(s)}/a^{[{\bf 0},k]}_{{\cal O}(s)}$ for ${\cal O} = \widetilde{X}_+$, $\widetilde{X}_-$, $T$, and ${\cal R}$ evaluated from the MB statistics ($a = 0$).
      The colored lines are the values for $s=0, \cdots, 5$.
      The dashed black line denotes the asymptotic line defined in Eq.(\ref{eq:a_rad}) with $b_{\pm(s)}=1/S = \theta_0/3$, $c_{+(s)}= -\rho_+/S \approx 3.390 \cdot \theta_0$, and $c_{-(s)} = - (\rho_+ + 2)/S \approx 2.723 \cdot \theta_0$ for all $s$.
      The relaxation scale is taken as $\theta_0=1$.
      The ratio for any $s$ behaves as the positive linear divergence, which is the typical feature of Borel nonsummable.
      $T$ and ${\cal R}$ also have the similar behavior.
      The asymptotic lines for $T$ and ${\cal R}$ are found by setting $b_{T,{\cal R}(s)}=1/S$, $c_{T(s)}=-(\rho_++3)/S \approx 2.390 \cdot \theta_0$, and $c_{{\cal R}(s)}=-(\rho_++1)/S \approx  3.057 \cdot \theta_0$.     
    }
    \label{fig:cofrad}
  \end{center}
\end{figure}

The question is if one can extract a value of $(b_{\pm(s)},c_{\pm(s)})$ which is consistent with the result of NP sectors from coefficients of the PT sector.
In order to see it, we plot the ratio of the coefficients given in Eq.(\ref{eq:a_rad}) for $s=0,\cdots,5$ evaluated from the MB statistics ($a=0$) in Fig.~\ref{fig:cofrad}(a)(b).
One can observe that $\widetilde{X}_{\pm(s)}$ is in general an alternating series in the lower order of $k$ because the ratio is negative, but it becomes a positive or negative series in the higher order.
We also estimate $(b_{\pm(s)},c_{\pm(s)})$ for the asymptotic lines from these plots, and the results are consistent with that $b_{\pm (s)}=1/S = \theta_0/3$, $c_{+ (s)} = -\rho_+/S \approx 3.390 \cdot \theta_0$, and $c_{- (s)} = -(\rho_++2)/S \approx 2.723 \cdot \theta_0$ for any $s \in {\mathbb N}_0$.
We also plot the same quantity of $(T,{\cal R})$ in Fig.~\ref{fig:cofrad}(c)(d).
One can observe that they have the similar behavior to $\widetilde{X}_\pm$, i.e. divergent series and Borel nonsummable.
The numerical fitting using the same function to Eq.(\ref{eq:a_rad}) shows that $b_{T,{\cal R}(s)}=1/S$, $c_{T(s)} = -(\rho_+ + 3)/S \approx 2.390 \cdot \theta_0$, and $c_{{\cal R}(s)} = -(\rho_++1)/S\approx 3.057 \cdot \theta_0$.
In summary, we found the following observations:
\begin{enumerate}[label=(\Roman*)]
\item According to Eqs.(\ref{eq:Xpnp_trans})(\ref{eq:Xmnp_trans}), the lowest order of $(\sigma_\beta \theta_0^2)^{-1}$ in the NP sectors depend on the order of $w^{-1}$.
  However, the fitting result using Eq.(\ref{eq:a_rad}) is independent on $s$.
\item The ratio of coefficients of all of $(T,{\cal R}, \widetilde{X}_\pm)$ have the similar large order behavior to each other,  such as exactly the same gradient, roughly the same intercept, and Borel nonsummable.
\item  $c_{-(s)}$ seems to be irrelevant to the value of $\rho_-$ or extremely far from it.
\end{enumerate}
The observation (III) can be easily solved by looking to the dominant part of $\widetilde{X}_{-{\rm np}}$.
According to Eq.(\ref{eq:Xmnp_trans}) and the value of $\rho_\pm$ in Eq.(\ref{eq:lam_rho0}), the lowest order of $\widetilde{X}_{-{\rm np}}$ is proportional to $\zeta_+/w^2$, and thus $(\rho_++2)$ in the result of $c_{-(s)}$ comes from this exponent of $w$.

The observations (I)(II) are more nontrivial and should be closely related to structure of the ODEs determining the resurgent structure.
As we can see later, however, both of them can be actually solved by focusing on the Borel transformed ODEs and introducing Stokes constants appropriately.
We will discuss these issues in detail in Sec.\ref{sec:resurgence}.

\subsection{Construction of the resurgent relation} \label{sec:resurgence}

In this section, we construct the resurgent relation{, i.e. the large order relation}.
In general, the form of resurgent relation is closely related to forms of given ODEs and structure of the Borel plane of Borel transformed variables.
We normally employ some methods such as Pade approximant to predict the resurgent structure, but those methods are not always helpful for the structure when a given problem is not so simple, e.g. multi-variables ODEs consisting of different types\footnote{
In general, existence of singularities on the real positive axis in the Borel plane is \textit{not} a sufficient condition to say that the theory contains nonperturbative effects.
In such a case, Borel resummed functions (or transseries) obtained by ${\cal S}_{0_\pm}$ include nonperturbative effects, but the effects vanish by eliminating discontinuity using the median resummation, ${\cal S}_{\rm med} := S_{0_+} \circ {\frak S}^{-1/2} = S_{0_-} \circ {\frak S}^{+1/2}$.
This special situation can be indeed observed in a quantum mechanics.
See Ref.~\cite{Kamata:2021jrs}, for example.
}.
Roughly, our problem consists of two types of ODE:
$\frac{d Y}{d w} = -S Y + \frac{1}{w} \left[ a +b Y \right] + O(Y^2, w^{-2})$ and $\frac{d Y}{d w} = \frac{Y}{w}  \left[ a + b Y \right] + O(Y^3 w^{-1},Y w^{-2})$.
For this reason, in our analysis we would take the following procedure: (1) identifying the number of Stokes constants by looking to the Borel transformed ODEs, (2) forming a conjecture of the resurgent relation which is consistent with the observations (I)(II) in Sec.\ref{sec:asymptotic}, and (3) numerically checking the resurgent relation.

Firstly, we consider the Borel transformed ODEs.
We define the Borel transform as ${\cal O}_B:= {\cal B}[{\cal O}]$.
Action of it to the ODEs (\ref{eq:ODE_b_w})-(\ref{eq:ODE_bp_w}) yields\footnote{
  In these equations, we used the simplified notation for the Landau symbol $O(\bullet)$ to omit the convolution with ${\cal R}_B$, such that  $O({\cal R}_B^{*n} * {\cal O}_B) \rightarrow O({\cal O}_B)$, because ${\cal R}_B = \sigma_{\cal R} \cdot {\rm id} + O({\mathbb I}_\xi)$ and the order of $\xi$ in $\widehat{\cal O}$ does not change by the convolution with ${\cal R}_B$.
}
\be
- \xi T_B &=& - 2 {\mathbb I}_\xi * T_B * \left[ {\rm id}  + O(T_B,\widetilde{X}_{\pm B} ) \right], \label{eq:ODE_T_w_B} \\
- \xi {\cal R}_B &=& - \frac{1}{8} {\mathbb I}_\xi * {\cal R}_B * \left[ \left( {\rm id} +  \frac{a}{2 \sqrt{2}} {\cal R}_B \right)  \right. \nl
  && \left. * \left\{ \left( 31 - \sqrt{1285}\right) \widetilde{X}_{+ B}  + \left(31 + \sqrt{1285}\right) \widetilde{X}_{- B} \right\} + O(T_B,\widetilde{X}_{\pm B}^{*2}) \right], \label{eq:ODE_R_w_B} \\
- \xi \widetilde{X}_{\pm B} &=& - S \widetilde{X}_{\pm B} + {\mathbb I}_\xi * \left[ \frac{4}{3} \left( 1 \mp \frac{\sqrt{1285}}{257} \right) \cdot {\rm id} + O(T_B, \widetilde{X}_{\pm B}) \right] + O(T_B,\widetilde{X}_{\pm B}^{*2}), \label{eq:ODE_X_w_B} 
\ee
where ${\cal B}[\frac{d {\cal O}}{d w}] = -\xi {\cal O}_B$, ${\mathbb I}_\xi:= {\cal B}[w^{-1}]$, $*$ denotes the convolution defined as $f_B * g_B(\xi):= \int_{0}^{\xi} d \xi^\prime \, f_B(\xi^\prime) g_B(\xi-\xi^\prime)$, and ${\cal O}_B^{*2} := {\cal O}_B * {\cal O}_B$.
In addition, ``${\rm id}$'' is the identity operator under the convolution\footnote{
The identity operator is normally defined as ${\rm id} :={\cal B}[1]$.
Additionally, we introduced ${\mathbb I}_\xi$ in order to explicitly avoid confusion which a number $c$ is a constant or $c \cdot {\cal B}[ w^{-1}]$.
In this notation, if $Y(w) = \sum_{k \in {\mathbb N}_0}  a_{Y}^{[k]} w^{-k}$, then its Borel transformation is denoted as ${\cal B}[Y](\xi)= a_Y^{[0]}  \cdot {\rm id} + a_Y^{[1]} {\mathbb I}_\xi + \sum_{k \in {\mathbb N}+1} \frac{a_Y^{[k]}}{\Gamma(k)} \xi^{k-1}$.},
i.e. ${\rm id} * {\cal O}_B = {\cal O}_B * {\rm id} = {\cal O}_B$.
The Borel transformed ODEs (\ref{eq:ODE_T_w_B})-(\ref{eq:ODE_X_w_B}) tell us several informations of the mechanism of the resurgence of our ODEs.
First,  $\widetilde{X}_{\pm B}$ has a singularity at $\xi = S$.
It can be easily seen by taking Eq.(\ref{eq:ODE_X_w_B}) the form that $\widetilde{X}_{\pm B} = (S-\xi)^{-1} \cdot \left[ \cdots \right]$.
Due to the singularity in $\widetilde{X}_{\pm B}$, quadratic terms of $\widetilde{X}_{\pm B}$ generate a new singularity at $\xi=2 S$\footnote{
Intuitively, it can be found in the following way:  
Consider $f_B(\xi) = \frac{\xi^s}{S-\xi}$ and $g_B(\xi) = \frac{\xi^t}{S-\xi}$ with $s,t \in {\mathbb N}_0$.
Then, $f_B * g_B = h(\xi)  \frac{\log (1-\xi/S )}{2S - \xi} + ({\rm reg}.)$ with a regular function $h(\xi)$ depending on $s,t$ at $\xi = 2 S$.
Thus, a new singularity appears at $\xi = 2 S$.
}.
By repeating the similar computation, one can find that singularities exist at $\xi = n S$ with $n \in {\mathbb N}$.
Second, the PF sector of $(T,{\cal R})$ is Borel summable (convergent series).
It can be shown from Eqs.(\ref{eq:ODE_T_w_B})(\ref{eq:ODE_R_w_B}) by setting $\widetilde{X}_{\pm B} = 0$ and considering if all terms contain the same type of singularities such as the location and the order\footnote{
One can find the observation in the following way: 
Eq.(\ref{eq:ODE_T_w_B}) can be written by $T_B =  2 \xi^{-1} {\mathbb I}_\xi * T_B * \left[ {\rm id} + O(T_B) \right]$.
Assume that $T_B$ has a term such as $\xi (1-\xi/A)^{-c}$ with real positive constants, $A$ and $c$, and that it does not have other type of singularities.
By easy computations, one can find the fact that $\xi^{-1} {\mathbb I}_\xi * T_B$ generates the $\log$-singularity at $\xi = A$, which contradicts to our assumption.
}.   
The regularity of ${\cal R}_B$ can be shown in the similar way.
Third, for the similar reason to the first, i.e. because of nonlinear terms with $\widetilde{X}_{\pm B}$, singularities of $(T_B,{\cal R}_B)$ appear at the same locations, $\xi = n S$ with $n \in {\mathbb N}$.
In summary, these considerations imply that the origin of Borel nonsummable for $(T, {\cal R},\widetilde{X}_\pm)$ comes from the singularities \textit{only} in $\widetilde{X}_{\pm B}$ and that $(T,{\cal R})$ becomes Borel nonsummable \textit{indirectly} through nonlinear terms with $\widetilde{X}_\pm$.

If this is the case, it means that the resurgence of $\widetilde{X}_\pm$ is essentially the same as that of
\be
\frac{d {\bf Y}}{d w} = - \Lambda {\bf Y} - \frac{1}{w} \left[ {\bf V} + {\frak B} {\bf Y} \right] + O({\bf Y}^2, w^{-2}), \label{eq:ODE_Y} 
\ee
where ${\bf V}$ is a constant vector, and $\Lambda$ and ${\frak B}$ are constant diagonal matrices.
The resurgence of this type of ODE has been known, for example, in Refs.~\cite{Costin2006TopologicalCO,Basar:2015ava,Aniceto:2015mto,Behtash:2019txb}, and the cancellation of discontinuity in the resulting Borel resummed function (or transseries) is obtained by introducing two Stokes constants as
\be
   {\cal S}_{0_+} [\widetilde{X}_\pm(w;\bm{\sigma} - \tfrac{1}{2}{\bf A})] = {\cal S}_{0_-} [\widetilde{X}_\pm(w;\bm{\sigma}+\tfrac{1}{2} {\bf A})], \label{eq:S0p_S0m}
\ee
where $\bm{\sigma}=(\sigma_+,\sigma_-) \in {\mathbb R}^2$ and ${\bf A} = (A_+,A_-) \in (i {\mathbb R})^2$ are vectorial forms of the integration constants and the Stokes constants, respectively.
Furthermore, by using this formula, the resurgent relation among the PT and NP sectors is obtained as 
\be
a_{\widetilde{X}_\pm}^{[{\bf 0},k]}&\simeq& \frac{1}{2 \pi i} \sum_{\substack{(n_+,n_-) \in {\mathbb N}_0^2  \\  |{\bf n}|>0}} \sum_{h \in {\mathbb N}_0} \frac{A_{+}^{n_+} A_{-}^{n_-}}{C(|{\bf n}|,n_-)} a_{\widetilde{X}_\pm}^{[{\bf n},h]}
\frac{\Gamma(k - {\bf n} \cdot \bm{\rho} -h)}{({\bf n} \cdot {\bf S})^{k-{\bf n} \cdot \bm{\rho}-h}}, \qquad (k \gg 1) \label{eq:res_rel0}
\ee
where $C(p,q)=\left( \begin{smallmatrix} p \\ q \end{smallmatrix} \right)$ is the binomial coefficient and ${\bf n}=(n_+,n_-)$.
In addition, ${\bf S}=(S_+,S_-)$ and $\bm{\rho}=(\rho_+,\rho_-)$ are defined by Eq.(\ref{eq:lam_rho0}).
The derivation is summarized in App.\ref{app:resurgence}.
If we take the similar form to Eq.(\ref{eq:ODE_Y}) for $\widetilde{X}_\pm$, then the constant parts should have the dependence of the integration constants, $(\sigma_\beta, \sigma_{\cal R})$, which also  means that the Stokes constants have the similar dependence.
From the fact that the coefficients of transseries are polynomials expanded by $(\sigma_\beta \theta_0^2)^{-1}$, we put the ansatz to the Stokes constants such that 
\be
A_\pm[\sigma_\beta, \sigma_{{\cal R}, a},\theta_0] = \sum_{s \in {\mathbb N}_0} A_{\pm (s)}[\sigma_{{\cal R}, a}, \theta_0] (\sigma_\beta \theta_0^{2})^{-s}. \qquad (A_\pm[\sigma_{{\cal R}, a}, \theta_0] \in i {\mathbb R}) \nn 
\ee 
Notice that this ansatz takes care of the observation(I) in Sec.\ref{sec:asymptotic} by supposing that the contribution from $A_{\pm(s=0)}$ is more dominant than $A_{\pm(s>0)}$.

It is straightforward to construct the resurgent relation of $(T,{\cal R})$ by using the facts that no singularity appears in their Borel transformed ODEs with $\widetilde{X}_{\pm B}=0$ and that the Borel resummation is homomorphism, i.e. ${\cal S}_{\theta}[f \cdot g] = {\cal S}_{\theta}[f] \cdot {\cal S}_{\theta}[g]$.
Since the cancellation of discontinuity (\ref{eq:S0p_S0m}) can be obtained by the shift of the integration constants which gives solutions of the same ODEs, the same equality should be satisfied for the r.h.s. of the ODEs, i.e. ${\cal S}_{0_+}[{\cal F}_{\cal O}](w;\bm{\sigma}-\tfrac{1}{2}{\bf A}) = {\cal S}_{0_-}[{\cal F}_{\cal O}](w;\bm{\sigma}+\tfrac{1}{2}{\bf A})$, where ${\cal O} \in \{T, {\cal R}\}$ and ${\cal F}_{T,{\cal R}}$ is defined in Eqs.(\ref{eq:ODE_b_w})(\ref{eq:ODE_R_w})\footnote{
${\cal F}_T$ is defined from ${\cal F}_\beta$ as ${\cal F}_T := - T^2 {\cal F}_{\beta}$.
},
which eventually gives ${\cal S}_{0_+}[{\cal O}](w;\bm{\sigma}-\tfrac{1}{2}{\bf A}) = {\cal S}_{0_-}[{\cal O}](w;\bm{\sigma}+\tfrac{1}{2}{\bf A})$ for ${\cal O} \in \{ T, {\cal R} \}$.
Therefore, the resurgent relation for $(T,{\cal R})$ has to be the same form to Eq.(\ref{eq:res_rel0}).
It is notable that this consideration is also extendable to variables which are functions of $(T,{\cal R},\widetilde{X}_\pm)$ using the property of homomorphism of the Borel resummation, i.e.
\be
{\cal S}_\theta[{\cal O}(T, {\cal R},\widetilde{X}_\pm)] = {\cal O}({\cal S}_\theta[T], {\cal S}_\theta[{\cal R}], {\cal S}_\theta[\widetilde{X}_\pm]). \label{eq:S_homomor}
\ee
In addition, the $\sigma_{{\cal R},a}$-dependence in the Stokes constants are given by the functional form of the transport coefficients.
Since in practice it is extremely hard to obtain their precise forms expanded around $\beta = +\infty$ for all orders, we would assume that they are analytic functions in a neighborhood of ${\cal R}=0$.

Let us form a conjecture by summarizing the above considerations.
We form the following conjecture of the resurgent relation which is consistent with the observations (I)(II) in Sec.\ref{sec:asymptotic}; \\ \\
\textit{
  \underline{Conjecture}: Suppose that ${\cal O}$ is a function of $(T,{\cal R},\widetilde{X}_\pm)$, and its transseries expanded around the equilibrium takes the form that
\be
   {\cal O}(w) \sim B w^{p} \sum_{k \in {\mathbb N_0}} a^{[{\bf n},k]}_{\cal O} \bm{\zeta}^{\bf n} w^{-k} \quad \mbox{as} \quad w \rightarrow +\infty, \qquad (B, p \in {\mathbb R}) \label{eq:conj3}
\ee
where B and $p$ are determined by the functional form of ${\cal O}$.
Then, ${\cal O}$ satisfies the resurgent relation given by
\be
a_{{\cal O}}^{[{\bf 0},k]}&\simeq& \frac{1}{2 \pi i} \sum_{\substack{(n_+,n_-) \in {\mathbb N}_0^2  \\  |{\bf n}|>0}} \sum_{h \in {\mathbb N}_0} \frac{A_{+}^{n_+} A_{-}^{n_-}}{C(|{\bf n}|,n_-)}  a_{\cal O}^{[{\bf n},h]}
\frac{\Gamma(k - {\bf n} \cdot \bm{\rho} -h)}{({\bf n} \cdot {\bf S})^{k-{\bf n} \cdot \bm{\rho}-h}}, \qquad (k \gg 1) \label{eq:conj1} 
\ee
where $A_\pm$ is the Stokes constant taking the form that
\be
A_\pm[\sigma_\beta, \sigma_{{\cal R}, a},\theta_0] = \sum_{s \in {\mathbb N}_0} A_{\pm (s)}[\sigma_{{\cal R}, a}, \theta_0](\sigma_\beta \theta_0^{2})^{-s}, \qquad (A_\pm[\sigma_{{\cal R}, a}, \theta_0] \in i {\mathbb R}) \label{eq:conj2} 
\ee
and $A_{\pm (s)}[\sigma_{{\cal R}, a}, \theta_0]$ is an analytic function in a neighborhood of $\sigma_{{\cal R}, a}=0$ for any $s \in {\mathbb N}_0$. \\
}
\noindent \\ \, \noindent
{
  The essential point of this conjecture is that, because of their dependence in the transseries of ${\cal O}$, the Stokes constant, $A_\pm$, is not just a constant but an analytic function of $(\sigma_\beta, \sigma_{{\cal R},a})$, i.e., Taylor expandable around $(\sigma_\beta, \sigma_{{\cal R},a}) = (+\infty, 0)$.  
Eq.(\ref{eq:conj2}) is the expanded form in terms only of $(\sigma_\beta \theta_0^2)^{-1}$ in accordance with analysis in Sec.\ref{sec:asymptotic}.
The similar analysis in the case of $A_\pm = {\rm const.}$ has been studied, e.g., in Ref.~\cite{Aniceto:2013fka,Aniceto:2018bis}, so that our conjecture is a sort of generalizations from the case.
Notice that Eq.(\ref{eq:conj1}) contains the contributions of all the branch-cuts, originated from the NP sectors labeled by $(n_+,n_-)$, along the positive real axis on the Borel plane 
}

{Since the Stokes constants are analytic functions,} the resurgent relation (\ref{eq:conj1}) can take a slightly different form by expanding $\sigma_\beta$ and $\sigma_{{\cal R},a}$.
For simplicity, let us consider the case that ${\cal O} \in \{ \beta, {\cal R}, \widetilde{X}_\pm \}$.
Substituting Eqs.(\ref{eq:a_para_deps0})(\ref{eq:a_para_deps_np0})(\ref{eq:conj2}) into Eq.(\ref{eq:conj1}) yields the resurgent relation for a fixed $s$ as
\be
a_{{\cal O}(s)}^{[{\bf 0},k]}  &\simeq& \frac{1}{2 \pi i} \sum_{\substack{{\bf n} \in {\mathbb N}_0^2  \\  |{\bf n}|>0}} \sum_{{\bf m} \in {\mathbb N}_0^2} \sum_{h \in {\mathbb N}_0} \frac{A_{+(m_+)}^{n_+} A_{-(m_-)}^{n_-}}{C(|{\bf n}|,n_-)} a_{{\cal O} (s-{\bf n} \cdot {\bf m})}^{[{\bf n},h]} \frac{\Gamma(k - {\bf n} \cdot \bm{\rho} -h)}{({\bf n} \cdot {\bf S})^{k-{\bf n} \cdot \bm{\rho} -h}}, \label{eq:res_rel022} 
\ee
where $a_{{\cal O} (t<0)}^{[{\bf n},h]} = 0$.
Additionally, the expanded form by both $\sigma_{\beta}$ and $\sigma_{{\cal R}, a}$ can be obtained by expanding the components in the resurgent relation as  $a^{[{\bf n},k]}_{{\cal O}(s)} = \sum_{t \in {\mathbb N}_0} a^{[{\bf n},k]}_{{\cal O}(s,t)} (\sigma_{{\cal R}, a})^t$, $A_{\pm(s)} = \sum_{t \in {\mathbb N}_0} A_{\pm(s,t)} (\sigma_{{\cal R}, a})^t$, and $\rho_{\pm} = \sum_{t \in {\mathbb N}_0} \rho_{\pm(t)} (\sigma_{{\cal R}, a})^t$.
The results for $t=0,1$ are expressed by
\be
a_{{\cal O}(s,0)}^{[{\bf 0},k]} &\simeq& \frac{1}{2 \pi i} \sum_{\substack{{\bf n} \in {\mathbb N}_0^2  \\  |{\bf n}|>0}} \sum_{{\bf m} \in {\mathbb N}_0^2} \sum_{h \in {\mathbb N}_0} \frac{ A_{+(m_+,0)}^{n_+} A_{-(m_-,0)}^{n_-}}{C(|{\bf n}|,n_-)} a_{{\cal O}(s-{\bf n} \cdot {\bf m},0)}^{[{\bf n},h]} \frac{\Gamma(k - {\bf n} \cdot \bm{\rho}_{(0)} -h)}{({\bf n} \cdot {\bf S})^{k-{\bf n} \cdot \bm{\rho}_{(0)} -h}}, \label{eq:res_rel023_t0}  \nl \\
a_{{\cal O}(s,1)}^{[{\bf 0},k]} &\simeq& \frac{1}{2 \pi i} \sum_{\substack{{\bf n} \in {\mathbb N}_0^2  \\  |{\bf n}|>0}} \sum_{{\bf m} \in {\mathbb N}_0^2} \sum_{h \in {\mathbb N}_0} \frac{ A_{+(m_+,0)}^{n_+} A_{-(m_-,0)}^{n_-}}{C(|{\bf n}|,n_-)} a_{{\cal O}(s-{\bf n} \cdot {\bf m},0)}^{[{\bf n},h]} \frac{\Gamma(k - {\bf n} \cdot \bm{\rho}_{(0)} -h)}{({\bf n} \cdot {\bf S})^{k-{\bf n} \cdot \bm{\rho}_{(0)} -h}} \nl
&& \cdot \left[ n_+ \frac{A_{+(m_+,1)}}{A_{+(m_+,0)}} + n_- \frac{A_{-(m_-,1)}}{A_{-(m_-,0)}} + \frac{a_{{\cal O} (s-{\bf n} \cdot {\bf m},1)}^{[{\bf n},h]}}{a_{{\cal O} (s-{\bf n} \cdot {\bf m},0)}^{[{\bf n},h]}} \right. \nl
  && \left. + \left( \log ({\bf n} \cdot {\bf S}) - \psi^{(0)}(k - {\bf n} \cdot \bm{\rho}_0 -h) \right) {\bf n} \cdot \bm{\rho}_{(1)} \right], \qquad (\rho_{\pm(1)} = 3/\sqrt{2}) \label{eq:res_rel023_t1} 
\ee
where $\psi^{(0)}(x)$ is the polygamma function\footnote{
  $\log ({\bf n} \cdot {\bf S})$ and $\psi^{(0)}(k - {\bf n} \cdot \bm{\rho}_0 -h)$ in Eq.(\ref{eq:res_rel023_t1}) imply that another type of transmonomials, $\log w$, appears due to expanding the higher transmonomial, $\zeta_{\pm}$, by $\sigma_{{\cal R}, a}$\cite{sauzin2014introduction}.
  It can be seen as
  \be
  \zeta_\pm = \frac{e^{- S_\pm w}}{w^{\rho_\pm}} = \frac{e^{- S_\pm w}}{w^{\rho_{\pm (0)}}}
  \sum_{n \in {\mathbb N}_0} \frac{\left( -\delta \rho_\pm \log w \right)^n}{n !},
\ee
where $\delta \rho_\pm:=\rho_\pm - \rho_{\pm(0)}$ is the $\sigma_{{\cal R}, a}$-dependent part in $\rho_\pm$.
Notice that $S_\pm$ is independent on $\sigma_{{\cal R}, a}$.
}. 
It is notable that the mass dependence is also contained in Eq.(\ref{eq:conj1}) by turning on $m$ as $\sigma_\beta \rightarrow m \sigma_\beta$.

Finally, we numerically check the above conjecture.
Because of the property of homomorphism of the Borel resummation (\ref{eq:S_homomor}), it is enough to check the conjecture for $(T,{\cal R},\widetilde{X}_\pm)$ only.
{
  Since the essential point of our conjecture is that the Stokes constants are analytic functions, we employ Eqs.(\ref{eq:res_rel023_t0})(\ref{eq:res_rel023_t1}) which are constructed by fully using the analyticity.
  These indicate that the resurgent relation individually holds for each the fixed $s$ and $t$ in $a^{[{\bf 0},k]}_{{\cal O}(s,t)}$.
}
For the numerical check, one firstly has to determine the value of the Stokes constants.
The values of them depend on the normalization of $\sigma_\pm$, so that we take $a^{[(1,0),0]}_{\widetilde{X}_+} = a^{[(0,1),0]}_{\widetilde{X}_-}=1$\footnote{
This normalization is the same choice taken in Eqs.(\ref{eq:b_np_ex})-(\ref{eq:bp_np_ex}).
}.
According to Eq.(\ref{eq:res_rel0}) and the fact that $\Gamma(k-\rho_+) \gg \Gamma(k-\rho_-)$, the contribution of $A_{-}$ can be explicitly seen in the $14 (\approx \rho_- - \rho_+)$-th order of the resurgent relation, that is sufficiently small {relative to that of $A_+$} in the small $h$.
For this reason, we do not consider $A_-$ in this paper.
We evaluate the value of the Stokes constant, $A_{+(s,t)}$, for $s=0,\cdots,5$ and $t=0,1$ from $\widetilde{X}_+$.
The result is shown in Tab.\ref{tab:Stokes_cnst}.
\begin{table}[t]
  \centering
\renewcommand{\arraystretch}{1.4}
\begin{tabular}{||c||c|c|c|c|c|c|} 
\hline
$\substack{{\rm Im} [A_{+(s,t)}] \\ {\footnotesize / (10^{-7} \theta_0^{\rho_{+(0)}})}}$
& $\substack{\mbox{\small $s=0$}}$ & $\substack{\mbox{\small $s=1$}}$ & $\substack{\mbox{\small $s=2$}}$ & $\substack{\mbox{\small $s=3$}}$ & $\substack{\mbox{\small $s=4$}}$ & $\substack{\mbox{\small $s=5$}}$ \\  [0.5ex]  \hline \hline
{\small $t=0$} & 1.2746 & 2.2381 & 1.5745 & 0.6423 & 0.1719 & 0.0333 \\ [0.5ex] \hline
{\small $t=1$} & $\substack{5.1356 \, + \, \\ 2.7038 \log \theta_0}$ & $\substack{9.4909 \, + \, \\ 4.7477 \log \theta_0}$ & $\substack{6.8066 \, + \, \\ 3.3400 \log \theta_0}$ & $\substack{2.8109 \, + \, \\ 1.3626 \log \theta_0}$ & $\substack{0.7530 \, + \, \\ 0.3648 \log \theta_0}$ & $\substack{0.1460 \, + \, \\ 0.0707 \log \theta_0}$ \\ [0.5ex] 
 \hline
\end{tabular}
\caption{The value of the Stokes constant, ${\rm Im}[A_{+(s,t)}]$ for $s=0,\cdots,5$ and $t=0,1$.
  These values are evaluated by substituting the coefficients of $\widetilde{X}_+$ into Eqs.(\ref{eq:res_rel023_t0})(\ref{eq:res_rel023_t1}).
  We used $k=100$ and truncated the summations by taking ${\bf n}_{\max}=(1,0)$, $h_{\max}=10$ for the upper bound.
$\rho_{+(0)}$ is given by $\rho_{+(0)} = -3 - \sqrt{1285}/5 \approx - 10.1694$.
}
\label{tab:Stokes_cnst}
\end{table}
Then, by using Eqs.(\ref{eq:res_rel023_t0})(\ref{eq:res_rel023_t1}) and the Stokes constants shown in Tab.\ref{tab:Stokes_cnst}, we evaluate $(T, {\cal R}, \widetilde{X}_-)$ and show the result in Tab.\ref{tab:TRXm}.
The result indicates that our conjecture works well.
Notice that the evaluated values of $a^{[0,100]}_{\widetilde{X}(s,t)}$ for $t=1$ involves larger errors than the others, but it can be interpreted as an artifact due to the truncation.

\begin{table}[t]
  \centering
\renewcommand{\arraystretch}{1.4}
\begin{tabular}{||c||c|c|c|c|c|c|} 
\hline
$\substack{a_{T(s,t)}^{[{\bf 0},100]} \\ {\footnotesize /( 10^{114} \theta_0^{100}})}$& $\substack{\mbox{\small $s=0$}}$ & $\substack{\mbox{\small $s=1$}}$ & $\substack{\mbox{\small $s=2$}}$ & $\substack{\mbox{\small $s=3$}}$ & $\substack{\mbox{\small $s=4$}}$ & $\substack{\mbox{\small $s=5$}}$ \\  [0.5ex]
\hline
    {\small $t=0$} & $\substack{7.4075  \\ (7.4075)}$  & $\substack{10.5722 \\ (10.5722)}$ & $\substack{5.2812 \\ (5.2812)}$ & $\substack{1.3933 \\ (1.3933)}$ & $\substack{0.1979 \\ (0.1979)}$ & $\substack{ 0.0201 \\ (0.0201)}$ \\ [0.5ex]
{\small $t=1$} & $\substack{-28.5119 \\ (-28.5119)}$& $\substack{-37.6836 \\ (-37.6836)}$ & $\substack{-18.1723 \\ (-18.1723)}$ & $\substack{-4.6171 \\ (-4.6171)}$ & $\substack{-0.6594 \\ (-0.6594)}$ & $\substack{-0.0617 \\ (-0.0617)}$ \\ [0.5ex] 
\hline
\hline
$\substack{a_{{\cal R}(s,t)}^{[{\bf 0},100]} \\ {\footnotesize /( 10^{114} \theta_0^{100}})}$ & $\substack{\mbox{\small $s=0$}}$ & $\substack{\mbox{\small $s=1$}}$ & $\substack{\mbox{\small $s=2$}}$ & $\substack{\mbox{\small $s=3$}}$ & $\substack{\mbox{\small $s=4$}}$ & $\substack{\mbox{\small $s=5$}}$ \\  [0.5ex]
\hline
    {\small $t=0$} & $\substack{-5.0428  \\ (-5.0428)}$  & $\substack{-7.2348 \\ (-7.2348)}$ & $\substack{-3.6486 \\ (-3.6486)}$ & $\substack{-0.9749 \\ (-0.9749)}$ & $\substack{-0.1416 \\ (-0.1416)}$ & $\substack{-0.0147 \\ (-0.0147)}$ \\ [0.5ex]
{\small $t=1$} & $\substack{21.9760 \\ (21.9760)}$& $\substack{29.4841 \\ (29.4841)}$ & $\substack{14.4240 \\ (14.4240)}$ & $\substack{3.7333 \\ (3.7333)}$ & $\substack{0.5446 \\ (0.5446)}$ & $\substack{0.0528 \\ (0.0528)}$ \\ [0.5ex] 
\hline
\hline
  $\substack{a_{\widetilde{X}_-(s,t)}^{[{\bf 0},100]} \\ {\footnotesize /( 10^{114} \theta_0^{100}})}$ & $\substack{\mbox{\small $s=0$}}$ & $\substack{\mbox{\small $s=1$}}$ & $\substack{\mbox{\small $s=2$}}$ & $\substack{\mbox{\small $s=3$}}$ & $\substack{\mbox{\small $s=4$}}$ & $\substack{\mbox{\small $s=5$}}$ \\  [0.5ex]  \hline 
{\small $t=0$} & $\substack{0.0342  \\ (0.0342)}$ & $\substack{-1.5235 \\ (-1.5235)}$ & $\substack{-2.1704 \\ (-2.1704)}$ & $\substack{-1.0462 \\ (-1.0462)}$ & $\substack{-0.2625 \\ (-0.2625)}$ & $\substack{-0.0340 \\ (-0.0340)}$ \\ [0.5ex]
{\small $t=1$} & $\substack{-0.0486 \\ (-0.0288)}$& $\substack{6.1488 \\ (6.1836)}$ & $\substack{7.9962 \\ (8.0207)}$ & $\substack{3.7095 \\ (3.7195)}$ & $\substack{0.8933 \\ (0.8958)}$ & $\substack{0.1170 \\ (0.1173)}$ \\ [0.5ex] 
\hline
\end{tabular}
\caption{The value of $a^{[{\bf 0},100]}_{T(s,t)}$, $a^{[{\bf 0},100]}_{{\cal R}(s,t)}$, $a^{[{\bf 0},100]}_{\widetilde{X}_-(s,t)}$ for $s=0,\cdots,5$ and $t=0,1$ evaluated from the resurgent relation (\ref{eq:res_rel023_t0})(\ref{eq:res_rel023_t1}).
  We truncated the summations by taking ${\bf n}_{\max}=(1,0)$, $h_{\max}=10$ as the upper bound and used the values of the Stokes constants in Tab.\ref{tab:Stokes_cnst}. 
The bracket, $(\cdots)$, denotes the exact value computed from the ODEs.
}
\label{tab:TRXm}
\end{table}

\section{Additional remarks} \label{sec:add_comments}
{
Our results have relations to various topics related to hydrodynamization of heavy-ion collisions and its resurgence.
In this section, we make comments on some issues related to  the transseries structure and the resurgence of the non-conformal Bjorken flow with FD and BE statics.
In Sec.~\ref{sec:U1broken}, we describe the case of broken U(1) symmetry and the difference of the transseries.
In Sec.~\ref{sec:massless}, we consider the transseries structure and the resurgence under the massless limit.
In Sec.~\ref{sec:gen_tau}, we briefly describe the transseries structure in the case of the generalized relaxation-time.
In Sec.~\ref{sec:attr}, we discuss attractor solutions from the aspect of the IR transseries.
}

\subsection{The case of broken ${\rm U(1)}$ symmetry} \label{sec:U1broken}
For comparison with our result, we briefly describe the case of broken ${\rm U(1)}$ symmetry.
Since imposing ${\rm U(1)}$ symmetry as ${\cal D}_{\mu} N^{\mu} = 0$ determines dynamics of the chemical potential, one can take $\mu = \alpha = 0$ when ${\cal D}_{\mu} N^{\mu} \ne 0$. 
In such a case, Eq.(\ref{eq:cons12}) gives
\be
- D \beta \cdot \Xi_{3}^{(1)} + ({\cal E} + P + \Pi) \theta - \pi^{\mu \nu} \sigma_{\mu \nu} = 0 \quad \Rightarrow \quad D \beta \approx \beta \chi_\beta \theta, \label{eq:chi_mu0}
\ee
and the speed of sound $(c_s^2 = \chi_\beta)$ is given by\cite{Jaiswal:2014isa}
\be
\chi_\beta &=&  \beta^{-1}\frac{({\cal E}_0 + P) \Xi_1^{(1)}}{\Xi_3^{(1)}} \nl
&\sim& z^{-1} - \frac{1}{2} z^{-2} + O(z^{-3}) \nl
&& +  \frac{ a e^{-z}}{2 \sqrt{2}} \left( z^{-1} - \frac{51}{16} z^{-2} + O(z^{-3})\right) + O(e^{-2 z}) \quad \mbox{as} \quad z \rightarrow +\infty.
\ee
The crucial point is that the leading order of $\chi_\beta$ is $O(z^{-1})$ in spite of $O(z^{0})$ in the case of conserved ${\rm U(1)}$ symmetry.
This difference makes a drastic change to the transseries structure.
The PF sector under the Bjorken symmetry is easily obtained from the expanded form of $c_s^2$ as 
\be
\beta_{\rm pf} &\sim& \log \tau \left[ 1 - \frac{\log_{(2)} \tau}{2 \log \tau} + \frac{\sigma_\beta}{\log \tau} + O((\log \tau)^{-2}) \right. \nl
  && \left. - \frac{a e^{-\sigma_\beta}}{2 \sqrt{2}} \cdot \frac{1}{\tau (\log \tau)^{1/2}} \left( 1 - \frac{43}{16 \log \tau} + O((\log \tau)^{-2}) \right) + O(\tau^{-2}) \right],
\ee
where $\log_{(2)} \tau := \log \log \tau$, and we took $m=1$ in this expression.
The set of transmonomials constructing the PF sector is $\{(\log \tau)^{-1}, \log_{(2)} \tau \}$ for the MB statistics $(a=0)$ and $\{\tau^{-1}, (\log \tau)^{-1/2}, \log_{(2)} \tau \}$ for the FD and BE statistics $(a = \pm 1)$.
Apparently, this structure is totally different from the case of conserved ${\rm U(1)}$ symmetry.
Since the structure of PF sector propagates to the PT and NP sectors, the whole structure of transseries entirely changes.
Transseries analysis for the MB statistics has been studied in Ref.~\cite{Kamata:2022jrc}.
The $a$-dependence is beginning with $O(\tau^{-1})$ in $\beta_{\rm pf}$ and becomes negligible near the equilibrium.
The translation between $\tau$ and $w$ can be performed by Eq.(\ref{eq:tautow}), and the leading order is obtained as $w \sim \tau (\log \tau)^{-1}$ or $\tau \sim w \log w$.
Hence, the transseries structure of $w$ is essentially the same as that of $\tau$.

\subsection{The massless case} \label{sec:massless}
We make comments on the massless case.
{
  The crucial point is that, since the massless limit is not commutative with the asymptotic limit, $w \rightarrow +\infty$, transseries in the massless theory can not be directly found from Eq.(\ref{eq:full_trans_massive}) by taking the massless limit.
  Conversely, this implies the fact that, as is also seen in Sec.\ref{sec:U1broken}, transseries structure is generally determined by symmetry of the equilibrium point  and works as far as the symmetry is not enhanced or broken by taking certain parameters in the theory.
}

Here, let us briefly see transseries and resurgence by beginning with the massless ODEs.
As is described in App.\ref{sec:massless_app}, the massless ODEs using $\tau_R= \beta \theta_0$ and $w=\tau/\beta$ are given by
\be
&& \frac{d \beta}{d w} = \frac{\beta}{w} \cdot \frac{\frac{1}{3} - \bar{\pi}}{\frac{2}{3}+\bar{\pi}}, \nl
&& \frac{d {\cal R}}{d w} = -\frac{3 {\cal R}}{w} \cdot \frac{{\rm Li}_3({\cal R}_a)}{{\rm Li}_2({\cal R}_a)} \cdot \frac{ \bar{\pi}}{\frac{2}{3}+\bar{\pi}}, \\
&& \frac{d \bar{\pi}}{d w} =  - \frac{\frac{\bar{\pi}}{\theta_0} + \frac{1}{w}\left[ - \frac{4}{3}  \bar{\beta}_{\pi}({\cal R}_a) + \frac{38}{21} \bar{\pi} + D({\cal R}_{a}) \bar{\pi}^2 \right]}{\frac{2}{3}+\bar{\pi}}, \nn
\ee
where
\be
&& \bar{\beta}_{\pi}({\cal R}_a) := \frac{4}{15} \cdot \frac{{\rm Li}_{2}({\cal R}_a) {\rm Li}_{4}({\cal R}_a)}{4 {\rm Li}_{2}({\cal R}_a) {\rm Li}_{4}({\cal R}_a) - 3 {\rm Li}_{3}({\cal R}_a)^2}, \nl
&& D({\cal R}_{a}):= 4 - \frac{3 {\rm Li}_3\left({\cal R}_a \right)^2 \left( 3 {\rm Li}_1\left({\cal R}_a \right) {\rm Li}_3({\cal R}_a) - 2 {\rm Li}_2\left({\cal R}_a \right)^2  \right)}{{\rm Li}_2 \left({\cal R}_a \right)^2 \left(4 {\rm Li}_2\left({\cal R}_a \right) {\rm Li}_4\left({\cal R}_a \right)-3 {\rm Li}_3\left({\cal R}_a \right)^2\right)}, \nn
\ee
with ${\cal R}:=e^{-\alpha}$ and ${\cal R}_a:=-a {\cal R}$.
Notice that the ODE of $\alpha$ is defined as
\be
\frac{d \alpha}{d w} = \frac{3}{w} \cdot \frac{{\rm Li}_3({\cal R}_a)}{{\rm Li}_2({\cal R}_a)} \cdot \frac{ \bar{\pi}}{\frac{2}{3}+\bar{\pi}}.
\ee
One can immediately see that the $\beta$-dependence does not exist in the ODEs of ${\cal R}$ and $\bar{\pi}$.
From these ODEs, the PF sector is obtained by taking $\bar{\pi}=0$, and thus,
\be
\beta_{\rm pf} \sim \sigma_\beta w^{1/2}, \qquad {\cal R}_{\rm pf} \sim \sigma_{\cal R}, \qquad \alpha_{\rm pf} \sim - \log \sigma_{\cal R},  \qquad (w \gg 1)
\ee
where $\sigma_\beta \in {\mathbb R}_+$ and $\sigma_{\cal R} \in {\mathbb R}_+$ are integration constants.
Because $w \sim (\sigma_\beta^{-1} \tau)^{2/3}$, the large order behavior of $\beta_{\rm pf}$ is $\beta_{\rm pf} \sim \sigma_{\beta}^{2/3} \tau^{-1/3}$.
The leading order of the chemical potential, $\mu = \alpha/\beta$, is given by $\mu_{\rm pf} = \alpha_{\rm pf}/\beta_{\rm pf} \sim - \sigma_{\beta}^{-1} \log \sigma_{\cal R} \cdot w^{-1/2}$, and thus, $\lim_{w \rightarrow +\infty} \mu_{\rm pf} = 0$.

The IR transseries essentially has the same form as that of the massive case but contains different values of $S$ and $\rho$:
\be
&& \beta \sim \sigma_\beta w^{1/2} \sum_{n \in {\mathbb N}_0} \sum_{k \in {\mathbb N}_0} a_{\beta}^{[n,k]} \zeta^n w^{-k}, \nl
&& {\cal R} \sim \sigma_{\cal R} \sum_{n \in {\mathbb N}_0} \sum_{k \in {\mathbb N}_0} a_{\cal R}^{[n,k]} \zeta^n w^{-k}, \\
&& \bar{\pi} \sim \sum_{n \in {\mathbb N}_0} \sum_{k \in {\mathbb N}_0} a_{\bar{\pi}}^{[n,k]} \zeta^n w^{-k},  \label{eq:full_transs_massless} \nn
\ee
where 
\be
&& \zeta :=  \sigma_{\bar{\pi}} \frac{e^{-S w}}{w^{\rho}}, \qquad S := \frac{3}{2 \theta_0}, \qquad \rho := \frac{19}{7} + 3 \bar{\beta}_{\pi}(\sigma_{{\cal R}, a}), \label{eq:lam_rho_m0}
\ee
with $\sigma_{{\cal R}, a}:=-a \sigma_{\cal R}$.
The normalization of the integration constants can be fixed by taking $a^{[0,0]}_\beta = a^{[0,0]}_{\cal R} = a^{[1,0]}_{\bar{\pi}}=1$ without the loss of generality.
The leading order of $\bar{\pi}$ can be directly computed as
\be
\bar{\pi} \sim \frac{4}{3} \bar{\beta}_\pi(\sigma_{{\cal R}, a}) \cdot \frac{\theta_0}{w} \quad \mbox{as} \quad w \rightarrow +\infty,
\ee
so that all of the coefficients, $(a_{\beta}^{[n,k]}, a_{\cal R}^{[n,k]}, a_{\bar{\pi}}^{[n,k]})$, are functions of $\sigma_{{\cal R}, a}$. 
In addition, $a^{[n,k]}_{\cal O} \propto \theta_0^k$ due to lack of the ${\sigma}_\beta$-dependence in ${\cal R}$ and $\bar{\pi}$.

In the similar way to the nonconformal case, the resurgent relation is obtained as
\be
a_{\bar{\pi}}^{[0,k]}&\simeq& \frac{1}{2 \pi i} \sum_{n \in {\mathbb N}} \sum_{h \in {\mathbb N}_0} A^n a_{\bar{\pi}}^{[n,h]} \frac{\Gamma(k - n \rho -h)}{(n S)^{k- n \rho-h}}, \qquad (k \gg 1) \label{eq:res_rel0_ml}
\ee
where $A \in i {\mathbb R}$ is the Stokes constant.
The coefficients, $a_{\bar{\pi}}^{[n,k]}$, and $\rho$ can be expanded in terms of $\sigma_{{\cal R}, a}$, so that the Stokes constant also depends on $\sigma_{{\cal R}, a}$.
Not only $\bar{\pi}$, but also $(\beta, {\cal R})$ are Borel nonsummable divergent series because of nonlinear terms with $\bar{\pi}$ in the ODEs.
In addition, the same resurgent relation (\ref{eq:res_rel0_ml}) is satisfied for $(\beta, {\cal R})$.

We would like to emphasize that, as well as the massive case, the IR transseries can not be continuously connected to that of broken ${\rm U(1)}$ symmetry, ${\cal D}_\mu N^{\mu} \ne 0$.
Naively thinking, it looks that transseries of broken ${\rm U(1)}$ symmetry can be realized by taking a certain limit to parameters. 
However, although transmonomials between the two cases are the same, this expectation is not correct.
The easiest way to see this fact is making sure if $\sigma_{\cal R}$ can be taken a value corresponding to $\mu=0$, i.e ${\cal R}=1$, for any $w$.
But, from the ODE of ${\cal R}$ in Eq.(\ref{eq:full_transs_massless}), one can immediately see that ${\cal R} \ne 1$ because $\bar{\pi} \sim O(w^{-1})$ and is nonzero.
Another way is evaluating UV convergent points to which flows converge in the UV limit, $w \rightarrow 0_+$. 
Indeed, the values between the cases with and without ${\rm U(1)}$ symmetry are different from each other.
It is because, when deriving the ODEs with ${\cal D}_\mu N^{\mu}=0$, contribution from the chemical potential has to be taken into account under the assumption that $\frac{d \alpha}{d \tau} \ne 0$.
Therefore, the ODE of $\bar{\pi}$ originally does not match with the one without ${\rm U(1)}$ symmetry.

\subsection{Generalized relaxation-time: $\tau_R(x,p) = (\beta E_{\bf p})^{\gamma} t_{R}(x)$} \label{sec:gen_tau}
We make comments on the transseries structure for the generalized relaxation-time.
One can generalize the relaxation-time to depend on the momentum $p$, for example, as\cite{Dusling:2009df,Dusling:2011fd,Kurkela:2017xis,Rocha:2021zcw,Dash:2021ibx,Mitra:2020gdk}
\be
\tau_R(x,p) = (\beta E_{\bf p})^{\gamma} t_{R}(x), \qquad (\gamma \in {\mathbb R}) \label{eq:gen_tau_R}
\ee
where $t_{R}(x)$ is a relaxation-time depending only on $x$.

The important fact for transseries is that the PF sector does not change at all.
It is because the PF sector is irrelevant to the collision part.
Thus, the large order behavior of the PF sector is also the same as $\beta_{\rm pf} \sim \sigma_\beta (m \tau)^{2/3}$ and $\alpha_{\rm pf} \sim - \sigma_{\beta} (m \tau)^{2/3}$.
When computing the PT sector, the effect of $\gamma$ needs to be taken care of.
If $\tau_R$ is given as Eq.(\ref{eq:gen_tau_R}), then the collision part in the ODEs replaced such as
\be
\frac{\bar{X}}{\tau_R} &\quad \rightarrow \quad & \frac{\bar{X}}{\tau_{R,\bar{X}}} = \frac{\bar{X}}{t_R z^{\gamma}} (1+O(z^{-1})) , \qquad (\bar{X} \in \{ \bar{\Pi},\bar{\pi} \}) \label{eq:col_gen_t}
\ee
where $\tau_{R,\bar{X}} \sim t_R z^{\gamma}(1+O(z^{-1}))$ is the effective relaxation-time associated with $\bar{\Pi}$ or $\bar{\pi}${, and $z := m \beta$}.
In contrast,  the non-collision part is essentially the same form as $\left[ C_{0}^{\bar{\beta}_{{\Pi},{\pi}}}({\cal R}_a)  + O(z^{-1},\bar{\Pi},\bar{\pi}) \right]/\tau$, where $C_{0}^{\bar{\beta}_{{\Pi},{\pi}}}({\cal R}_a)$ denotes the $O(z^{0})$ part of $\bar{\beta}_{\Pi,\pi}$ depending on $\gamma$.
Notice that the explicit form of transport coefficients depends on $\gamma$, but only the coefficients change when taking the asymptotic expansion around $z = +\infty$.
It is because $\Xi_{k}$ and $\Xi^{(n)}_{k}$ are $O(z^{-3/2})$ for any $n \in {\mathbb N}$ and $k \in {\mathbb R}$.

The convergence to the equilibrium is naturally obtained under the assumption that the collision part is dominant relative to the non-collision part around the equilibrium, so that it gives a constraint to the value of $\gamma$ depending on the definition of $t_R$.
Here, we suppose that $t_R \approxprop \tau^{\delta}$ with $\delta \in {\mathbb R}$ around the equilibrium.
One can obtain the constraint as
\be
\frac{1}{t_{R} z^\gamma} \approxprop \tau^{-(\delta+\frac{2}{3} \gamma)} \gg \tau^{-1} \quad \mbox{as} \quad \tau \rightarrow +\infty \quad \Rightarrow \quad  \gamma < \frac{3}{2} (1 - \delta).
\ee
Notice that taking $t_{R} = \beta \theta_0$ gives $\gamma<1/2$.
The asymptotics of $(\bar{\Pi},\bar{\pi})$ also depends on $\gamma$, and the leading order is given by
\be
\bar{\Pi} \sim O(\tau^{2(\delta + \frac{2}{3} \gamma -1)}), \qquad \bar{\pi} \sim O(\tau^{\delta + \frac{2}{3} \gamma -1}).
\ee

When thinking of transseries analysis 
including the PT and NP sectors, the choice of the flow time $w$ might be crucial.
In this case, there is no flow time to make the collision part as ${\rm const.} \times \bar{X}$ for both $\bar{\Pi}$ and $\bar{\pi}$ simultaneously, but the definition of $w$ that we used in the above analysis essentially works for construction of transseries.
Let us see this fact by taking $t_{R} = \beta \theta_0$ $(\delta = 2/3)$ and the flow time as $w=\tau (\beta z^{\gamma})^{-1}$.
Below, we take the mass unit, $m=1$.
Since the derivative for the variable transformation is given by
\be
\frac{d w}{d \tau} &=& \frac{1 - ( 1 + \gamma)  \left( \chi_\beta + \bar{\Pi} - \bar{\pi} \right)}{\beta^{1+\gamma}
},
\ee
the ODEs take the form as
\be
&& \frac{d \bar{\Pi}}{d w} = - \frac{3}{1-2\gamma} \left[  \frac{\bar{\Pi}}{\theta_0} +  \frac{1}{w} \left\{ C^{\bar{\beta}_{\Pi}}_0({\cal R}_a) + O(\beta^{-1},\bar{X}) \right\} \right] + O(\beta^{-1} \bar{X}, \bar{X}^2), \\ 
&& \frac{d \bar{\pi}}{d w} = - \frac{3}{1-2\gamma} \left[ \frac{\bar{\pi}}{\theta_0} + \frac{1}{w} \left\{ - \frac{4}{3} C^{\bar{\beta}_{\pi}}_0({\cal R}_a) + O(\beta^{-1},\bar{X}) \right\} \right] + O(\beta^{-1} \bar{X}, \bar{X}^2). 
\ee
From these ODEs, the leading orders of variables are expressed as $\beta_{\rm pf} \sim \sigma_\beta w^{2/(1-2 \gamma)}$, $\alpha_{\rm pf} \sim - \sigma_\beta w^{2/(1-2 \gamma)}$, $\bar{\Pi} \sim a_{\bar{\Pi}}[\sigma_{{\cal R},a}] (\frac{\theta_0}{w})^2$, and $\bar{\pi} \sim a_{\bar{\pi}} [\sigma_{{\cal R},a}] \frac{\theta_0}{w}$, where $a_{\bar{\Pi},\bar{\pi}}[\sigma_{{\cal R},a}]$ is a function of $\sigma_{{\cal R},a}$ depending on $\gamma$.
In addition, if the exponent of the exponential decay in the NP transmonomial is proportional to $w$ such as $e^{- S w}$, then it gives another constraint to $\gamma$, that comes from $O(\beta^{-1} \bar{X})$ in the ODEs\footnote{
  If $\gamma <-1/2$, then the exponential decay is $\exp \left[ -\frac{3}{(1-2 \gamma)\theta_0} w - S_1 w^{-\frac{1 + 2 \gamma}{1-2 \gamma}} \right]$ with  a real constant $S_1$.
}:
\be
\lim_{w \rightarrow +\infty} \int \frac{d w}{\beta_{\rm pf}} = 0 \ \ \mbox{or} \ \ O(\log w) \quad \Rightarrow \quad \gamma  \ge - \frac{1}{2}.
\ee
Hence, when $\gamma$ is in the range that $- 1/2 \le \gamma < 1/2$, the transseries can be obtained as the similar form to Eq.(\ref{eq:full_trans_massive}) but with different $S_\pm$ and $\rho_\pm$.
The transmonomial generating the power expansion is relevant to cardinality of $\gamma$, i.e. $\gamma \in {\mathbb Q}$ or $\gamma \in {\mathbb R} \setminus {\mathbb Q}$\cite{edgar2010transseries,Costin2006TopologicalCO}.
The PT sector is a single expansion generated by $\{ w^{-1/q} \}$ with $q = {\rm lcm}(1-2 \gamma,2)/2 \in {\mathbb N}$ if $\gamma \in {\mathbb Q}$ but a double expansion by $\{ w^{-1}, w^{-2/(1-2 \gamma) }\}$, otherwise.
Therefore, the transseries are given by
\be
&&\beta \sim w^{2/(1-2 \gamma)}  \sigma_\beta \sum_{{\bf n} \in {\mathbb N}_0^2} \sum_{k \in {\mathbb N}_0} a^{[{\bf n},k]}_\beta \bm{\zeta}^{\bf n} w^{-k/q}, \nl
&& {\cal R}\sim  \sigma_{{\cal R}} \sum_{{\bf n} \in {\mathbb N}_0^2} \sum_{k \in {\mathbb N}_0} a^{[{\bf n},k]}_{\cal R} \bm{\zeta}^{\bf n} w^{-k/q} , \label{eq:trans_gam1} \\
&& \bar{X} \sim  \sum_{{\bf n} \in {\mathbb N}_0^2} \sum_{k \in {\mathbb N}_0} a^{[{\bf n},k]}_{\bar{X}} \bm{\zeta}^{\bf n} w^{-k/q},  \qquad (\bar{X} \in \{ \bar{\Pi}, \bar{\pi} \}) \nn
\ee
for $\gamma \in {\mathbb Q}$, or 
\be
&&\beta \sim w^{2/(1-2 \gamma)}  \sigma_\beta \sum_{{\bf n} \in {\mathbb N}_0^2} \sum_{{\bf k} \in {\mathbb N}_0^2} a^{[{\bf n},{\bf k}]}_\beta \bm{\zeta}^{\bf n} w^{-k_1 - 2 k_2/(1-2 \gamma)}, \nl
&& {\cal R}\sim  \sigma_{{\cal R}} \sum_{{\bf n} \in {\mathbb N}_0^2} \sum_{{\bf k} \in {\mathbb N}_0^2} a^{[{\bf n},{\bf k}]}_{\cal R} \bm{\zeta}^{\bf n} w^{-k_1 - 2 k_2/(1-2 \gamma)} , \label{eq:trans_gam2}  \\
&& \bar{X} \sim  \sum_{{\bf n} \in {\mathbb N}_0^2} \sum_{{\bf k} \in {\mathbb N}_0^2} a^{[{\bf n}, {\bf k}]}_{\bar{X}} \bm{\zeta}^{\bf n} w^{-k_1 - 2 k_2/(1-2 \gamma)},  \qquad (\bar{X} \in \{ \bar{\Pi}, \bar{\pi} \}) \nn
\ee
otherwise.

It is possible to construct the resurgent relation with $\gamma \in {\mathbb Q}$ 
in the similar way described in App.\ref{app:resurgence}, and one can obtain it as the same form as our conjecture.
In contrast, for the resurgent relation with $\gamma \in {\mathbb R} \setminus {\mathbb Q}$,
one has to take care of branch-cuts caused by the fractional power in the double expansion, and the form of the relation should be more nontrivial.
This construction is beyond the scopes of our main purpose, so that we do not argue this issue in this paper.

\subsection{Attractor solution from the aspect of IR transseries} \label{sec:attr}
We consider attractor solution from the aspect of IR transseries.
An attractor is known as an invariant subspace of flow (ISOF) defined on the flow subspace of dissipative hydro variables to which flows deviating from the ISOF converge. 
A convergent rate to the attractor is expected as a key point of \textit{universality} characterizing a given system and for the memory-loss effect of the UV domain.
Those attractors have been well-studied in both the conformal and nonconformal systems, e.g. in Refs.~\cite{Berges:2013fga,Blaizot:2020gql,Heller:2015dha,Basar:2015ava,Kamata:2022jrc,Soloviev:2021lhs,Casalderrey-Solana:2017zyh,Spalinski:2018mqg,Spalinski:2017mel,Du:2021fok,Denicol:2018pak,Jaiswal:2019cju,Romatschke:2017vte,Strickland:2018ayk,Alalawi:2022pmg}.

When considering a nonautonomous system, it is convenient to prepare a trivial bundle defined as ${\cal M} = {\cal F} \times {\cal T}$, where the fiber is a space of the dynamical variables, i.e. $(\beta, {\cal R}, \bar{\Pi},\bar{\pi}) \in {\cal F} = {\mathbb R}_+^2 \times {\mathbb R}^2$, and the base space is defined as a space of the flow time, $w \in {\cal T} = {\mathbb R}_{\ge 0}$\cite{kloeden,caraballo2017applied}\footnote{Since our ODEs are nonautonomous system, the shift symmetry of a flow time as $w \rightarrow w_0$ is broken.
  In such a case, flows has to be defined on a tangent space including the time axis.
}.
 In high energy nuclear collision hydro physics, people normally look at its projected subspace spanned by viscous variables and a flow time, i.e. $\widehat{\cal M} = \widehat{\cal F} \times {\cal T}$, where $(\bar{\Pi},\bar{\pi}) \in \widehat{\cal F}={\mathbb R}^2$, and investigate existence of an attractor and its property.
On this projected subspace, at least in the conformal case, an attractor solution is considered as a one-dimensional invariant subspace of flows, but in the nonconformal case this aspect is not always correct in the sense that the initial condition of $(\beta, {\cal R})$ changes the flow structure on $\widehat{\cal M}$.
This means that $(\beta_0, {\cal R}_0) := (\beta(w_0), {\cal R}(w_0))$ at a fixed $w_0$ are control parameters of flows on $\widehat{\cal M}$, and an attractor (in the physical sense) are defined as a three-dimensional invariant subspace of flows embedded on the original total space, ${\cal M}$.
In contrast, the conformal case without ${\rm U(1)}$ symmetry is special.
In this case, the attractor is originally a two-dimensional invariant subspace of flows on the total space defined as ${\cal M} = {\cal F} \times {\cal T}$ with $(\beta, \bar{\pi}) \in {\cal F}={\mathbb R}_+ \times {\mathbb R}$ and $w \in {\cal T}={\mathbb R}_{\ge 0}$, but when using $w=\tau/\tau_R$ it is enough to look at its projected subspace $\widehat{\cal M} = \widehat{\cal F} \times {\cal T}$ with $\bar{\pi} \in \widehat{\cal F}$ for seeing flows of $\bar{\pi}$  because the dynamics of $\beta$ is decoupled in the ODE of $\bar{\pi}$ due to the existence of D-symmetry\cite{Kamata:2022jrc}.
Therefore, the conformal attractor without ${\rm U(1)}$ symmetry can be projected onto $\widehat{\cal M}$ without missing its information and regarded as a one-dimensional object on $\widehat{\cal M}$.
However, in the nonconformal case, even if ${\rm U(1)}$ symmetry is not imposed, this logic does not work in general.

One can investigate convergent rates of the attractor in the IR regime, i.e. (local) forward attractor,  by using the IR transseries (\ref{eq:full_trans_massive}) and considering perturbation from the attractor\footnote{
  We consider only the IR regime.
  So that we suppose that this attractor is a local forward attractor but do not need to impose the conditions of pullback attractor for this discussion\cite{kloeden,caraballo2017applied}.
  As we considered in App.\ref{sec:CP_UV}, any convergent points do not appear in the UV limit when using $w$ for the flow time.
}.
Suppose that we prepared $Y_{0} := (\beta_0, {\cal R}_0)= (\beta(w_0), {\cal R}(w_0))$ at a fixed time, $w=w_0$.
Then, an attractor can be found by tuning the initial condition of $\bar{\bf X} =( \bar{\Pi}, \bar{\pi})$ and solving the ODEs, which we denote as $\bar{\bf X}_{\rm att}(w;Y_0)$.
If one slightly perturbs $\bar{\bf X}$ from $\bar{\bf X}_{\rm att}$ at $w=w_0$ as $\bar{\bf X}(w_0;Y_0) = \bar{\bf X}_{\rm att}(w_0;Y_0) + \delta \bar{\bf X}(w_0;Y_0)$, it corresponds to perturbation of the integration constants in the IR transseries.
The point is that the perturbation of two variables possibly changes all of the integration constants as $(\sigma_{\beta, {\rm att}} + \delta \sigma_{\beta}, \sigma_{{\cal R}, {\rm att}} + \delta \sigma_{\cal R}, \sigma_{\pm, {\rm att}} + \delta \sigma_{\pm})$, where $\sigma_{{\cal O},{\rm att}}:= \sigma_{{\cal O}}(Y_0,\bar{\bf X}_{{\rm att}}(w_0;Y_0))$, because the initial condition at a fixed time relates to the integration constants through a nontrivial relationship.
Thus, for $w \gg 1$, one can easily compute the perturbation as
\be
\delta \beta &\sim& \delta \sigma_\beta w^2, \nl
 \delta {\cal R} &\sim& \delta \sigma_{\cal R}, \nl
 \delta \bar{\Pi} &\sim&  0.5500 \cdot a \delta \sigma_{\cal R} \theta_0^2 w^{-2} \nl
&& + \left( 9.3333 + a \sigma_{{\cal R},{\rm att}} \right) \cdot \frac{\delta \sigma_{\beta}}{\sigma_{\beta,{\rm att}}^2} \theta_0^2 w^{-4} - 0.8569 \cdot \delta \sigma_+ \frac{e^{-S_+ w}}{w^{\rho_{+(0)}}}, \nl
 \delta \bar{\pi} &\sim&  0.2357 \cdot a \delta \sigma_{\cal R} \theta_0 w^{-1} \nl
 && + \left( 3.1111 + a \sigma_{{\cal R},{\rm att}} \right)\cdot \frac{\delta \sigma_{\beta}}{\sigma_{\beta,{\rm att}}^2} \theta_0 w^{-3} + 1.0000 \cdot \delta \sigma_+ \frac{e^{-S_+ w}}{w^{\rho_{+(0)}}},
\ee
where $\rho_{\pm(0)}$ is the zero-th order of $\rho_\pm$ expanded around  $\sigma_{{\cal R},a}=0$.
Here, we omitted the contribution from $\delta \sigma_-$ because the effect of $\delta \sigma_+$ is generally more dominant. 
Notice that $w \sim (\sigma_{\beta, {\rm att}}^{-1} \tau)^{1/3}$ in the nonconformal case.
The result implies that the convergent rate near the equilibrium is a power law because of the existence of chemical potential and/or particle mass.
The exponential decay coupling to $\delta \sigma_+$ is expected to be relevant to a single-line effect and memory-loss of initial conditions.
In order to see the effect, one has to choose an initial condition such that $ | a \delta \sigma_{\cal R} | \ll 1$ and $|\delta \sigma_{\beta}/\sigma_{\beta,{\rm att}}^2| \ll 1$.
When taking the MB statistics ($a=0$), the first condition is trivially satisfied.
One can satisfy the second condition by taking a sufficiently large $\sigma_{\beta, {\rm att}}$, i.e. setting a sufficiently small temperature in the initial condition compared with the mass unit.

The above consideration based on asymptotic analysis tells us that special behaviors such as a single line effect and globally defined ``universality'' generally do not exist in the nonconformal Bjorken flow.
It is because the perturbation from the attractor holds various decay rates even just around the equilibrium and extracting a certain decay rate from them is essentially a fine-tuning problem of the initial condition.
This statement is unchanged for any variable which is a function of $(\beta, {\cal R},\bar{\Pi},\bar{\pi})$.
It is also important to note that, in a nonautonomous system, changing only $w_0$ as fixing values of the initial condition changes the flow structure on $\widehat{\cal M}$ because $Y_0$ has the same value but is given at a different $w_0$.

The massless case can be also obtained by using Eq.(\ref{eq:full_transs_massless}).
In the similar way, one can find
\be
&& \delta \beta \sim \delta \sigma_\beta w^{1/2}, \nl
&& \delta {\cal R} \sim \delta \sigma_{\cal R}, \nl
&& \delta \bar{\pi} \sim  0.0667 \cdot a \delta \sigma_{\cal R} \theta_0 w^{-1} + \delta \sigma_{\bar{\pi}} \frac{e^{-S w}}{w^{\rho_{(0)}}}. 
\ee
In the massless case, the leading order of $w$ in terms of $\tau$ is written as $w \sim (\sigma_{\beta, {\rm att}}^{-1} \tau)^{2/3}$.
When taking the FD or BE statistics, i.e. $a = \pm 1$, the change of initial condition affects $\sigma_{\cal R}$, so that it gives a power law.
In the MB case ($a=0$), in contrast, the leading order is exponential decay, which is thought to be related to a single-line effect.

\section{Summary and outlook} \label{sec:summary}

\begin{figure}[tp]
  \begin{center}
    \begin{tabular}{c}
      \begin{minipage}{1.\hsize}
        \begin{center} 
          \includegraphics[clip, width=140mm]{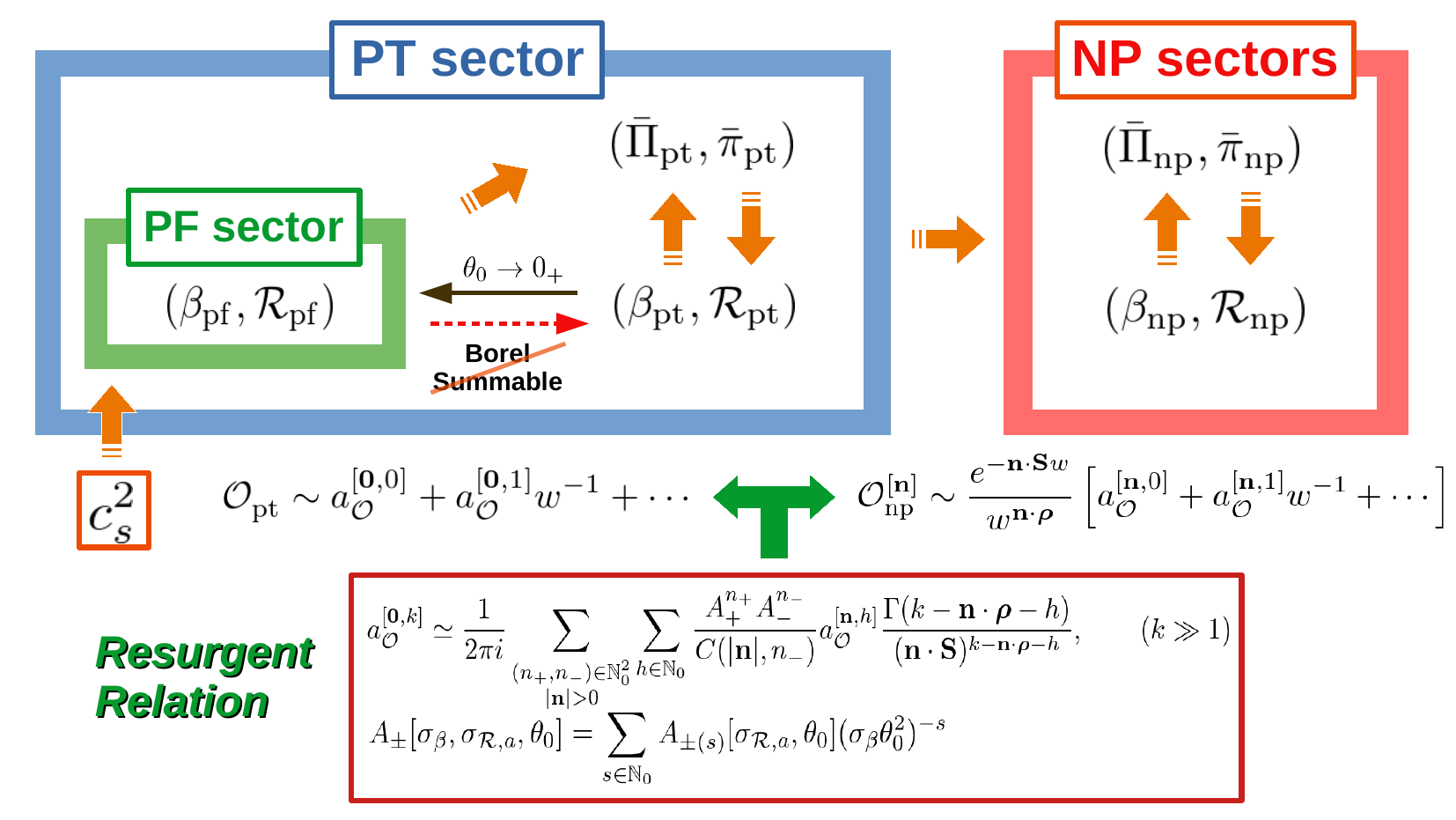}
        \end{center}
      \end{minipage}      
    \end{tabular} 
    \caption{
      Schematic figure of transseries structure and resurgence.
      The transseries structure is beginning with the speed of sound, $c_s^2$, that determines the PF sector.
      By using the result of the PF sector, the leading order of $(\bar{\Pi}_{\rm pt},\bar{\pi}_{\rm pt})$ is obtained, and then, all coefficients of the PT sector can be recursively computed order by order.
      The PT sector is a formal power expansion.
      Whereas $(\beta, {\cal R})$ is Borel summable in the PF sector, it becomes Borel nonsummable in the PT sector due to nonlinear terms with $(\bar{\Pi}_{\rm pt}, \bar{\pi}_{\rm pt})$ which is Borel nonsummable.
      After computation of the PT sector, the NP sectors can be recursively obtained in the similar way.
      The ${\bf n}$-th NP sector is labeled by ${\bf n}$ in the higher transmonomial as $e^{- {\bf n} \cdot {\bf S} w}$.
      Although our dynamical system consists of four variables, $(\beta, {\cal R}, \bar{\Pi}, \bar{\pi})$, their resurgent relation is the same form derived from the type of ODE as $\frac{d {\bf Y}}{d w} = - S {\bf Y} + \frac{1}{w} \left[V + {\frak B} {\bf Y} \right] + O({\bf Y}^2,w^{-2})$ and contains two Stokes constants, $A_{\pm}$.
      These Stokes constants are functions of $(\sigma_\beta,\sigma_{\cal R})$.
    }
    \label{fig:summary}
  \end{center}
\end{figure}

In this paper, we have considered transseries analysis and resurgence of the nonconformal Bjorken flow with Fermi-Dirac (FD) and Bose-Einstein (BE) statistics on the relaxation-time approximation by imposing conservation laws to both the energy-momentum tensor and the current density. 

We have firstly obtained the full formal transseries expanded around the equilibrium using the flow time defined as $w=\tau/\beta$ in Sec.\ref{sec:trans_IRdomain}.
The conservation law of current density drastically changes the physics around equilibrium in the nonconformal case.
The effect of ${\rm U(1)}$ symmetry enters not only into the transport coefficients but also into the speed of sound which determines the explicit form of the transmonomials in the transseries.
In our setup, the large order behavior of $(\beta,\alpha,\bar{\Pi},\bar{\pi})$ around the equilibrium is a power law with respect to $w=\tau/\beta (\sim \tau^{1/3})$.
In particular, the temperature behaves as $\beta \sim w^{2} \sim \tau^{2/3}$, and the value of exponent correctly deviates from $1/3$, which is the feature of the conformal symmetry breaking.
This statement is also true for the Maxwell-Boltzmann statistics that the chemical potential is decoupled in the ODEs of viscosities, $(\bar{\Pi},\bar{\pi})$.
The PT sector is a formal power expansion, and the NP sectors include exponential decay as $e^{-{\bf n} \cdot {\bf S} w}/w^{{\bf n} \cdot \bm{\rho}}$.
The coefficients in all the sectors depend on the integration constants $(\sigma_\beta,\sigma_{{\cal R}})$, and the particle mass appears only as coupling with $\sigma_\beta$ as $m \sigma_\beta$.
The transseries structure is sensitive to symmetry of the equilibrium, and emerging/breaking symmetry causes a drastic change to the structure.
These IR transseries derived from the equilibrium imposed different symmetry, i.e. with/without a particle mass and/or ${\rm U(1)}$ symmetry, are not continuously connected to each other even by taking the limit for parameters in the theories. 
  
We have considered the resurgence by forming a conjecture of the resurgent relation structure of the Borel transformed ODEs in Sec.\ref{sec:Resurgence_analysis}.
Whereas the Borel transformed ODEs of $(\bar{\Pi},\bar{\pi})$ (or $\widetilde{X}_\pm$) explicitly contain singularities on the positive real axis of the Borel plane, discontinuity of $(T, {\cal R})$ by Borel resummation are indirectly caused by nonlinear terms with $(\bar{\Pi},\bar{\pi})$ in the ODEs of them. 
As a result, the PF sector which is Borel summable becomes Borel nonsummable in the PT sector.
This situation affects the number of Stokes constants in the resurgent relation, i.e. the resurgent relation needs two Stokes constants associated with $(\bar{\Pi},\bar{\pi})$ that depend on the initial conditions and the particle mass.
We have numerically checked that all of $(T, {\cal R}, \widetilde{X}_\pm)$ satisfy the conjectured resurgent relation provided in Eq.(\ref{eq:conj1}) by explicitly evaluating the value of the dominant Stokes constant.
We summarized the transseries structure and the resurgence in Fig.~\ref{fig:summary}.
 
We have mentioned additional remarks related to transseries structure and resurgent relation in Sec.\ref{sec:add_comments}.
We made comments on the change of transseries and the resurgence in some particular cases such as broken ${\rm U(1)}$ symmetry, the massless case, and generalized relaxation-time in Secs.\ref{sec:U1broken}-\ref{sec:gen_tau}.
In addition, we have mentioned the (local forward) attractor solution from the aspect of the IR transseries in Sec.\ref{sec:attr}.
The memory-loss effect of the UV domain characterized by the exponential decay does not appear due to a physical scale such as a particle mass and a chemical potential.

Not only just as a method of approximation, transseries analysis and resurgence theory are powerful mathematical methods to provide rich informations of hydrodynamics and kinetic theoretical approach by looking to their structure.
If a local equilibrium is consistently defined, then constructions of IR transmonomials and transseries by labeling ``sectors" can be performed consistently  with an underlying method for formulating hydrodynamics such as Chapman-Enskog expansion, as we can see in Fig.~\ref{fig:summary}\footnote{
  One of the exceptions is Gubser flow.
  In this model, the shear viscosity does not converges to zero in the IR limit due to the nontrivial background geometry, and the NP sector does not exist.
  See Ref.~\cite{Behtash:2019qtk}
}.
This lesson suggests large possibility of application of transseries analysis and resurgence theory to a more generic setup of hydrodynamics and kinetic theory motivated by QCD.
One of the interesting issues is a relationship between transseries structure and symmetry.
In this work we imposed strong symmetry, i.e. Bjorken symmetry, to the theory, but one can expect that a more interesting feature can be seen by relaxing symmetry, e.g. by introducing magnetic source fields.
In such a case, the Boltzmann equation generally becomes PDE, and its transseries structure and resurgent relation must become much more nontrivial.
We would study these formal constructions based on resurgence theory as future works.


\acknowledgments
We would thank X.-G. Huang for helpful discussions for extended relaxation-time approximation and comments on the case of broken ${\rm U(1)}$ symmetry.
We also would thank A. Behtash and N. Sueishi for helpful discussions for resurgent relation.
S.~K. is supported by the Polish National Science Centre grant 2018/29/B/ST2/02457.

\appendix

\section{Transport coefficients and ODEs} \label{sec:trans_ODE}
{In this appendix, we summarize the transport coefficients and the ODEs in our setup.
  We introduce the FD and BE distributions and find transport coefficients using the Chapman-Enskog-like expansion in App.~\ref{sec:prep}, and then derive the ODEs in App.~\ref{sec:derive_ODEs}.
  In App.~\ref{sec:asy_exp}, we write down asymptotic expansions of the transport coefficients in the small temperature.
  We also derive ODEs of the massless limit in App.~\ref{sec:massless_app}.
  See, e.g., Ref.~\cite{Florkowski:2015lra,Dash:2021ibx,Ambrus:2022vif,Denicol:2010xn} if more details are necessary.
}
\subsection{Distributions and transport coefficients} \label{sec:prep}

We start with the Maxwell-Boltzmann (MB), Fermi-Dirac (FD) and the Bose-Einstein (BE) distribution 
given by
\be
f_{{\rm eq}}(x,p;a) = \frac{1}{\exp \left[ \beta(x) E_{\bf p}(x,p) 
    + \alpha(x) \right] + a},
\qquad
a:=
\begin{cases}
  0 & \mbox{for MB} \\
  +1 & \mbox{for FD} \\
  -1 & \mbox{for BE}
\end{cases},
\label{eq:fd_dis}
\ee
where $\beta=1/T$ is the inverse temperature, $\alpha$ is defined as $\alpha:= \mu \beta$ with a chemical potential $\mu$, and $E_{\bf p}$ is defined as $E_{\bf p}:=u \cdot p$ with the fluid velocity $u^{\mu} = u^{\mu}(x)$ and the particle momentum $p$ satisfying the on-shell condition, $p^2=m^2$.
Eq.(\ref{eq:fd_dis}) can be expanded using the Maxwell-Boltzmann (MB) distribution as
\be
f_{\rm eq} &=& \sum_{n \in {\mathbb N}_0} (-a)^{n} ( f_{\rm MB})^{n+1}, \qquad f_{\rm MB} = \exp \left( - \beta E_{\bf p} - \alpha \right). \label{eq:fd_dis_ex} 
\ee
Then, we consider the derivative to $f_{\rm eq}$ with $\alpha$ as
\be
f^{(n)}_{{\rm eq}} := (-\pd_\alpha )^n f_{{\rm eq}}, \qquad (n \in {\mathbb N})
\ee
and it can be written in terms of $f_{\rm eq}$ as
\be
f^{(1)}_{{\rm eq}} &=&  f_{{\rm eq}} \left( 1- a f_{{\rm eq}} \right), \nl
f^{(2)}_{{\rm eq}}  &=& f_{{\rm eq}} \left( 1 - a f_{{\rm eq}} \right) \left( 1- 2 a f_{{\rm eq}} \right), \\
f^{(3)}_{{\rm eq}} &=& f_{{\rm eq}} (1 - a f_{{\rm eq}} ) \left(1 -6 a f_{{\rm eq}} + 6 a^2 f_{{\rm eq}}^2 \right), \nl
\cdots && \nn
\ee
The derivative of $\beta$ and $E_{\bf p}$ can be also expressed by $f^{(n)}_{\rm eq}$ as
\be
\pd_\beta f_{{\rm eq}} &=& - E_{\bf p} f^{(1)}_{{\rm eq}}, \qquad \pd_{E_{\bf p}} f_{{\rm eq}} = - \beta {f}^{(1)}_{{\rm eq}}.
\ee
Then, we define\footnote{
In this expression, the energy dimensions are given by
\be
&& [z]=[\alpha] = [a] =0, \nl
&& [E_{\bf p}]=[1/\beta]= [m] = [\mu]=1, \\
&& [\Xi_{k}] = [\Psi^n_k] =[\Xi^{(n)}_k]= [\aleph_{k}^{(n,s)}] = k+2. \nn
\ee
}
\be
\Psi^n_k(z,\alpha)  &:=&  \int_{E_{\bf p}} E_{\bf p}^k \, e^{- n \left( E_{\bf p} \beta + \alpha \right)},\qquad (n \in {\mathbb N}, k \in {\mathbb Z}) \nl
&=& \frac{m^{k+2}}{2 \pi^2} \int^{\infty}_1 dp \, p^k \sqrt{p^2-1} e^{- n \left( p z + \alpha \right)} = \Psi^1_k(n z,n \alpha), \\
 \Xi_{k}(z,\alpha) &:=& \int_{E_{\bf p}} E_{\bf p}^k \, f_{\rm eq} = \sum_{n \in {\mathbb N}_0} ( -a)^n \Psi^{n+1}_k(z,\alpha) 
, \qquad (k \in {\mathbb Z}), \\
\Xi^{(n)}_{k}(z,\alpha)  &:=& (-\pd_\alpha)^n \Xi_{k}(z,\alpha) = \int_{E_{\bf p}} E_{\bf p}^k \, f_{\rm eq}^{(n)}, \qquad (n \in {\mathbb N}, \, k \in {\mathbb Z}), \\
\aleph^{(n,s)}_{k}(z,\alpha) &:=& \sum_{t=0}^s
\begin{pmatrix}
  s \\
  t
\end{pmatrix} (- m^2)^t \, \Xi^{(n)}_{k-2t}(z,\alpha), \qquad (n,s \in {\mathbb N}, \, k \in {\mathbb Z}), 
\ee
where {
\be
z := \beta m, \qquad \int_{E_{\bf p}} := \frac{1}{2 \pi^2} \int_{m}^\infty d E_{\bf p} \, \sqrt{E_{\bf p}^2 -m^2}.
\ee
}
It is notable that, since the integration measure of $E_{\bf p}$ has a cut-off in the lower bound due to the mass, the mode $k$ can be taken as a negative integer\cite{Ambrus:2022vif}.

By using these functions, the transport coefficients are given by
\be
\delta_{\Pi \Pi} &:=& -\frac{1}{9} \sum_{k=-1}^{1} \Phi_{\Pi,k} \left( k \aleph^{(1,2)}_{k+2} - \beta \aleph^{(2,2)}_{k+3} \right), \\
\lambda_{\Pi \pi} &:=& \frac{2}{45} \Phi_{\pi,-1} \left( \aleph^{(1,3)}_{3} + \beta \aleph^{(2,3)}_{4} \right), \\
\tau_{\pi \pi} &:=& \frac{8}{105} \Phi_{\pi,-1} \left( \aleph^{(1,3)}_{3} + \beta \aleph^{(2,3)}_{4} \right),  \\
\delta_{\pi \pi} &:=& \frac{2}{45} \Phi_{\pi,-1} \left( \aleph^{(1,3)}_{3} + \beta \aleph^{(2,3)}_{4} \right), \\
\lambda_{\pi \Pi} &:=& - \frac{2}{15} \sum_{k=-1}^{1} \Phi_{\Pi,k} \left( k \aleph^{(1,2)}_{k+2} - \beta \aleph^{(2,2)}_{k+3} \right), \\
\beta_\pi &:=& \frac{\beta}{15} \aleph^{(1,2)}_{3}, \\
\widetilde{\beta}_\Pi &:=& - \frac{1}{3} \left[ \beta \left( \chi_\beta - \frac{1}{3} \right) \aleph^{(1,1)}_{3} + \chi_\alpha \aleph^{(1,1)}_{2} + \frac{m^2 \beta}{3} \aleph^{(1,1)}_{1}  \right],
\ee
where
\be
&& \Phi_{\Pi,1} = -\frac{m^2 \beta}{3 \beta_\Pi} \cdot \frac{(\Xi^{(1)}_{1})^2 - \Xi^{(1)}_{2} \Xi^{(1)}_{0}}{(\Xi^{(1)}_{2})^2 - \Xi^{(1)}_{3} \Xi^{(1)}_{1}}, \qquad  \Phi_{\Pi,0} = \frac{m^2 \beta}{3 \beta_\Pi} \cdot \frac{\Xi^{(1)}_{2} \Xi^{(1)}_{1} - \Xi^{(1)}_{3} \Xi^{(1)}_{0}}{(\Xi^{(1)}_{2})^2 - \Xi^{(1)}_{3} \Xi^{(1)}_{1}}, \nl
&& \Phi_{\Pi,-1} = - \frac{m^2 \beta}{3 \beta_\Pi}, \\ 
&& \Phi_{\pi,-1} = \frac{\beta}{2 \beta_\pi}, \\
&& \Phi_{n,0} = - \frac{1}{\kappa_n} \cdot \frac{\aleph^{(1,1)}_{2}}{\aleph^{(1,1)}_{3}}, \qquad \Phi_{n,-1} = \frac{1}{\kappa_n}, \\ \nl
&& \kappa_n := - \frac{1}{3} \left[ \aleph^{(1,1)}_{1} - \frac{( \aleph^{(1,1)}_{2} )^2}{\aleph^{(1,1)}_{3}} \right], \\
&& \beta_\Pi := \frac{m^4 \beta}{9}  \left[ \frac{(\Xi^{(1)}_{1})^3 - 2 \Xi^{(1)}_{2}  \Xi^{(1)}_{1} \Xi^{(1)}_{0} + \Xi^{(1)}_{3}  (\Xi^{(1)}_{0})^2}{(\Xi^{(1)}_{2})^2 - \Xi^{(1)}_{3} \Xi^{(1)}_{1}} + \Xi^{(1)}_{-1} \right], \\
&& \chi_\beta :=  - \beta^{-1} \frac{\left( {\cal E}_0 + P \right)  \Xi^{(1)}_{1} - n_0 \Xi^{(1)}_{2}}{(\Xi^{(1)}_{2})^2 - \Xi^{(1)}_{3} \Xi^{(1)}_{1}}, \label{eq:chi_b_ap} \\
&& \chi_\alpha := \frac{\left( {\cal E}_0 + P \right)  \Xi^{(1)}_{2} - n_0 \Xi^{(1)}_{3}}{(\Xi^{(1)}_{2})^2 - \Xi^{(1)}_{3} \Xi^{(1)}_{1}}. \label{eq:chi_a_ap} 
\ee
$E_0$, $P$, and $n_0$ in Eqs.(\ref{eq:chi_b_ap})(\ref{eq:chi_a_ap}) are the energy density, bulk pressure, and charge density, respectively, given by
\be
   {\cal E}_0 = \Xi_{2}, \qquad P = \frac{\Xi_{2} - m^2 \Xi_{0}}{3},  \qquad n_0 = \Xi_1.
\ee

The derivations of these quantities are summarized in App.\ref{sec:derive_ODEs}.

\subsection{Derivation of ODEs} \label{sec:derive_ODEs}
Let us start with the standard BGK kernel defined as
\be
C[f] = - \frac{E_{\bf p}}{\tau_R} \left( f - f_{\rm eq}\right), 
\ee
and the gradient expansion by introducing the Knudsen number $\epsilon$ as
\be
&& p^\mu \pd_\mu f  = \frac{1}{\epsilon} C[f], \qquad f = \sum_{n=0}^\infty \epsilon^n f^{[n]}, \qquad f^{[0]} = f_{\rm eq}.
\ee
It gives the recursion relation in terms of $f^{[n]}$ as
\be
f^{[n]} = - \frac{\tau_R}{E_{\bf p}} p^{\mu} \pd_\mu  f^{[n-1]} = \left( - \frac{\tau_R}{E_{\bf p}} p^{\mu} \pd_\mu \right)^{n} f^{[0]}.
\ee
Here, we truncate the higher order of $\epsilon$ and define the distribution up to $O(\epsilon)$, as
\be
f \approx f_{\rm eq} + \epsilon f^{[1]}, \qquad f^{[1]} = - \frac{\tau_R}{E_{\bf p}} p^{\mu} \pd_\mu  f^{[0]}.
\ee
When taking the Landau frame, the EM tensor and current density can be written by
\be
&& T^{\mu \nu} = {\cal E} u^{\mu} u^{\nu} - (P + \Pi) \Delta^{\mu \nu} + \pi^{\mu \nu}, \\
&& N^{\mu} = n u^{\mu} + n^{\mu},
\ee
The energy density and the charge density are evaluated by the local equilibrium, $f_{\rm eq}$, which we denote ${\cal E}={\cal E}_0$ and $n=n_0$.
Imposing the conservation laws of the EM tensor and the current density yields
\be
&& D {\cal E} + \left( {\cal E} + P + \Pi \right) \theta - \pi^{\mu \nu} \sigma_{\mu \nu}= 0, \label{eq:cons1} \\
&& \left( {\cal E} + P + \Pi \right) D u^{\mu} - \nabla^\mu (P+\Pi) + \Delta^{\mu}_{\ \nu} {\cal D}_{\rho} \pi^{\rho \nu} = 0, \label{eq:cons2} \\
&& D n + n \theta - {\cal D}_{\mu} n^{\mu} = 0, \label{eq:cons3} 
\ee
where ${\cal D}_{\mu}$ is the covariant derivative, and
\be
&& D:= u^{\mu} {\cal D}_{\mu}, \\ 
&& \nabla^{\mu} := \Delta^{\mu \nu} {\cal D}_{\nu}, \\ 
&& \theta:= {\cal D}_{\mu} u^{\mu}, \\ 
&& \sigma^{\mu \nu} := \frac{1}{2} \left( \nabla^{\mu} u^{\nu} + \nabla^{\nu} u^{\mu} \right) - \frac{1}{3} \Delta^{\mu \nu} \theta. 
\ee
The Navier-Stokes expression implies that
\be
\Pi =- \zeta \theta, \qquad \pi^{\mu \nu} = 2 \eta \sigma^{\mu \nu}, \qquad n^\mu = \kappa_n \nabla^\mu \alpha.
\ee
 
We define the hydro variables by $f \approx f_{\rm eq} + \epsilon f^{[1]}$
as
\be
&& {\cal E} := u_{\alpha} u_{\beta} \int_p p^\alpha p^\beta f, \qquad \quad \ \ \ \
 P :=  - \frac{1}{3} \Delta_{\alpha \beta} \int_p p^\alpha p^\beta f_{\rm eq}, \nl
&& \Pi := - \frac{1}{3} \Delta_{\alpha \beta} \int_p p^\alpha p^\beta f^{[1]}, \qquad \pi^{\mu \nu} :=  \Delta_{\alpha \beta}^{\mu \nu} \int_p p^\alpha p^\beta f^{[1]}, \label{eq:def_Pipi} \\ 
 && n := u_{\alpha} \int_p p^\alpha f, \qquad \qquad \qquad  \ n^\mu := \Delta_{\ \alpha}^{\mu} \int_p p^\alpha f^{[1]}, \nn 
\ee
with the momentum integration defined by
\be
&& \int_p := \int \frac{d^4 p}{(2 \pi)^3 \sqrt{- \det g}} \delta(p^2) \cdot 2 \theta(p^0), \qquad p^2 = m^2,
\ee
using the delta function $\delta(x)$ and the step function $\theta(x)$.
Notice that $\Delta_{\alpha \beta} p^{\alpha} p^{\beta} = - E_{\bf p}^2 + m^2$.

Imposing the Landau matching condition, $u_\nu T^{\mu \nu} = {\cal E} u^{\mu}$, and the matching condition yields
\be
{\cal E} = {\cal E}_0 = \Xi_{2}, \qquad P = \frac{\Xi_{2} - m^2 \Xi_{0}}{3}, \qquad n = n_0 = \Xi_{1}, \label{eq:E0Pn0}
\ee
where the subscript ``$0$'' denotes the quantity evaluated by $f_{\rm eq}$.
Taking derivatives gives
\be
D {\cal E} &=& \left( D \beta \cdot \pd_\beta + D \alpha \cdot \pd_\alpha \right) \Xi_{2} \nl
&=& - D \beta \cdot  \Xi^{(1)}_{3} - D \alpha \cdot  \Xi^{(1)}_{2}, \\
\nabla^\mu P &=& \left( \nabla^\mu \beta \cdot \pd_\beta + \nabla^\mu \alpha \cdot \pd_\alpha \right) \frac{\Xi_{2} - m^2 \Xi_{0}}{3} \nl
&=& -\frac{1}{3}\left[ \aleph^{(1,1)}_{3} \nabla^\mu \beta  + \aleph^{(1,1)}_{2} \nabla^\mu \alpha  \right], \\
D n &=& \left( D \beta \cdot \pd_\beta + D \alpha \cdot \pd_\alpha \right) \Xi_{1} \nl
&=& - D \beta \cdot  \Xi^{(1)}_{2} - D \alpha \cdot  \Xi^{(1)}_{1}.
\ee
Using Eqs.(\ref{eq:cons1})-(\ref{eq:cons3}), one obtains
\be
&& - D \beta \cdot  \Xi^{(1)}_{3} - D \alpha \cdot  \Xi^{(1)}_{2} + \left( {\cal E} + P + \Pi \right) \theta - \pi^{\mu \nu} \sigma_{\mu \nu} = 0, \label{eq:cons12} \\
&& \left( {\cal E} + P + \Pi \right) D u^{\mu} +\frac{1}{3}\left[ \aleph^{(1,1)}_{3} \nabla^\mu \beta + \aleph^{(1,1)}_{2} \nabla^\mu \alpha \right] - \nabla^\mu \Pi + \Delta^{\mu}_{\ \nu} {\cal D}_{\rho} \pi^{\rho \nu} = 0, \label{eq:cons22} \\
&& - D \beta \cdot  \Xi^{(1)}_{2} - D \alpha \cdot  \Xi^{(1)}_{1} + n \theta - {\cal D}_{\mu} n^{\mu} = 0, \label{eq:cons32} 
\ee
and thus,
\be
D \beta &=& - \frac{[\left( {\cal E} + P + \Pi \right) \theta - \pi^{\mu \nu} \sigma_{\mu \nu} ] \Xi^{(1)}_{1} - (n \theta -{\cal D}_\mu n^\mu ) \Xi^{(1)}_{2}}{(\Xi^{(1)}_{2})^2 - \Xi^{(1)}_{3} \Xi^{(1)}_{1}} \nl
&\approx& - \frac{\left( {\cal E}_0 + P \right)  \Xi^{(1)}_{1} - n_0 \Xi^{(1)}_{2}}{(\Xi^{(1)}_{2})^2 - \Xi^{(1)}_{3} \Xi^{(1)}_{1}} \theta \quad = \beta \chi_\beta \theta, \label{eq:Db} \\
D \alpha &=&  \frac{[\left( {\cal E} + P + \Pi \right) \theta - \pi^{\mu \nu} \sigma_{\mu \nu} ] \Xi^{(1)}_{2} - (n \theta -{\cal D}_\mu n^\mu ) \Xi^{(1)}_{3}}{(\Xi^{(1)}_{2})^2 - \Xi^{(1)}_{3} \Xi^{(1)}_{1}} \nl
&\approx& \frac{\left( {\cal E}_0 + P \right)  \Xi^{(1)}_{2} - n_0 \Xi^{(1)}_{3}}{(\Xi^{(1)}_{2})^2 - \Xi^{(1)}_{3} \Xi^{(1)}_{1}} \theta \quad = \chi_\alpha \theta, \label{eq:Da} \\
\nabla^\mu \beta &=& -\frac{\aleph^{(1,1)}_{2} \nabla^\mu \alpha + 3 \left[  \left( {\cal E} + P + \Pi \right) D u^{\mu} - \nabla^\mu \Pi + \Delta^{\mu}_{\ \nu} {\cal D}_{\rho} \pi^{\rho \nu} \right]}{\aleph^{(1,1)}_{3}} \nl
&\approx& -\frac{\aleph^{(1,1)}_{2}}{\aleph^{(1,1)}_{3}} \nabla^\mu \alpha - \frac{3({\cal E}_0 + P)}{\aleph^{(1,1)}_{3}} D u^{\mu}, \label{eq:nab_beta}
\ee
where ``$\approx$'' denotes the equilibrium approximation, and
\be
   \chi_\beta &:=&  - \beta^{-1} \frac{\left( {\cal E}_0 + P \right)  \Xi^{(1)}_{1} - n_0 \Xi^{(1)}_{2}}{(\Xi^{(1)}_{2})^2 - \Xi^{(1)}_{3} \Xi^{(1)}_{1}}, \\
   \chi_\alpha &:=& \frac{\left( {\cal E}_0 + P \right)  \Xi^{(1)}_{2} - n_0 \Xi^{(1)}_{3}}{(\Xi^{(1)}_{2})^2 - \Xi^{(1)}_{3} \Xi^{(1)}_{1}}. 
\ee
In addition, Eq.(\ref{eq:nab_beta}) can be reexpressed as
\be
D u^{\mu} &\approx& - \frac{\aleph^{(1,1)}_{3} \nabla^\mu \beta + \aleph^{(1,1)}_{2} \nabla^\mu \alpha}{3 \left( {\cal E}_0 + P \right)}.
\ee
Thus, Eqs.(\ref{eq:Db})(\ref{eq:Da}) including viscous variables can be written as
\be
\frac{d \beta}{d \tau} =  \frac{\beta}{\tau} \left[\chi_\beta + \gamma_\beta  ( \Pi - \widehat{\pi}) \right], \qquad \frac{d \alpha}{d \tau} =  \frac{1}{\tau} \left[ \chi_\alpha - \gamma_\alpha  ( \Pi - \widehat{\pi} ) \right],
\ee
where
\be
\gamma_\beta := - \frac{\beta^{-1} \Xi^{(1)}_{1} }{(\Xi^{(1)}_{2})^2 - \Xi^{(1)}_{3} \Xi^{(1)}_{1}}, \qquad \gamma_\alpha := - \frac{\Xi^{(1)}_{2} }{(\Xi^{(1)}_{2})^2 - \Xi^{(1)}_{3} \Xi^{(1)}_{1}}.
\ee

Then, we introduce the equilibrium distribution equipping ``thermodynamic frame'', $f^*_{\rm eq}$ to be consistent between the conservation laws of hydrodynamics and kinetic theory\cite{Hoult:2021gnb,Banerjee:2012iz,Jensen:2012jh,Teaney:2013gca,Kovtun:2019hdm,Dash:2021ibx}.
In this frame, the thermodynamic variables and the fluid velocity has the gradient as ${\cal O}_* = {\cal O} + \pd {\cal O} + O(\pd^2)$ under the assumption that ${\cal O}_* \rightarrow {\cal O}$ as closing to the equilibrium.
Since we consider the NS hydro, we define ${\cal O}_* = {\cal O} + \delta {\cal O}$, where $\delta {\cal O} = O(\pd)$, and ignore the higher orders, $O(\pd^2)$.
Because of the gradient, $\delta {\cal O}$ can be interpreted as the first order effect in $f_{\rm eq}^*$, and we decompose it as $f_{\rm eq}^* := f_{\rm eq} + f^{[1]}_*$.
In contrast, one also has the first order effect defined by performing the CE expansion to $f_{\rm eq}$, and we denotes it as $f_{\epsilon}^{[1]}$.
Therefore, the first order distribution function can be decomposed into two parts, $f^{[1]} = f^{[1]}_* + f^{[1]}_\epsilon$.
From the definitions, these can be written down as
\be
f_*^{[1]} &=& - \left[ E_{\bf p} \delta \beta + \delta \alpha + \beta p^{\mu} \delta u_{\mu} \right] f^{(1)}_{{\rm eq}}, \\
f^{[1]}_{\epsilon} &=& - \frac{\tau_R}{E_{\bf p}} p^\mu \pd_{\mu} f_{\rm eq} \nl
&=& \tau_R \left[ \beta E_{\bf p} \left( \chi_\beta - \frac{1}{3} \right) \theta + \chi_\alpha \theta + \frac{\beta}{E_{\bf p}}  \left( p^\mu p^\nu \sigma_{\mu \nu} + \frac{m^2}{3} \theta \right) \right. \nl
  && \left. + \left( 1 - \frac{\beta \aleph^{(1,1)}_{3}}{3 \left( {\cal E}_0 + P \right)} \right) p^\mu \nabla_\mu \beta + \left( \frac{1}{E_{\bf p}} -\frac{\beta \aleph^{(1,1)}_{2}}{3 \left( {\cal E}_0 + P \right)} \right) p^\mu \nabla_\mu \alpha \right] {f}^{(1)}_{{\rm eq}}. 
\ee
Here, we assume that $\delta {\cal O}$ is sufficiently small around the equilibrium.
So that $\delta u^{\mu}$ is transverse to $u^\mu$:
\be
&& (u+\delta u)^\mu (u+\delta u)_\mu = 1 \quad \Rightarrow \quad u \cdot \delta u \approx 0.
\ee
$(\delta \beta, \delta \alpha, \delta u^{\mu})$ can be easily determined from the consistency with the Landau matching condition.
Since
\be
n_* &=& - \Xi^{(1)}_{2} \delta \beta - \Xi^{(1)}_{1} \delta \alpha, \\
n_{\epsilon} &=& \tau_R \left[ \beta \left\{ \left( \chi_\beta - \frac{1}{3} \right) \Xi^{(1)}_{2} + \frac{m^2}{3} \Xi^{(1)}_{0} \right\} + \chi_\alpha \Xi^{(1)}_{1} \right] \theta, \\
{\cal E}_* &=& - \Xi^{(1)}_{3} \delta \beta - \Xi^{(1)}_{2} \delta \alpha, \\
{\cal E}_{\epsilon} &=& \tau_R \left[ \beta \left\{ \left( \chi_\beta - \frac{1}{3} \right) \Xi^{(1)}_{3} + \frac{m^2}{3} \Xi^{(1)}_{1} \right\} + \chi_\alpha \Xi^{(1)}_{2} \right] \theta, \\
q^\mu_{*} &=&  \frac{\beta}{3} \aleph^{(1,1)}_{3} \delta u^{\mu}, \\
q^\mu_{\epsilon} &=& - \frac{\tau_R}{3} \left[ \beta \aleph^{(1,1)}_{3} - 3 \left( {\cal E}_0 + P \right) \right]  D u^{\mu},
\ee
where $q^{\mu}$ is the heat flow, imposing the Landau condition gives ${\cal O}_* + {\cal O}_\epsilon=0$ for these variables and
\be
&& \delta \beta = \tau_R C_\beta \theta, \qquad \delta \alpha = \tau_R C_\alpha \theta, \qquad \delta u^{\mu} = \tau_R C_u D u^\mu, \label{eq:deltabau}
\ee
where
\be
C_\beta &:=& \beta \left[ \left( \chi_\beta - \frac{1}{3} \right) - \frac{m^2}{3} \cdot \frac{(\Xi^{(1)}_{1})^2 - \Xi^{(1)}_{2} \Xi^{(1)}_{0}}{(\Xi^{(1)}_{2})^2 - \Xi^{(1)}_{3} \Xi^{(1)}_{1}} \right] = 0, \label{eq:Cb}\\
C_{\alpha} &:=&  \chi_\alpha + \frac{m^2 \beta}{3} \cdot  \frac{ \Xi^{(1)}_{2} \Xi^{(1)}_{1} - \Xi^{(1)}_{3}  \Xi^{(1)}_{0}}{(\Xi^{(1)}_{2})^2 - \Xi^{(1)}_{3} \Xi^{(1)}_{1}}, \label{eq:Ca} \\
 C_u &:=&  1 - \frac{3 \left( {\cal E}_0 + P \right)}{\beta \aleph^{(1,1)}_{3}} =0. \label{eq:Cu}
\ee
Thus, only $C_\alpha$ is non-zero and the others are identically zero through nontrivial relations among $\Xi_{k}^{(n)}$.
From the result, the first order distribution is obtained as
\be
f^{[1]} &=& f_{*}^{[1]} + f^{[1]}_{\epsilon} \nl
  &=& \tau_R \left[   \left\{ E_{\bf p} \beta \left( \chi_\beta - \frac{1}{3} \right) + \chi_\alpha - C_\alpha + \frac{m^2 \beta}{3} \frac{1}{E_{\bf p}} \right\} \theta \right. \nl
  && \qquad \left.  + \frac{\beta}{E_{\bf p}} p^\mu p^\nu \sigma_{\mu \nu} + \left( \frac{1}{E_{\bf p}} - \frac{\aleph^{(1,1)}_{2}}{\aleph^{(1,1)}_{3}}  \right) p^\mu \nabla_\mu \alpha  \right] f^{(1)}_{{\rm eq}}, \label{eq:f1_form}
\ee
and the variables can be expressed as
\be
n_*^{\mu} &=& 0, \\
n_\epsilon^{\mu} &=& - \frac{\tau_R}{3}  \left(  \aleph^{(1,1)}_{1} - \frac{( \aleph^{(1,1)}_{2} )^2}{\aleph^{(1,1)}_{3}}\right) \nabla^\mu \alpha, \\
\Pi_* &=&  -\frac{\tau_R}{3}  \aleph^{(1,1)}_{2} C_\alpha \theta, \\
\Pi_\epsilon &=& \frac{\tau_R}{3} \left[ \beta \left( \chi_\beta - \frac{1}{3} \right) \aleph^{(1,1)}_{3} + \chi_\alpha \aleph^{(1,1)}_{2} + \frac{m^2 \beta}{3} \aleph^{(1,1)}_{1}  \right] \theta,  \\
\pi^{\mu \nu}_* &=& 0, \\
\pi^{\mu \nu}_\epsilon &=&  \tau_R \frac{2 \beta}{15} \aleph^{(1,2)}_{3} \sigma^{\mu \nu},
\ee
and thus,
\be
&& n^{\mu} = n^{\mu}_* + n^{\mu}_\epsilon =  \tau_R \kappa_n \nabla^\mu \alpha, \nl
&& \Pi = \Pi_* + \Pi_\epsilon = - \tau_R \beta_\Pi \theta, \\
&& \pi^{\mu \nu} = \pi^{\mu \nu}_* + \pi^{\mu \nu}_\epsilon = 2 \tau_R \beta_\pi \sigma^{\mu \nu} \nl \nl
&\Rightarrow \quad& \nabla^\mu \alpha = \frac{n^\mu}{\tau_R \kappa_n},  \qquad \theta = -\frac{\Pi}{\tau_R \beta_\Pi}, \qquad  \sigma^{\mu \nu} = \frac{\pi^{\mu \nu}}{2 \tau_R \beta_\pi},
\ee
where
\be
\kappa_n &:=& - \frac{1}{3} \left[ \aleph^{(1,1)}_{1} - \frac{( \aleph^{(1,1)}_{2} )^2}{\aleph^{(1,1)}_{3}} \right], \\
\beta_\Pi &:=& - \frac{1}{3} \left[ \beta \left( \chi_\beta - \frac{1}{3} \right) \aleph^{(1,1)}_{3} + (\chi_\alpha - C_\alpha) \aleph^{(1,1)}_{2} + \frac{m^2 \beta}{3} \aleph^{(1,1)}_{1}  \right] \nl
&=&\frac{m^4 \beta}{9}  \left[ \frac{(\Xi^{(1)}_{1})^3 - 2 \Xi^{(1)}_{2}  \Xi^{(1)}_{1} \Xi^{(1)}_{0} + \Xi^{(1)}_{3}  (\Xi^{(1)}_{0})^2}{(\Xi^{(1)}_{2})^2 - \Xi^{(1)}_{3} \Xi^{(1)}_{1}} + \Xi^{(1)}_{-1} \right], \\
\beta_\pi &:=& \frac{\beta}{15} \aleph^{(1,2)}_{3}.
\ee
The distribution (\ref{eq:f1_form}) can be expressed by
\be
f^{[1]} &=& \left[ \sum_{k=-1}^1 \Phi_{\Pi,k} E_{\bf p}^k \Pi + \Phi_{\pi,-1} E_{\bf p}^{-1} p^\mu p^\nu \pi_{\mu \nu} +  \sum_{k=-1}^0 \Phi_{n,k} E_{\bf p}^k p^\mu n_\mu  \right] f^{(1)}_{{\rm eq}}, \nl \label{eq:f1_form2}
\ee
where\footnote{
The energy dimensions of $\Phi_{{\cal O},k}$ is given by
  \be
     [\Phi_{\Pi,k}] = -4-k, \qquad [\Phi_{\pi,-1}] = -5, \qquad [\Phi_{n,k}] = -4-k.
  \ee
  }
\be
&& \Phi_{\Pi,1} = - \frac{\beta}{\beta_\Pi} \left( \chi_\beta - \frac{1}{3} \right) = -\frac{m^2 \beta}{3 \beta_\Pi} \cdot \frac{(\Xi^{(1)}_{1})^2 - \Xi^{(1)}_{2} \Xi^{(1)}_{0}}{(\Xi^{(1)}_{2})^2 - \Xi^{(1)}_{3} \Xi^{(1)}_{1}}, \nl
&& \Phi_{\Pi,0} = - \frac{\chi_\alpha - C_\alpha}{\beta_\Pi} = \frac{m^2 \beta}{3 \beta_\Pi} \cdot \frac{\Xi^{(1)}_{2} \Xi^{(1)}_{1} - \Xi^{(1)}_{3} \Xi^{(1)}_{0}}{(\Xi^{(1)}_{2})^2 - \Xi^{(1)}_{3} \Xi^{(1)}_{1}}, \\
&& \Phi_{\Pi,-1} = - \frac{m^2 \beta}{3 \beta_\Pi}, \nl
&& \Phi_{\pi,-1} = \frac{\beta}{2 \beta_\pi}, \\
&& \Phi_{n,0} = - \frac{1}{\kappa_n} \cdot \frac{\aleph^{(1,1)}_{2}}{\aleph^{(1,1)}_{3}}, \qquad \Phi_{n,-1} = \frac{1}{\kappa_n}.
\ee

Then, we derive the ODE of dissipative quantities.
By rewriting the Boltzmann equation as
\be
p^{\mu} \pd_\mu f = -\frac{E_{\bf p}}{\tau_R}  f^{[1]} &\quad \Rightarrow \quad &    D  f^{[1]} = -\frac{1}{\tau_R} f^{[1]} - D  f_{\rm eq} - \frac{1}{E_{\bf p}} p^\mu \nabla_\mu f,
\ee
and using the definitions in Eq.(\ref{eq:def_Pipi}), one finds that\cite{Denicol:2010xn},
\be
&& D n^\mu = -\frac{n^\mu}{\tau_R} - \Delta^{\mu}_{\ \alpha} \int_p p^\alpha \left(D  f_{\rm eq} + \frac{1}{E_{\bf p}} p^\nu \nabla_\nu f \right), \\
&& D \Pi = -\frac{\Pi}{\tau_R}  + \frac{\Delta_{\alpha \beta}}{3} \int_p p^\alpha p^\beta \left(D  f_{\rm eq} + \frac{1}{E_{\bf p}} p^\mu \nabla_\mu f \right), \\
&& D \pi^{\mu \nu} = -\frac{\pi^{\mu \nu}}{\tau_R}  - \Delta^{\mu \nu}_{\alpha \beta} \int_p p^\alpha p^\beta \left(D  f_{\rm eq} + \frac{1}{E_{\bf p}} p^\rho \nabla_\rho f \right).
\ee
The leading order is obtained as
\be
D f_{\rm eq} &=& - \left[ \beta E_{\bf p} \chi_\beta \theta  + \chi_\alpha \theta + \beta p^\mu D u_{\mu} \right] {f}^{(1)}_{{ \rm eq}}, \\
\frac{1}{E_{\bf p}} p^\mu \nabla_\mu f_{\rm eq} &=&  \left[ \frac{3 \left({\cal E}_0 + P \right)}{\aleph^{(1,1)}_{3}} p^{\mu} D u_{\mu} -  \left( \frac{1}{E_{\bf p}} - \frac{\aleph^{(1,1)}_{2}}{\aleph^{(1,1)}_{3}} \right) p^{\mu} \nabla_\mu \alpha \right. \nl
  && \left. + \frac{\beta}{3}  \left( E_{\bf p} - \frac{m^2}{E_{\bf p}} \right) \theta - \frac{\beta}{E_{\bf p}} p^{\mu} p^\nu \sigma_{\mu \nu}   \right] f^{(1)}_{{ \rm eq}},
\ee
and 
\be
D f_{\rm eq} + \frac{1}{E_{\bf p}} p^\mu \nabla_\mu f_{\rm eq} &=& - \left[ \left\{ \beta E_{\bf p} \left( \chi_\beta  - \frac{1}{3} \right) + \chi_\alpha + \frac{m^2 \beta}{3 E_{\bf p}} \right\} \theta + \frac{\beta}{E_{\bf p}} p^{\mu} p^\nu \sigma_{\mu \nu}  \right. \nl
  && \left. + \left( \beta -  \frac{3 \left({\cal E}_0 + P \right)}{\aleph^{(1,1)}_{3}} \right) p^{\mu} D u_{\mu} + \left( \frac{1}{E_{\bf p}} - \frac{\aleph^{(1,1)}_{2}}{\aleph^{(1,1)}_{3}}  \right) p^{\mu} \nabla_\mu \alpha \right] f^{(1)}_{{ \rm eq}}. \label{eq:sum_rhs_0} \nl
\ee
Thus,
\be
-\Delta^{\mu}_{\ \alpha} \int_p p^\alpha \left( D f_{\rm eq} + \frac{1}{E_{\bf p}} p^\mu \nabla_\mu f_{\rm eq}\right) &=& \kappa_n \nabla^\mu \alpha, \\
\frac{\Delta_{\alpha \beta}}{3} \int_p p^\alpha p^\beta \left( D f_{\rm eq} + \frac{1}{E_{\bf p}} p^\mu \nabla_\mu f_{\rm eq}\right) &=& - \widetilde{\beta}_\Pi \theta, \\
-\Delta^{\mu \nu}_{\alpha \beta} \int_p p^\alpha p^{\beta}
\left( D f_{\rm eq} + \frac{1}{E_{\bf p}} p^\mu \nabla_\mu f_{\rm eq}\right) &=& 2 \beta_\pi \sigma^{\mu \nu},
\ee
where
\be
\widetilde{\beta}_{\Pi} := -\frac{1}{3} \left[ \beta \left( \chi_\beta - \frac{1}{3} \right) \aleph^{(1,1)}_{3} + \chi_\alpha \aleph^{(1,1)}_{2} + \frac{m^2 \beta}{3} \aleph^{(1,1)}_{1}  \right].
\ee
The next leading order is obtained as
\be
&& \frac{1}{E_{\bf p}} p^\mu \nabla_\mu f^{[1]} \nl
&=& \left[ \sum_{{\cal O} \in \{ \beta, \alpha \}} \left( \sum_{k=-1}^1  \pd_{\cal O}  \Phi_{\Pi,k} \cdot \nabla_\mu {\cal O} \cdot E_{\bf p}^{k-1} p^\mu \Pi + \pd_{\cal O} \Phi_{\pi,-1} \cdot \nabla_\mu {\cal O} \cdot E_{\bf p}^{-2} p^\mu p^\nu p^\rho \pi_{\nu \rho} \right. \right. \nl
  && \left. \left. + \sum_{k=-1}^0 \pd_{\cal O} \Phi_{n,k} \cdot \nabla_\mu {\cal O} \cdot E_{\bf p}^{k-1} p^\mu p^\nu n_\nu \right) \right. \nl
  &&  \left. + \left( \sum_{k=-1}^{1} k \Phi_{\Pi,k} E_{\bf p}^{k-2} p^\mu p^{\lambda} \Pi - \Phi_{\pi,-1} E_{\bf p}^{-3}  p^\mu p^\nu p^\rho p^\lambda \pi_{\nu \rho} \right. \right. \nl
  && \left. \left. - \Phi_{n,-1} E_{\bf p}^{-3} p^\mu p^\nu p^\lambda n_\nu \right) \cdot \left( \sigma_{\mu \lambda} + \frac{1}{3} \Delta_{\mu \lambda} \theta \right)  \right. \nl
  && \left. + \sum_{k=-1}^{1} \Phi_{\Pi,k} E_{\bf p}^{k-1} p^\mu \nabla_\mu \Pi + \Phi_{\pi,-1} E_{\bf p}^{-2} p^\mu p^\nu p^\rho \nabla_{\mu} \pi_{\nu \rho} + \sum_{k=-1}^0 \Phi_{n,k} E_{\bf p}^{k-1} p^\mu p^\nu \nabla_\mu n_\nu  \right] {f}^{(1)}_{{\rm eq}} \nl
&& - \left[ \sum_{k=-1}^1 \Phi_{\Pi,k} E_{\bf p}^{k-1} p^\mu \Pi + \Phi_{\pi,-1} E_{\bf p}^{-2} p^\mu p^\nu p^\rho \pi_{\nu \rho} \right. \nl
  && \left. + \sum_{k=-1}^0 \Phi_{n,k} E_{\bf p}^{k-1} p^\mu p^\nu n_\nu  \right] \cdot \left[ E_{\bf p} \nabla_{\mu} \beta + \nabla_{\mu} \alpha + \beta p^\lambda \left( \sigma_{\mu \lambda} + \frac{1}{3} \Delta_{\mu \lambda} \theta \right) \right] {f}^{(2)}_{{\rm eq}}, \label{eq:sum_rhs_1}
\ee
Thus, one finds that
\be
&& -\int_p  \frac{p^{\langle \mu \rangle}}{E_{\bf p}} p^\mu \nabla_\mu f^{[1]} \nl
 &\xrightarrow{\scriptsize \mbox{B-sym}}& - \frac{2}{15} \left[ \Phi_{\pi,-1} \aleph^{(1,2)}_{2} \left( \nabla^{\nu} \pi_{\nu}^{\ \mu} + u^\mu \pi_{\nu \rho} \sigma^{\nu \rho} \right) \right] = 0, \nl \\ 
&& \frac{\Delta_{\alpha \beta}}{3} \int_p \frac{p^\alpha p^\beta}{E_{\bf p}} p^\mu \nabla_\mu f^{[1]} \nl
\nl
 &\xrightarrow{\scriptsize \mbox{B-sym}}&  \frac{1}{9} \left[ \sum_{k=-1}^{1} \Phi_{\Pi,k} \left( k \aleph^{(1,2)}_{k+2} - \beta \aleph^{(2,2)}_{k+3} \right) \Pi \theta + \frac{2}{5}  \Phi_{\pi,-1} \left( \aleph^{(1,3)}_{3} + \beta \aleph^{(2,3)}_{4} \right) \left( \pi_{\zeta}^{\ \zeta}/\tau \right) \right], \nl \\
&& - \frac{p^{\langle \mu} p^{\nu \rangle}}{E_{\bf p}} p^\mu \nabla_\mu f^{[1]} \nl
 &\xrightarrow{\scriptsize \mbox{B-sym}}&  - \frac{2}{15} \left[ \sum_{k=-1}^{1} \Phi_{\Pi,k} \left( k  \aleph^{(1,2)}_{k+2} - \beta \aleph^{(2,2)}_{k+3} \right) \Pi  \sigma^{\mu \nu} + \frac{11}{14}  \Phi_{\pi,-1} \left( \aleph^{(1,3)}_{3} + \beta \aleph^{(2,3)}_{4} \right) \pi_{\zeta}^{\ \zeta}  \sigma^{\mu\nu}  \right], \nl
\ee
where ``B-sym'' denotes the expression under the Bjorken symmetry.
Therefore, under the Bjorken symmetry, the ODEs of viscosities are obtained as
\be
\frac{d \Pi}{d \tau} &=& - \frac{\Pi}{\tau_R}  - \widetilde{\beta}_\Pi \theta - \delta_{\Pi \Pi} \Pi \theta + \lambda_{\Pi \pi} \pi_{\mu \nu} \sigma^{\mu \nu}, \\
\frac{d \pi^{\mu \nu}}{d \tau} &=& - \frac{\pi^{\mu \nu}}{\tau_R} +  2 \beta_\pi  \sigma^{\mu \nu} - \tau_{\pi \pi} \pi_{\rho}^{\ \langle \mu} \sigma^{\nu \rangle \rho} - \delta_{\pi \pi} \pi^{\mu \nu} \theta + \lambda_{\pi \Pi} \Pi \sigma^{\mu \nu},
\ee
where
\be
\delta_{\Pi \Pi} &:=& -\frac{1}{9} \sum_{k=-1}^{1} \Phi_{\Pi,k} \left( k \aleph^{(1,2)}_{k+2} - \beta \aleph^{(2,2)}_{k+3} \right), \nl
\lambda_{\Pi \pi} &:=& \frac{2}{45} \Phi_{\pi,-1} \left( \aleph^{(1,3)}_{3} + \beta \aleph^{(2,3)}_{4} \right), \nl
\tau_{\pi \pi} &:=& \frac{8}{105} \Phi_{\pi,-1} \left( \aleph^{(1,3)}_{3} + \beta \aleph^{(2,3)}_{4} \right),  \\
\delta_{\pi \pi} &:=& \frac{2}{45} \Phi_{\pi,-1} \left( \aleph^{(1,3)}_{3} + \beta \aleph^{(2,3)}_{4} \right), \nl
\lambda_{\pi \Pi} &:=& - \frac{2}{15} \sum_{k=-1}^{1} \Phi_{\Pi,k} \left( k \aleph^{(1,2)}_{k+2} - \beta \aleph^{(2,2)}_{k+3} \right). \nn
\ee
Notice that the Bjorken symmetry gives
\be
n^{\mu} = 0.
\ee
In summary, the ODEs are given by $(\widehat{\pi}:=\pi_{\zeta}^{\ \zeta})$
\be
\frac{d \beta}{d \tau} &=&  \frac{\beta}{\tau} \left[\chi_\beta + \gamma_\beta  ( \Pi - \widehat{\pi}) \right], \\
\frac{d \alpha}{d \tau} &=&  \frac{1}{\tau} \left[ \chi_\alpha - \gamma_\alpha  ( \Pi - \widehat{\pi} ) \right], \\
\frac{d \Pi}{d \tau} &=& - \frac{\Pi}{\tau_R}  - \frac{1}{\tau} \left[ \widetilde{\beta}_\Pi + \delta_{\Pi \Pi} \Pi - \lambda_{\Pi \pi} \widehat{\pi} \right], \\
\frac{d \widehat{\pi}}{d \tau} &=& - \frac{\widehat{\pi}}{\tau_R} + \frac{1}{\tau} \left[ \frac{4}{3} \beta_\pi  - \left( \frac{1}{3} \tau_{\pi \pi}  + \delta_{\pi \pi} \right) \widehat{\pi} + \frac{2}{3} \lambda_{\pi \Pi} \Pi \right].
\ee

For convenience, we redefine the above ODEs by using the following new variables:
\be
\bar{\Pi} := \gamma_\beta \Pi, \qquad \bar{\pi} := \gamma_\beta \widehat{\pi}.
\ee
The modified ODEs are obtained as\footnote{
  In order to derive the modified ODEs, we used the following result:
\be
\frac{d \log \gamma_\beta}{d \tau} &=& \left( F_{\beta} \pd_\beta  + F_\alpha \pd_\alpha \right) \log \gamma_\beta \nl
&=& \left[ \frac{\beta}{\tau} \left( \chi_\beta +  \bar{\Pi} - \bar{\pi} \right) \pd_\beta  + \frac{1}{\tau} \left(\chi_\alpha - \beta \frac{\Xi^{(1)}_{2}}{\Xi^{(1)}_{1}}  ( \bar{\Pi} - \bar{\pi} ) \right) \pd_\alpha \right] \log \gamma_\beta \nl
&=& \frac{1}{\tau} \left[ \left( \beta  \chi_\beta  \pd_\beta + \chi_\alpha \pd_\alpha \right) \log \gamma_\beta  + \left( \bar{\Pi} - \bar{\pi} \right) \beta \left( \pd_\beta   - \frac{\Xi^{(1)}_{2}}{\Xi^{(1)}_{1}} \pd_\alpha \right) \log \gamma_\beta \right].
\ee
  }
\be
\frac{d \beta}{d \tau} &=& \frac{\beta}{\tau} \left[ \chi_\beta +  \bar{\Pi} - \bar{\pi} \right] \quad =: F_\beta, \label{eq:ODE_b0} \\
\frac{d \alpha}{d \tau} &=& \frac{1}{\tau} \left[ \chi_\alpha - \beta \frac{\Xi^{(1)}_{2}}{\Xi^{(1)}_{1}}  ( \bar{\Pi} - \bar{\pi} ) \right] \quad =: F_\alpha, \label{eq:ODE_a0} \\
\frac{d \bar{\Pi}}{d \tau} &=& -  \frac{\bar{\Pi}}{\tau_R} - \frac{1}{\tau} \left[ \bar{\beta}_\Pi + C_\Pi \bar{\Pi} - \lambda_{\Pi \pi} \bar{\pi} - D  \left( \bar{\Pi} - \bar{\pi} \right) \bar{\Pi} \right] \quad =: F_{\bar{\Pi}}, \label{eq:ODE_bP0} \\
\frac{d \bar{\pi}}{d \tau} &=& - \frac{\bar{\pi}}{\tau_R} - \frac{1}{\tau} \left[ - \frac{4}{3} \bar{\beta}_\pi  + C_\pi \bar{\pi} - \frac{2}{3} \lambda_{\pi \Pi} \bar{\Pi} - D \left(\bar{\Pi} - \bar{\pi} \right) \bar{\pi} \right] \quad =: F_{\bar{\pi}}, \label{eq:ODE_bp0}
\ee
where
\be
&& \bar{\beta}_\Pi := \gamma_\beta \widetilde{\beta}_\Pi, \qquad \bar{\beta}_\pi := \gamma_\beta \beta_\pi, \nl
&& C_{\Pi} := \delta_{\Pi \Pi} - \left( \beta  \chi_\beta  \pd_\beta + \chi_\alpha \pd_\alpha \right) \log \gamma_\beta, \nl
&& C_{\pi} := \frac{1}{3} \tau_{\pi \pi} + \delta_{\pi \pi} - \left( \beta  \chi_\beta  \pd_\beta + \chi_\alpha \pd_\alpha \right) \log \gamma_\beta, \nl
&& D := \beta \left( \pd_\beta   - \frac{\Xi^{(1)}_{2}}{\Xi^{(1)}_{1}} \pd_\alpha \right) \log \gamma_\beta.
\ee

\subsection{Asymptotic expansion around $z= \infty$ for transport coefficients} \label{sec:asy_exp}
{
  The asymptotic expansions around $z = +\infty$ for the transport coefficients in the ODEs (\ref{eq:ODE_b0})-(\ref{eq:ODE_bp0}) are obtained as 
}
\be
\chi_\beta &\sim& 0.6667 -1.6667 z^{-1} + 6.6667 z^{-2} \nl
&& + {\cal R}_a \left(  0.2578 z^{-1} -1.9312 z^{-2} \right) + O({\cal R}_a^2, z^{-3}), \label{eq:chi_b_asym_zinf} \\ \nl
\chi_\alpha &\sim& -0.6667 z + 1.6667 - 5.4167 z^{-1} + 19.1667 z^{-2}
\nl
&& + {\cal R}_a \left(-  0.2578  + 1.3524 z^{-1} - 6.7174 z^{-2} \right) + O({\cal R}_a^2, z^{-3}), \label{eq:chi_a_asym_zinf} \\ \nl
\beta \frac{\Xi^{(1)}_2}{\Xi^{(1)}_1} &\sim& z + 1.5000 + 1.8750 z^{-1} - 1.8750 z^{-2} \nl
&& + {\cal R}_a \left( -0.5303 - 0.4972 z^{-1} + 1.4864 z^{-2} \right) + O({\cal R}_a^2, z^{-3}), \\ \nl
\bar{\beta}_{\Pi}  &\sim& 0.5556 z^{-2}  - {\cal R}_a \cdot 0.0522 z^{-2} + O({\cal R}_a^2, z^{-3}), \\ \nl
\bar{\beta}_\pi &\sim& 0.6667 - 2.3333 z^{-1} + 11.3333 z^{-2} \nl
&& + {\cal R}_a \left( -0.1768 + 0.7697 z^{-1} - 4.7190 z^{-2} \right) + O({\cal R}_a^2, z^{-3}), \\ \nl
\lambda_{\Pi \pi} &\sim& 2.3333 - 2.3333 z^{-1} + 15.1667 z^{-2} \nl
&& + {\cal R}_a \left( 0.2062  z^{-1} - 2.1011 z^{-2} \right) + O({\cal R}_a^2, z^{-3}), \\ \nl
\lambda_{\pi \Pi} &\sim& 3.6000 - 16.8000 z^{-1} + 159.6000 z^{-2} \nl
&& + {\cal R}_a \left( 1.5313  z^{-1} - 18.4455 z^{-2} \right) + O({\cal R}_a^2, z^{-3}), \\ \nl
C_{\Pi} &\sim& 1.3333 - 14.0000 z^{-1} + 139.6667 z^{-2} \nl
&& + {\cal R}_a \left( 0.8894  z^{-1} - 14.8891 z^{-2} \right) + O({\cal R}_a^2, z^{-3}), \\ \nl
C_{\pi} &\sim& 2.0000 - 3.6667 z^{-1} + 30.5000 z^{-2} \nl
&& + {\cal R}_a \left( - 0.0626  z^{-1} - 2.8195 z^{-2} \right) + O({\cal R}_a^2, z^{-3}), \\ \nl
D &\sim& 1.0000 + 2.5000 z^{-1} - 13.7500 z^{-2} \nl
&& + {\cal R}_a \left( - 0.1626 -0.1795 z^{-1} + 0.6064 z^{-2} \right) + O({\cal R}_a^2, z^{-3}).
\ee
In addition, the energy density, bulk pressure, and charge density behave as
\be
   {\cal E}_0 &\sim& m^4 \frac{{\cal R}}{\sqrt{z^3}} \left[ 0.0635 + 0.2143 z^{-1} + 0.3497 z^{-2}  \right. \nl
     && \left. + {\cal R}_a \left( 0.0224 + 0.0379 z^{-1} + 0.0309 z^{-2} \right) +  O({\cal R}_a^2, z^{-3}) \right], \\ \nl
   P &\sim& m^4 \frac{{\cal R}}{\sqrt{z^3}} \left[ 0.0635 z^{-1} + 0.1191 z^{-2}  \right. \nl
     && \left. + {\cal R}_a \left( 0.0112 z^{-1} + 0.0105 z^{-2} \right) +  O({\cal R}_a^2, z^{-3}) \right], \\ \nl
   n_0 &\sim& m^3 \frac{{\cal R}}{\sqrt{z^3}} \left[ 0.0635 + 0.1191 z^{-1} + 0.0521 z^{-2} \right. \nl
     && \left. + {\cal R}_a \left( 0.0224 + 0.0210 z^{-1} + 0.0046 z^{-2} \right) +  O({\cal R}_a^2, z^{-3}) \right].
\ee

\subsection{The massless limit} \label{sec:massless_app}
We obtain ODEs of the massless Bjorken flow.
Since the massless limit gives
\be
&& \Xi_{k} \rightarrow -\frac{\Gamma(k+2) \beta^{-(k+2)}}{2 \pi^2 a} {\rm Li}_{k+2}\left({\cal R}_a \right), \nl
&& \Xi^{(n)}_{k} \rightarrow -\frac{\Gamma(k+2) \beta^{-(k+2)}}{2 \pi^2 a} {\rm Li}_{k+2-n}\left({\cal R}_a \right), \\
&& \aleph^{(n,s)}_{k} \rightarrow -\frac{\Gamma(k+2) \beta^{-(k+2)}}{2 \pi^2 a} {\rm Li}_{k+2-n}\left({\cal R}_a \right), 
\ee
where ${\cal R}_a := - a {\cal R}$ with ${\cal R}:=e^{- \alpha}$, one can find
\be
&& {\cal E}_0 \rightarrow - \frac{3 \beta^{-4}}{\pi^2 a} {\rm Li}_{4}\left({\cal R}_a\right) \qquad  P \rightarrow - \frac{\beta^{-4}}{\pi^2 a} {\rm Li}_{4}\left({\cal R}_a \right), \nl
&& \chi_\beta \rightarrow \frac{1}{3}, \qquad \chi_\alpha \rightarrow 0, \nl
&& \kappa_n \rightarrow \frac{\beta^{-3}}{\pi^2 a} \left[ \frac{{\rm Li}_{2}\left({\cal R}_a \right)}{3} - \frac{{\rm Li}_{3}\left({\cal R}_a \right)^2}{4 {\rm Li}_{4}\left({\cal R}_a \right)} \right], \nl
&& \beta_\Pi \rightarrow 0, \qquad \beta_\pi \rightarrow - \frac{4 \beta^{-4}}{5 \pi^2 a} {\rm Li}_{4}\left({\cal R}_a \right),
\ee
and $\Pi \propto \beta_\Pi \rightarrow 0$.
Thus, the well-defined construction of ODEs for the massless case needs to begin with Eq(\ref{eq:f1_form}) and reconstruct the ODEs from the exactly massless distribution:
\be
f^{[1]} &=& f_{*}^{[1]} + f^{[1]}_{\epsilon} \nl
&\rightarrow& \tau_R \left[  \frac{\beta}{E_{\bf p}} p^\mu p^\nu \sigma_{\mu \nu} + \left( \frac{1}{E_{\bf p}} - \frac{\beta {\rm Li}_3\left( {\cal R}_a \right)}{4 {\rm Li}_4\left({\cal R}_a \right)} \right) p^\mu \nabla_\mu \alpha  \right] f^{(1)}_{{\rm eq}} \nl
&=& \left[ \Phi_{\pi,-1} E_{\bf p}^{-1} p^\mu p^\nu \pi_{\mu \nu} +  \sum_{k=-1}^0 \Phi_{n,k} E_{\bf p}^k p^\mu n_\mu  \right] f^{(1)}_{{\rm eq}}. \label{eq:f1_form_ml}
\ee
It is notable that in the massless limit $C_{\alpha}$ given in Eq.(\ref{eq:Ca}) becomes
\be
C_{\alpha} \rightarrow 0,
\ee
and thus, thermodynamic frame naturally vanishes, i.e.
\be
f^{[1]}_* \rightarrow 0.
\ee
By repeating the same construction, the ODEs are obtained as
\be
\frac{d \beta}{d \tau} &=&  \frac{\beta}{\tau} \left[ \frac{1}{3}  -   \bar{\pi} \right], \\
\frac{d {\cal R}}{d \tau} &=&  - \frac{3 {\cal R}}{\tau} \cdot \frac{{\rm Li}_3({\cal R}_a)}{{\rm Li}_2({\cal R}_a)} \bar{\pi}, \\
\frac{d \bar{\pi}}{d \tau} &=&  - \frac{\bar{\pi}}{\tau_R} - \frac{1}{\tau} \left[ - \frac{4}{3} \bar{\beta}_\pi({\cal R}_a)  + C_{\pi}  \bar{\pi}  + D({\cal R}_a) \bar{\pi}^2 \right],
\ee
where
\be
&& C_{\pi} := \frac{38}{21}, \nl
&& \bar{\beta}_\pi({\cal R}_a):=  \frac{4}{15}  \cdot \frac{{\rm Li}_{2}({\cal R}_a) {\rm Li}_{4}({\cal R}_a)}{4 {\rm Li}_{2}({\cal R}_a) {\rm Li}_{4}({\cal R}_a) - 3 {\rm Li}_{3}({\cal R}_a)^2},\\
&& D({\cal R}_a):= 4 - \frac{3 {\rm Li}_3\left({\cal R}_a \right)^2 \left( 3 {\rm Li}_1\left({\cal R}_a \right) {\rm Li}_3({\cal R}_a) - 2 {\rm Li}_2\left({\cal R}_a \right)^2 \right)}{{\rm Li}_2\left({\cal R}_a \right)^2 \left(4 {\rm Li}_2\left({\cal R}_a \right) {\rm Li}_4\left({\cal R}_a \right)-3 {\rm Li}_3\left({\cal R}_a \right)^2\right)}. \nn
\ee
Setting $\tau_R= \beta \theta_0$ and $w=\tau/\beta$ yields
\be
 \frac{d \beta}{d w} &=& \frac{\beta}{w} \cdot \frac{\frac{1}{3} - \bar{\pi}}{\frac{2}{3}+\bar{\pi}} \\
 \frac{d {\cal R}}{d w} &=& -\frac{3 {\cal R}}{w} \cdot \frac{{\rm Li}_3({\cal R}_a)}{{\rm Li}_2({\cal R}_a)} \cdot \frac{\bar{\pi}}{\frac{2}{3}+\bar{\pi}}, \\
 \frac{d \bar{\pi}}{d w} &=&  - \frac{ \frac{\bar{\pi}}{\theta_0} + \frac{1}{w} \left[ - \frac{4}{3} \bar{\beta}_\pi({\cal R}_a) + \frac{38}{21} \bar{\pi} + D({\cal R}_a)  \bar{\pi}^2 \right]}{\frac{2}{3}+\bar{\pi}}.
\ee

\section{Transseries and Borel resummation}
In this appendix, we summarize topics related to transseries, Borel resummation,  and resurgence.
{
  In App.~\ref{sec:transseries}, we briefly explain construction of the transmonomials, which are ingredients of the full transseries solutions, from the ODEs (\ref{eq:ODE_b_w})-(\ref{eq:ODE_a_w}) in the large $w$.
  In App.~\ref{sec:trans_XX_Obs}, we summarize explicit forms of the results of the first few orders of other variables such as the energy density and the bulk pressure.
  We give a brief review of Borel resummation in App.~\ref{app:borel_resum}, and then derive the resurgent relation in App.~\ref{app:resurgence}.
}
See Refs.~\cite{edgar2010transseries,Marino:2012zq,Aniceto:2013fka,Dorigoni:2014hea,Aniceto:2018bis,Costin2006TopologicalCO,sauzin2014introduction,Mitschi2016,LodayRichaud2016}, for example, in more detail.

\subsection{Construction of transmonomials} \label{sec:transseries}

In this part, we derive transmonomials expanded around $w=+\infty$.
We take $m=1$ for simplicity.
Let us start with ODEs given by Eqs.(\ref{eq:ODE_b_w})-(\ref{eq:ODE_bp_w}).
We firstly assume that $\lim_{w \rightarrow +\infty}\bar{\Pi} = \lim_{w \rightarrow +\infty}\bar{\pi}=0$ and $\lim_{w \rightarrow +\infty} \beta = +\infty$.
From Eqs.(\ref{eq:ODE_b_w})(\ref{eq:ODE_R_w}), the leading order of $(\beta,\alpha)$ is given by
\be
\frac{d \beta}{d w} \sim \frac{2}{w} \beta, \qquad \frac{d {\cal R}}{d w} \sim  \frac{1}{w} O(\beta^{-1}) \quad \Rightarrow \quad \beta \sim \sigma_\beta w^2, \quad {\cal R} \sim \sigma_{\cal R}.
\ee
The leading order of $(\beta, {\cal R})$ and Eqs.(\ref{eq:ODE_bP_w})(\ref{eq:ODE_bp_w}) give the one of $(\bar{\Pi}, \bar{\pi})$ as
\be
\bar{\Pi} \sim \frac{28}{9} C^{\bar{\beta}_{\pi}}_0(\sigma_{{\cal R}, a}) \left( \frac{\theta_0}{w} \right)^2, \qquad \bar{\pi} \sim \frac{4}{3} C^{\bar{\beta}_{\pi}}_0(\sigma_{{\cal R}, a}) \frac{\theta_0}{w}.
\ee
The higher orders can be recursively obtained by each of the ODEs, and their formal power expansion with respect to $w^{-1}$ is closed under all operations in the ODEs.
Notice that these are consistent with our assumption.

Then, we construct higher transmonomials in the NP sectors.
It can be derived from the linearized equation of $(\bar{\Pi}, \bar{\pi})$ given by
\be
&& \frac{d \delta \bar{\bf X}}{d w} = - \left[\Lambda + \frac{1}{w} {\frak B} \right] \delta \bar{\bf X} + O(\delta \bar{\bf X}^2, \delta \bar{\bf X} w^{-2}), 
\ee
where
\be
&& \Lambda:= S {\mathbb I}_2, \qquad S := \frac{3}{\theta_0} \nl
&&   {\frak B}:=
\begin{bmatrix}
  4 \left( 1 - 3 C^{\bar{\beta}_{\pi}}_0(\sigma_{{\cal R}, a}) \right)& - 7 \\
 -\frac{36}{5} & 6 \left( 1 - 2 C^{\bar{\beta}_{\pi}}_0(\sigma_{{\cal R}, a}) \right)
\end{bmatrix}, \label{eq:Lam_B} \\
&& \delta \bar{\bf X}:= (\delta \bar{\Pi},\delta \bar{\pi})^{\top}. \nn
\ee
Solving the linearized equation gives
\be
\delta \bar{\bf X}
= U^{-1}
\begin{pmatrix}
  \zeta_+ \\
  \zeta_-
\end{pmatrix},
\ee
where
\be
&& \zeta_\pm :=  \sigma_{\pm} \frac{e^{-S_\pm w}}{w^{\rho_{\pm}}}, \qquad S_{\pm} = S := \frac{3}{\theta_0}, \qquad 
\rho_{\pm} := 5 - 12 C^{\bar{\beta}_{\pi}}_0(\sigma_{{\cal R}, a}) \mp \frac{\sqrt{1285}}{5}, \label{eq:lam_rho} \\
&&
U = \begin{pmatrix}
  \frac{18}{\sqrt{1285}} & \frac{1}{2} - \frac{5}{2 \sqrt{1285}} \\
- \frac{18}{\sqrt{1285}} & \frac{1}{2} + \frac{5}{2 \sqrt{1285}}
\end{pmatrix},
\qquad
U^{-1} = \begin{pmatrix}
 \frac{5+\sqrt{1285}}{36} & \frac{5-\sqrt{1285}}{36} \\
  1 & 1
\end{pmatrix}, \label{eq:U_Uinv}
\ee
and $\sigma_\pm \in {\mathbb R}$ is the integration constant.

\subsection{Explicit form of transseries for other variables} \label{sec:trans_XX_Obs}

\subsubsection{Transseries of $\widetilde{X}_\pm$} \label{sec:trans_XX}
\be
\widetilde{X}_{+{\rm pt}} &\sim& (0.3825 - 0.1014 \sigma_{{\cal R},a} ) \theta_0 w^{-1} + (0.4040 - 0.5128 \sigma_{{\cal R},a} ) \theta_0^2 w^{-2}  \nl
&& + \left\{ 0.5615 - 0.1267 \sigma_{{\cal R},a} + ( -1.3386 + 0.2514 \sigma_{{\cal R},a} ) (\sigma_\beta \theta_0^2)^{-1} \right\}  \theta_0^3 w^{-3} \nl
&& + O(w^{-4},\sigma_{{\cal R},a}), \\ \nl
\widetilde{X}_{-{\rm pt}} &\sim& (0.5064 - 0.1343 \sigma_{{\cal R},a} ) \theta_0 w^{-1} + (-1.8855 - 0.0372 \sigma_{{\cal R},a} ) \theta_0^2 w^{-2}  \nl
&& + \left\{ 6.3916 + 0.9019 \sigma_{{\cal R},a} + ( -1.7725 + 0.3329 \sigma_{{\cal R},a} ) (\sigma_\beta \theta_0^2)^{-1} \right\}  \theta_0^3 w^{-3} \nl
&& + O(w^{-4},\sigma_{{\cal R},a}),
\ee
\be
\widetilde{X}_{+{\rm np}} &\sim& \zeta_+ \left[ 1.0000 \right. \nl
  && \left. + \left\{ 61.7850 - 13.9493 \sigma_{{\cal R},a} + ( -15.0000 + 2.3202 \sigma_{{\cal R},a}) (\sigma_\beta \theta^{2}_0)^{-1} \right\} \theta_0 w^{-1} \right. \nl
  && \left.  + \left\{ 1713.7587 - 817.8193 \sigma_{{\cal R},a} + ( -811.7237 + 311.5972 \sigma_{{\cal R},a}) (\sigma_\beta \theta^{2}_0)^{-1} \right. \right. \nl
  && \left. \left. + (112.5000 - 34.8029 \sigma_{{\cal R},a}) (\sigma_\beta \theta^{2}_0)^{-2} \right\} \theta_0^{2} w^{-2} + O(w^{-3},\sigma_{{\cal R},a}^2) \right] \nl
&& + \zeta_- \left[ 0.0368 \sigma_{{\cal R},a} \theta_0 w^{-1} \right. \nl
  && \left.  + \left\{ -0.1378 + 0.7902 \sigma_{{\cal R},a} + ( 0.2733 - 0.5776 \sigma_{{\cal R},a}) (\sigma_\beta \theta^{2}_0)^{-1} \right\} \theta_0^{2} w^{-2}  \right. \nl
  && \left.  + O(w^{-3},\sigma_{{\cal R},a}^2) \right]  + O(\zeta_\pm^2), \label{eq:Xpnp_trans} \\ \nl
\widetilde{X}_{-{\rm np}} &\sim&  \zeta_+ \left[ \left\{ 0.0617 + 0.2641 \sigma_{{\cal R},a} + ( -2.9069 + 0.2062 \sigma_{{\cal R},a}) (\sigma_\beta \theta^{2}_0)^{-1} \right\} \theta_0^{2} w^{-2}  \right. \nl
  && \left.  + O(w^{-3},\sigma_{{\cal R},a}^2) \right] \nl
  && + \zeta_- \left[ 1.0000 \right. \nl
  && \left.  + \left\{ 214.9524 - 56.7540 \sigma_{{\cal R},a} + ( - 291.7763 + 66.1111 \sigma_{{\cal R},a}) (\sigma_\beta \theta^{2}_0)^{-1} \right. \right. \nl
  && \left. \left. + (112.5000 - 34.8029 \sigma_{{\cal R},a}) (\sigma_\beta \theta^{2}_0)^{-2} \right\} \theta_0^{2} w^{-2} + O(w^{-3},\sigma_{{\cal R},a}^2) \right] + O(\zeta_\pm^2). \nl \label{eq:Xmnp_trans}
\ee

\subsubsection{Transseries of $({\cal E}_0,P,n_0)$ and $(P_{\perp}, P_{||})$} \label{sec:trans_obs}
\be
   {\cal E}_{0 \, {\rm pt}} &\sim& \frac{\sigma_{\cal R}}{\sigma_\beta^{3/2}} w^{-3}  \left[ 0.0635 + 0.0224 \sigma_{{\cal R},a}  + \left( - 0.5079 - 0.0449 \sigma_{{\cal R},a} \right) \theta_0 w^{-1}  \right. \nl
     && \left. + \left\{ 3.7250  - 0.1197 \sigma_{{\cal R},a} + \left( -0.3810 -0.0779 \sigma_{{\cal R},a} \right) (\sigma_\beta \theta_0^2)^{-1} \right\} \theta_0^2 w^{-2} \right. \nl
     && \left. + O(w^{-3}, \sigma_{{\cal R},a}^2) \right], \\ \nl
   P_{{\rm pt}} &\sim& \frac{\sigma_{\cal R}}{\sigma_\beta^{5/2}} w^{-5} \left[ 0.0635 + 0.0112 \sigma_{{\cal R},a} + \left(-1.0159 + 0.0449 \sigma_{{\cal R},a} \right) \theta_0 w^{-1}  \right. \nl
     && \left. + \left\{ 11.5135  - 2.2149 \sigma_{{\cal R},a} + \left( -0.9524 -0.0316 \sigma_{{\cal R},a} \right) (\sigma_\beta \theta_0^2)^{-1} \right\} \theta_0^2 w^{-2} \right. \nl
     && \left. + O(w^{-3}, \sigma_{{\cal R},a}^2) \right], \\ \nl
   n_{0 \, {\rm pt}} &\sim& \frac{\sigma_{\cal R}}{\sigma_\beta^{3/2}} w^{-3} \left[ \left( 0.0635 + 0.0224 \sigma_{{\cal R},a} \right) + \left( -0.5079 - 0.0449 \sigma_{{\cal R},a} \right) \theta_0 w^{-1} \right. \nl
     && \left. + \left\{ 3.7250  - 0.1197 \sigma_{{\cal R},a} + \left( -0.4762 -0.0947 \sigma_{{\cal R},a} \right) (\sigma_\beta \theta_0^2)^{-1} \right\} \theta_0^2 w^{-2} \right. \nl
     && \left. + O(w^{-3}, \sigma_{{\cal R},a}^2) \right], \\ \nl
P_{\perp \, {\rm pt}} &\sim& \frac{\sigma_{\cal R}}{\sigma_\beta^{5/2}} w^{-5} \left[ 0.0635 + 0.0112 \sigma_{{\cal R},a}  + \left(-0.9736 + 0.0524 \sigma_{{\cal R},a} \right) \theta_0 w^{-1} \right. \nl
  && \left. + \left\{ 10.9632  - 2.1625 \sigma_{{\cal R},a} + \left( -0.9524 -0.0316 \sigma_{{\cal R},a} \right) (\sigma_\beta \theta_0^2)^{-1} \right\} \theta_0^2 w^{-2} \right. \nl
  && \left. + O(w^{-3}, \sigma_{{\cal R},a}^2) \right], \\ \nl 
P_{\parallel \, {\rm pt}} &\sim& \frac{\sigma_{\cal R}}{\sigma_\beta^{5/2}} w^{-5} \left[ 0.0635 + 0.0112 \sigma_{{\cal R},a}  + \left(-1.1006 + 0.0299 \sigma_{{\cal R},a} \right) \theta_0 w^{-1} \right. \nl
  && \left. + \left\{ 13.2067  - 2.2149 \sigma_{{\cal R},a} + \left( -0.9524 -0.0316 \sigma_{{\cal R},a} \right) (\sigma_\beta \theta_0^2)^{-1} \right\} \theta_0^2 w^{-2} \right. \nl
  && \left. + O(w^{-3}, \sigma_{{\cal R},a}^2) \right], 
\ee
\\
\be 
   {\cal E}_{0 \, {\rm np}} &\sim& \frac{\sigma_{\cal R}}{\sigma_{\beta}^{3/2}} w^{-3} \zeta_+ \left[ \left( -0.3537 - 0.1251 \sigma_{{\cal R},a} \right) \theta_0 w^{-1}    \right. \nl
&&     \left. + \left\{  - 3.0035 - 0.9928 \sigma_{{\cal R},a} + \left( 5.3054 + 1.0551 \sigma_{{\cal R},a} \right) (\sigma_\beta \theta_0^2)^{-1} \right\} \theta_0^2 w^{-2}  \right. \nl
   &&  \left. + \left\{- 8.7911 - 4.9825 \sigma_{{\cal R},a} + \left( 26.6861 + 12.8558 \sigma_{{\cal R},a} \right) (\sigma_\beta \theta_0^2)^{-1}  \right. \right. \nl
   && \left. \left.  + \left( -39.7908 - 1.7585 \sigma_{{\cal R},a} \right)(\sigma_\beta \theta_0^2)^{-2}   \right\} \theta_0^3 w^{-3}   + O(w^{-4},\sigma_{{\cal R},a}^2)  \right] \nl
&& + \frac{\sigma_{\cal R}}{\sigma_{\beta}^{3/2}} w^{-3} \zeta_- \left[ \left( 0.0256 + 0.0091 \sigma_{{\cal R},a} \right) \theta_0 w^{-1}    \right. \nl
&&     \left. + \left\{ 1.3210 + 0.1804 \sigma_{{\cal R},a} + \left( -0.3847 - 0.0765 \sigma_{{\cal R},a} \right) (\sigma_\beta \theta_0^2)^{-1} \right\} \theta_0^2 w^{-2}  \right. \nl
   &&  \left. + \left\{  30.0234 - 3.2081 \sigma_{{\cal R},a} + \left( - 16.0793 + 0.5731 \sigma_{{\cal R},a} \right) (\sigma_\beta \theta_0^2)^{-1}  \right. \right. \nl
   && \left. \left.  + \left( 2.8851 + 0.1275 \sigma_{{\cal R},a} \right)(\sigma_\beta \theta_0^2)^{-2}   \right\} \theta_0^3 w^{-3}   + O(w^{-4},\sigma_{{\cal R},a}^2)  \right] + O(\zeta_\pm^2), \\ \nl
 P_{{\rm np}} &\sim& \frac{\sigma_{\cal R}}{\sigma_{\beta}^{5/2}} w^{-5} \zeta_+ \left[ \left( -0.7074 - 0.1563 \sigma_{{\cal R},a} \right) \theta_0 w^{-1}    \right. \nl
  &&  \left. + \left\{  - 0.3479 - 1.0641 \sigma_{{\cal R},a} + \left( 10.6109 + 0.7034 \sigma_{{\cal R},a} \right) (\sigma_\beta \theta_0^2)^{-1} \right\} \theta_0^2 w^{-2}  \right. \nl
   &&  \left. + \left\{- 11.0259 - 4.3437 \sigma_{{\cal R},a} + \left( - 24.0872 + 30.3412 \sigma_{{\cal R},a} \right) (\sigma_\beta \theta_0^2)^{-1}  \right. \right. \nl
   && \left. \left.  + \left( - 79.5816 + 7.0341 \sigma_{{\cal R},a} \right)(\sigma_\beta \theta_0^2)^{-2}   \right\} \theta_0^3 w^{-3}   + O(w^{-4},\sigma_{{\cal R},a}^2)  \right] \nl
&& + \frac{\sigma_{\cal R}}{\sigma_{\beta}^{5/2}} w^{-5} \zeta_- \left[ \left( 0.0513 + 0.0113 \sigma_{{\cal R},a} \right) \theta_0 w^{-1}    \right. \nl
&&  \left. + \left\{ 2.2316 + 0.0014 \sigma_{{\cal R},a} + \left( -0.7694 - 0.0510 \sigma_{{\cal R},a} \right) (\sigma_\beta \theta_0^2)^{-1} \right\} \theta_0^2 w^{-2}  \right. \nl
   &&  \left. + \left\{ 41.9206 - 10.4999 \sigma_{{\cal R},a} + \left( - 26.5421 + 4.5612 \sigma_{{\cal R},a} \right) (\sigma_\beta \theta_0^2)^{-1}  \right. \right. \nl
   && \left. \left.  + \left( 5.7703 - 0.5100 \sigma_{{\cal R},a} \right)(\sigma_\beta \theta_0^2)^{-2}   \right\} \theta_0^3 w^{-3}   + O(w^{-4},\sigma_{{\cal R},a}^2)  \right] + O(\zeta_\pm^2), \\ \nl
  n_{0 \, {\rm np}} &\sim& \frac{\sigma_{\cal R}}{\sigma_{\beta}^{3/2}} w^{-3} \zeta_+ \left[ \left( -0.3537 - 0.1251 \sigma_{{\cal R},a} \right) \theta_0 w^{-1}    \right. \nl
&& \left. + \left\{ - 3.0035 - 0.9928 \sigma_{{\cal R},a} + \left( 5.3054 + 1.0551 \sigma_{{\cal R},a} \right) (\sigma_\beta \theta_0^2)^{-1} \right\} \theta_0^2 w^{-2}  \right. \nl
   &&  \left. + \left\{- 8.7911 - 4.9825 \sigma_{{\cal R},a} + \left( 27.7472 + 13.0902 \sigma_{{\cal R},a} \right) (\sigma_\beta \theta_0^2)^{-1}  \right. \right. \nl
   && \left. \left.  + \left( -39.7908 - 1.7585 \sigma_{{\cal R},a} \right)(\sigma_\beta \theta_0^2)^{-2}   \right\} \theta_0^3 w^{-3}   + O(w^{-4},\sigma_{{\cal R},a}^2)  \right] \nl
&& + \frac{\sigma_{\cal R}}{\sigma_{\beta}^{3/2}} w^{-3} \zeta_- \left[ \left( 0.0256 + 0.0091 \sigma_{{\cal R},a} \right) \theta_0 w^{-1}    \right. \nl
&&     \left. + \left\{ 1.3210 + 0.1804 \sigma_{{\cal R},a} + \left( -0.3847 - 0.0765 \sigma_{{\cal R},a} \right) (\sigma_\beta \theta_0^2)^{-1} \right\} \theta_0^2 w^{-2}  \right. \nl
   &&  \left. + \left\{  30.0234 - 3.2081 \sigma_{{\cal R},a} + \left( - 16.1562 + 0.5561 \sigma_{{\cal R},a} \right) (\sigma_\beta \theta_0^2)^{-1}  \right. \right. \nl
   && \left. \left.  + \left( 2.8851 + 0.1275 \sigma_{{\cal R},a} \right)(\sigma_\beta \theta_0^2)^{-2}   \right\} \theta_0^3 w^{-3}   + O(w^{-4},\sigma_{{\cal R},a}^2)  \right] + O(\zeta_\pm^2), \\ \nl 
 P_{\perp{\rm np}} &\sim& \frac{\sigma_{\cal R}}{\sigma_{\beta}^{5/2}} w^{-5} \zeta_+ \left[ -0.0340 - 0.0150 \sigma_{{\cal R}, a}   \right. \nl
  &&  \left. + \left\{  - 0.9639 - 0.3175 \sigma_{{\cal R},a} + \left( 0.5098 + 0.1464 \sigma_{{\cal R},a} \right) (\sigma_\beta \theta_0^2)^{-1} \right\} \theta_0 w^{-1}  \right. \nl
   &&  \left. + \left\{- 1.5042 - 1.9009 \sigma_{{\cal R},a} + \left(  12.8380 + 2.8703 \sigma_{{\cal R},a} \right) (\sigma_\beta \theta_0^2)^{-1}  \right. \right. \nl
   && \left. \left.  + \left( - 3.8236  - 0.5069 \sigma_{{\cal R},a} \right) (\sigma_\beta \theta_0^2)^{-2}   \right\} \theta_0^2 w^{-2}   + O(w^{-3},\sigma_{{\cal R},a}^2)  \right] \nl
&& + \frac{\sigma_{\cal R}}{\sigma_{\beta}^{5/2}} w^{-5} \zeta_- \left[ 0.1557 + 0.0688 \sigma_{{\cal R},a}   \right. \nl
&&  \left. + \left\{ 7.1793 + 1.7052 \sigma_{{\cal R},a} + \left( - 2.3353 - 0.6708 \sigma_{{\cal R},a} \right) (\sigma_\beta \theta_0^2)^{-1} \right\} \theta_0 w^{-1}  \right. \nl
   &&  \left. + \left\{ 143.3957 + 0.7239 \sigma_{{\cal R},a} + \left( - 91.6245 - 7.8216 \sigma_{{\cal R},a} \right) (\sigma_\beta \theta_0^2)^{-1}  \right. \right. \nl
   && \left. \left.  + \left( 17.5144 + 2.3221 \sigma_{{\cal R},a} \right)(\sigma_\beta \theta_0^2)^{-2}   \right\} \theta_0^2 w^{-2}   + O(w^{-3},\sigma_{{\cal R},a}^2)  \right] + O(\zeta_\pm^2), \\ \nl
 P_{\parallel{\rm np}} &\sim& \frac{\sigma_{\cal R}}{\sigma_{\beta}^{5/2}} w^{-5} \zeta_+ \left[ -0.1768 - 0.0782 \sigma_{{\cal R}, a}   \right. \nl
  &&  \left. + \left\{  - 2.0423 - 1.0243 \sigma_{{\cal R},a} + \left( 2.6527 + 0.7620 \sigma_{{\cal R},a} \right) (\sigma_\beta \theta_0^2)^{-1} \right\} \theta_0 w^{-1}  \right. \nl
   &&  \left. + \left\{- 2.8573 - 5.0047 \sigma_{{\cal R},a} + \left(  21.9815 + 12.3426 \sigma_{{\cal R},a} \right) (\sigma_\beta \theta_0^2)^{-1}  \right. \right. \nl
   && \left. \left.  + \left( - 19.8954  - 2.6378 \sigma_{{\cal R},a} \right) (\sigma_\beta \theta_0^2)^{-2}   \right\} \theta_0^2 w^{-2}   + O(w^{-3},\sigma_{{\cal R},a}^2)  \right] \nl
&& + \frac{\sigma_{\cal R}}{\sigma_{\beta}^{5/2}} w^{-5} \zeta_- \left[ 0.0128 + 0.0057 \sigma_{{\cal R},a}   \right. \nl
&&  \left. + \left\{ 0.6384 + 0.1501 \sigma_{{\cal R},a} + \left( - 0.1923 - 0.0553 \sigma_{{\cal R},a} \right) (\sigma_\beta \theta_0^2)^{-1} \right\} \theta_0 w^{-1}  \right. \nl
   &&  \left. + \left\{ 13.7766 + 0.0105 \sigma_{{\cal R},a} + \left( - 7.7467 - 0.5033 \sigma_{{\cal R},a} \right) (\sigma_\beta \theta_0^2)^{-1}  \right. \right. \nl
   && \left. \left.  + \left( 1.4426 + 0.1913 \sigma_{{\cal R},a} \right)(\sigma_\beta \theta_0^2)^{-2}   \right\} \theta_0^2 w^{-2}   + O(w^{-3},\sigma_{{\cal R},a}^2)  \right] + O(\zeta_\pm^2). 
\ee

\subsection{Review of Borel resummation} \label{app:borel_resum}
In this part, we briefly review Borel resummation theory.
We suppose the following transseries expanded around $w =+\infty$:
\be
&& X(w) \sim \sum_{n \in {\mathbb N}_0} \zeta^n X^{[n]}(w), \qquad X^{[n]}(w) \sim \sum_{k \in {\mathbb N}_0} a^{[n,k]} w^{-k}, \qquad  \zeta = \sigma \frac{e^{- S w}}{w^b},  \label{eq:Xw}
\ee
where $w \in {\mathbb R}$, $a^{[n,k]} \in {\mathbb C}, b \in {\mathbb C} \setminus {\mathbb Z}_{0-}, S \in \{ x \in {\mathbb C} \, | \, {\rm Re} \, x > 0 \}, \sigma \in {\mathbb C}$.
In the main text, we call the PT sector for $n=0$ and the $n$-the NP sector if $n \ne 0$.
For simplicity, we assume that $a^{[0,0]}=0$.
For the technical reason, we redefine Eq.(\ref{eq:Xw}) as
\be
\widecheck{X}^{[n]}(w) = w^{-nb} X^{[n]}(w) \label{eq:Xw_ch}
\ee
such that $\zeta^{n}X^{[n]}(w) = (\sigma e^{-S w})^n \widecheck{X}^{[n]}(w)$.
The Borel transform, ${\cal B}$, to $\widecheck{X}^{[n]}(w)$ is defined as
\be
\widehat{X}^{[n]}(\xi) :=  {\cal B}[\widecheck{X}^{[n]}](\xi) = \sum_{k \in {\mathbb N}_0} \frac{a^{[n,k]}}{\Gamma(k + nb)} \xi^{k+nb-1}. \qquad (\xi \in {\mathbb C})
\ee
In order to avoid confusion, we also define ${\mathbb I}_\xi := {\cal B}[w^{-1}]$.
It is a homomorphism, and the multiplication of ${\cal B}[X](\xi)$ and ${\cal B}[Y](\xi)$ is defined using the convolution $*$ as
\be
&& {\cal B}[X \cdot Y](\xi) = {\cal B}[X] * {\cal B}[Y](\xi) := \int_{0}^{\xi} d \xi^\prime \, {\cal B}[X](\xi^\prime) \, {\cal B}[Y](\xi - \xi^\prime).
\ee
The Laplace integral using the integration ray with a complex phase $\theta$, ${\cal L}_\theta$, is defined as
\be
\widecheck{X}^{[n]}_{{\rm ex},\theta}(w) :=  {\cal L}_{\theta}[\widehat{X}^{[n]}](w) := \int^{+\infty e^{i \theta}}_0 d \xi e^{-\xi w} \, \widehat{X}^{[n]}(\xi).
\ee
In this paper, we consider the case that $\theta=0$ or $0_\pm$.
Taking the asymptotic limit, $|w| \rightarrow +\infty$, to $\widecheck{X}^{[n]}_{{\rm ex},0}(w)$ reduces to $\widecheck{X}^{[n]}(w)$ in Eq.(\ref{eq:Xw_ch}).
The combination of the Borel transform and the Laplace integral, ${\cal S}_\theta := {\cal L}_\theta \circ {\cal B}$, is called as Borel resummation.
It is important that the Borel resummation is homomorphism, i.e. ${\cal S}_\theta[f \cdot g] = {\cal S}_\theta[f] \cdot {\cal S}_\theta[g]$.

If $\widecheck{X}^{[n]}(w)$ is Borel nonsummable along $\theta=0$, meaning that the Laplace integral is not performable because of singular points on the real positive axis on the Borel plane, we introduce the infinitesimal complex phase, $\theta={0_\pm}$, to avoid them in the Laplace integral.
In such a case, due to the singular points, the resulting Borel resummed function (or transseries) has discontinuity, 
\be
{\cal S}_{0_+}[\widecheck{X}^{[n]}](w) \ne {\cal S}_{0_-}[\widecheck{X}^{[n]}](w). \label{eq:SpSmneq}
\ee
Here, we formally introduce the Stokes automorphism, ${\frak S}_\theta$, to make Eq.(\ref{eq:SpSmneq}) to be equality as
\be
{\cal S}_{0_+}[\widecheck{X}^{[n]}](w) = {\cal S}_{0_-} \circ {\frak S}_0[\widecheck{X}^{[n]}](w). \label{eq:SpSmeq}
\ee
If $\widecheck{X}^{[n]}(w)$ is Borel summable along $\theta=0$, then ${\frak S}_0 = {\rm id}$ and ${\cal S}_{0_+}[\widecheck{X}^{[n]}](w) = {\cal S}_{0_-}[\widecheck{X}^{[n]}](w)$.
From now on, we omit the subscript $\theta$ in ${\frak S}_\theta$ because we only consider the case that $\theta=0$.
The Stokes automorphism is a mapping from a transseries to a transseries generally including exponential decay.
So that it can be extended to group transformation by introducing a real parameter, $\nu \in {\mathbb R}$, as ${\frak S} \rightarrow {\frak S}^{\nu}$ defined as
\be
 {\frak S}^{\nu} = \exp \left[ \nu \sum_{\xi_* \in \Gamma} \bul{\Delta}_{\xi_*}  \right], \qquad  \bul{\Delta}_{\xi_*} := e^{-\xi_* w} \Delta_{\xi_*}, 
\ee
where $\Gamma$ is a set of singular points on the positive real axis of the Borel plane and $\Delta_{\xi_*}$ is the alien derivative with respect to the singular point, $\xi_*$.
The alien derivative is a mapping from a formal transseries to the same type of a formal transseries,
\be
\Delta_{\xi_*}: w^{-b} {\mathbb C}[[w^{-1}]] \rightarrow w^{-b^\prime} {\mathbb C}[[w^{-1}]]
\ee
with some $b^\prime \in {\mathbb C}$.
Notice that ${\frak S}^{\nu=0}= {\rm id}$ and that ${\frak S}$ in Eq.(\ref{eq:SpSmeq}) is identical to ${\frak S}^{\nu=1}$.
The transformation of Stokes automorphism depends on a given problem.
We construct the resurgent relation, which is a relationship among sectors, by defining the transformation based on the structure of our ODE in the next section.

For the $i$-dimensional vectorial expression, ${\bf X}=(X_1, \cdots, X_i)$, the above definition is applicable by the extension such as $n \rightarrow {\bf n} \in {\mathbb N}^i, \zeta^{n} \rightarrow \bm{\zeta}^{\bf n} = \prod_{j=1}^{i} \zeta_{j}^{n_j}$ with $\zeta_{i} = \sigma_i e^{-S_i w}/w^{b_i}$ on the real positive axis.

\subsection{Derivation of the resurgent relation} \label{app:resurgence}
In this part, we review the construction of the resurgent relation.
In this paper, we only consider the case reproducing the PT sector from the NP sectors, but other cases are also possible. See Refs.~\cite{Costin2006TopologicalCO,Aniceto:2013fka,Aniceto:2018bis,Marino:2012zq,Dorigoni:2014hea} in detail.

We define $\widetilde{\bf X}:= U \bar{\bf X}$, where $\widetilde{\bf X}= (\widetilde{X}_{+}, \widetilde{X}_{-})^{\top}$ and $\bar{\bf X}= (\bar{\Pi}, \bar{\pi})^{\top}$, which has the following asymptotic form:
\be
\widetilde{X}_\pm(w) \sim \sum_{{\bf n}\in {\mathbb N}_0^2} \bm{\zeta}^{\bf n} \widetilde{X}_\pm^{[{\bf n}]}(w), \qquad \widetilde{X}_\pm^{[{\bf n}]}(w) = \sum_{k \in {\mathbb N}_0} a_{\widetilde{X}_\pm}^{[{\bf n},k]} w^{-k}. 
\ee
Notice that $a_{\widetilde{X}_+}^{[(0,1),0]} = a_{\widetilde{X}_-}^{[(1,0),0]}=0$.
For the simplified notation, we define
\be
\widecheck{X}_\pm^{[{\bf n}]}(w) = w^{-{\bf n}\cdot \bm{\rho}} \widetilde{X}^{[{\bf n}]}_\pm(w). \label{eq:Xt_ch}
\ee

As is seen in App.\ref{sec:transseries}, $\widetilde{\bf X}$ is the asymptotic solution of the type of ODE given by $\frac{d \widetilde{\bf X}}{d w} = -  \Lambda \widetilde{\bf X} - \frac{1}{w} \left[ {\bf V} + {\frak B} \widetilde{\bf X}\right] + O(\widetilde{\bf X}^2, w^{-2})$, where $\Lambda$ and ${\frak B}$ are defined in Eq.(\ref{eq:Lam_B}) and ${\bf V}$ is a constant vector with nonzero-components.
Acting the Borel transform to the ODE yields
\be
- \xi \widetilde{{\bf X}}_{B} = - \Lambda \widetilde{{\bf X}}_B
- {\mathbb I}_\xi {\bf V} - {\frak B} {\mathbb I}_\xi * \widetilde{{\bf X}}_B + O(\widetilde{{\bf X}}_B^{*2},\xi), \label{eq:BdwdX}
\ee
where ${\cal B}[{\frac{d \widetilde{\bf X}}{dw}}] =- \xi \widetilde{\bf X}_B$, $\widetilde{\bf X}_B:={\cal B}[\widetilde{\bf X}]$, $\widetilde{\bf X}_B^{*2}:=\widetilde{\bf X}_B*\widetilde{\bf X}_B$, and ${\mathbb I}_\xi := {\cal B}[w^{-1}]$, and it can be written as
\be
\widetilde{{\bf X}}_B =   - ( \Lambda - \xi)^{-1} \left[ {\mathbb I}_\xi {\bf V} + O({\mathbb I}_\xi * \widetilde{{\bf X}}_B, \widetilde{{\bf X}}_B^{*2},\xi) \right]. \label{eq:BdwdX2}
\ee
As one can see immediately, the singularity appears at $\xi = S = 3/\theta_0$.
In this type of problem, the information of the $n$-th sector propagate to the  $(n+1)$-th sector, but not \textit{directly} to the $(n+2)$-th sector or higher.
By using the alien derivative, it can be expressed by
\be
{\Delta}_{\ell S}[\widecheck{X}_\pm^{[{\bf 0}]}](w) = 
\begin{cases}
 \sum_{s=\pm} A_{s} \widecheck{X}_\pm^{[\bm{\delta}_s]}(w)  & \mbox{if $\ell=1$} \\
0  & \mbox{otherwise}
\end{cases}, \qquad
\bm{\delta}_\pm :=
\begin{cases}
  (1,0) & \mbox{for $\bm{\delta}_+$} \\
  (0,1) & \mbox{for $\bm{\delta}_-$}
\end{cases},
\ee
where $A_\pm \in i {\mathbb R}$ is the Stokes constant.
For a generic ${\bf n}=(n_+,n_-)$,
\be
{\Delta}_{\ell S} [\widecheck{X}_\pm^{[{\bf n}]}](w) &=&
\begin{cases}
  \sum_{s = \pm} (n_s + 1) A_{s} \widecheck{X}_\pm^{[{\bf n}+\bm{\delta}_s]}(w)  & \mbox{if $\ell=1$} \\
0  & \mbox{otherwise}
\end{cases}, \\ \nl
(\Delta_S)^p [\widecheck{X}_\pm^{[{\bf n}]}](w) &=&  \sum_{q=0}^p A_{+}^q A_{-}^{p-q} \frac{(n_++q)! (n_-+p-q)!}{n_+! n_-!} \widecheck{X}_\pm^{[{\bf n} + (q, p-q)]}(w). \label{eq:Delp_w} 
\ee
Therefore, the action of the Stokes automorphism is obtained as
\be
{\frak S}^{\nu} [\widetilde{\bf X}](w;\bm{\sigma}) = \widetilde{\bf X}(w;\bm{\sigma}+\nu {\bf A}),
\ee
where $\bm{\sigma}:=(\sigma_+,\sigma_-)$ and ${\bf A}:=(A_+,A_-)$.
Additionally, the discontinuity (imaginary ambiguity) cancellation is realized by the median resummation given by
\be
&& {\cal S}_{0_+} \circ {\frak S}^{-1/2} [\widetilde{\bf X}](w;\bm{\sigma})  = {\cal S}_{0_-} \circ {\frak S}^{+1/2} [\widetilde{\bf X}](w;\bm{\sigma}) \nl
&\Rightarrow \quad& {\cal S}_{0_+} [\widetilde{\bf X}](w;\bm{\sigma}-\tfrac{1}{2} {\bf A}) = {\cal S}_{0_-} [\widetilde{\bf X}](w;\bm{\sigma}+\tfrac{1}{2} {\bf A}).  
\ee

Let us obtain the resurgent relation of $a^{[{\bf 0},k]}_{\widetilde{X}_\pm}$ for $k \gg 1$.
In order to do it, we start with
\be
a^{[{\bf 0},k]}_{\widetilde{X}_\pm} &=& \frac{\Gamma(k)}{(2 \pi i)^2} \oint_{|w| \ll 1} \frac{d w}{w} \oint_{|\xi| \ll 1} \frac{d \xi}{\xi^{k}} \widecheck{X}_{\pm,B}^{[{\bf 0}]}(\xi) e^{-\xi w} \nl
&=& \frac{\Gamma(k)}{(2 \pi i)^2} \oint_{|w| \ll 1} \frac{d w}{w} (-{\rm Int}_w)^k\oint_{|\xi| \ll 1} d \xi \widecheck{X}_{\pm,B}^{[{\bf 0}]}(\xi) e^{-\xi w},
\ee
where ${\rm Int}_w$ is the integration with respect to $w$.
Since\footnote{We assume that singular points exist only on the positive real axis on the Borel plane.}
\be
\oint_{|\xi| \ll 1} d \xi \widecheck{X}_{\pm,B}^{[{\bf 0}]}(\xi) e^{-\xi w} &=& ( {\cal S}_{0_+} - {\cal S}_{0_-}) [\widecheck{X}_\pm^{[{\bf 0}]}](w)  \nl
 &=&  \sum_{p \in {\mathbb N}} \frac{(\bul{\Delta}_\lambda)^p}{p!}[\widecheck{X}_\pm^{[{\bf 0}]}](w) \, = \, \sum_{\substack{{\bf n} \in {\mathbb N}_0^2 \\ |{\bf n}| >0}} \frac{A_+^{n_+} A_-^{n_-}}{C(|{\bf n}|,n_-)}  e^{-{\bf n}\cdot {\bf S} w} \widecheck{X}_\pm^{[{\bf n}]}(w),
\ee
where $|{\bf n}|:=n_+ + n_-$, ${\bf n} \cdot {\bf S} = n_+ S_+ + n_- S_-$, and $C(p,q)=\left( \begin{smallmatrix} p \\ q \end{smallmatrix} \right)$ is the binomial coefficient, one finds that
\be
&& (-{\rm Int}_w)^k \oint_{|\xi| \ll 1} d \xi \widecheck{X}_{\pm,B}^{[{\bf 0}]}(\xi) e^{-\xi w} \nl
&=& \sum_{\substack{{\bf n} \in {\mathbb N}_0^2 \\ |{\bf n}| >0}}
\sum_{h \in {\mathbb N}_0} \frac{A_+^{n_+} A_-^{n_-}}{C(|{\bf n}|,n_-)} 
a_{\widetilde{X}_\pm}^{[{\bf n},h]} \frac{({\bf n} \cdot {\bf S})^{{\bf n} \cdot \bm{\rho} +h-k}}{(k-1)!} \sum_{s=0}^{k-1} 
\begin{pmatrix}
  k-1 \\
  s
\end{pmatrix} \Gamma(k - {\bf n} \cdot \bm{\rho} -h -s) \nl
&& \cdot (-1)^{s} \left[ ( {\bf n} \cdot \bm{\rho} \, w)^{s} - e^{- {\bf n} \cdot {\bf S} w}({\bf n} \cdot {\bf S} \,  w)^{k - {\bf n} \cdot \bm{\rho} -h} \sum_{r \in {\mathbb N}_0} \frac{({\bf n} \cdot {\bf S} \,  w)^{r}}{\Gamma(k - {\bf n} \cdot \bm{\rho} -h -s + r + 1)} \right], \nl \label{eq:-IntwInt}
\ee
where ${\bf n} \cdot \bm{\rho} = n_+ \rho_+ + n_- \rho_-$. 
Therefore, the resurgent relation is obtained as
\be
a_{\widetilde{X}_\pm}^{[{\bf 0},k]} &\simeq& \frac{1}{2 \pi i}  \sum_{\substack{{\bf n} \in {\mathbb N}_0^2 \\ |{\bf n}| >0}} \sum_{h \in {\mathbb N}_0} \frac{A_{+}^{n_+} A_{-}^{n_-}}{C(|{\bf n}|,n_-)}  a_{\widetilde{X}_\pm}^{[{\bf n},h]} \frac{\Gamma(k - {\bf n} \cdot \bm{\rho} -h)}{({\bf n} \cdot {\bf S})^{k-{\bf n} \cdot \bm{\rho}-h}}. \qquad (k \gg 1) \label{eq:res_rel}
\ee
Here, we took the most dominant part of Eq.(\ref{eq:-IntwInt}) for the large $k$.

\section{Convergent point in the UV limit} \label{sec:CP_UV}

We briefly describe finding convergent points (CPs) the UV limit, $\tau \rightarrow 0_+$, by beginning with ODEs in Eqs.(\ref{eq:ODE_b0})-(\ref{eq:ODE_bp0}) with $\tau_R = \beta \theta_0$.
We assume that the temperature is divergent ($\beta \rightarrow 0_+$) in this limit, so that the transport coefficients has the following asymptotic forms:
\be
&& \chi_\beta \sim \frac{1}{3} + O(z^{2}), \qquad \chi_\alpha \sim O(z^2), \qquad  z \frac{\Xi^{(1)}_{2}}{\Xi^{(1)}_{1}} \sim \frac{3 {\rm Li}_{3}({\cal R}_a)}{{\rm Li}_{2}({\cal R}_a)} + O(z^{2}),  \nl
&& \bar{\beta}_{\Pi} \sim O(z^{4}), \qquad C_{\Pi} \sim -\frac{1}{3} + O(z), \qquad \lambda_{\Pi \pi} \sim 2 + O(z^{2}), \nl
 && \bar{\beta}_{\pi} \sim C^{\bar{\beta}_\pi}_0({\cal R}_a) + O(z^2), \qquad C_\pi \sim \frac{38}{21}, \qquad \lambda_{\pi \Pi} \sim \frac{6}{5} + O(z), \nl
 && D \sim C^{D}_0({\cal R}_a) + O(z^2), \label{eq:leading_trans1_UV}
\ee
where ${\cal R}_a:=-a e^{-\alpha}$, and
\be
&& C^{\bar{\beta}_{\pi}}_0({\cal R}_a) =  \frac{4 {\rm Li}_{2}({\cal R}_a) {\rm Li}_{4}({\cal R}_a)}{60 {\rm Li}_{2}({\cal R}_a) {\rm Li}_{4}({\cal R}_a) - 45 {\rm Li}_{3}({\cal R}_a)^2}, \\
&& C^{D}_0({\cal R}_a) = 4 - \frac{3 {\rm Li}_3\left({\cal R}_a \right)^2 \left( 3 {\rm Li}_1\left({\cal R}_a \right) {\rm Li}_3({\cal R}_a) - 2 {\rm Li}_2\left({\cal R}_a \right)^2 \right)}{{\rm Li}_2\left({\cal R}_a \right)^2 \left(4 {\rm Li}_2\left({\cal R}_a \right) {\rm Li}_4\left({\cal R}_a \right)-3 {\rm Li}_3\left({\cal R}_a \right)^2\right)}.
\ee
The approximated ODEs 
are given by
\be
\widehat{F}_{\beta} &\sim& -\frac{\beta}{s} \left[ \frac{1}{3} + \bar{\Pi} - \bar{\pi} \right], \label{eq:beta_UV_app} \\
\widehat{F}_{\alpha} &\sim& \frac{3}{s} \cdot \frac{{\rm Li}_3({\cal R}_a)}{{\rm Li}_2({\cal R}_a)} (\bar{\Pi}-\bar{\pi}), \label{eq:alp_UV_app} \\
\widehat{F}_{\bar{\Pi}}&\sim&  \frac{\bar{\Pi}}{\tau_R s^2} - \frac{1}{s} \left[  \frac{\bar{\Pi}}{3} + 2 \bar{\pi} + C^{D}_0({\cal R}_a) \left( \bar{\Pi} - \bar{\pi} \right) \bar{\Pi} \right], \nl \\
\widehat{F}_{\bar{\pi}}&\sim&  \frac{\bar{\pi}}{\tau_R s^2} - \frac{1}{s} \left[ \frac{4}{3} C^{\bar{\beta}_\pi}_0({\cal R}_a)
  - \frac{38}{21} \bar{\pi} + \frac{4}{5} \bar{\Pi} + C^{D}_0({\cal R}_a) \left(\bar{\Pi} - \bar{\pi} \right) \bar{\pi} \right],
\ee
where $\widehat{F}_{\cal O} := -F_{\cal O}/s^2$ with $s:=1/\tau$, and thus, $\frac{d {\cal O}}{d s}=\widehat{F}_{\cal O}$.

\subsubsection*{Case (I): the non-collision part is dominant}
We firstly consider the situation that the non-collision term is dominant in this limit.
In this case, the CPs are given by solving the following equations:
\be
&& \frac{\bar{\Pi}_{\rm UV}}{3} + 2 \bar{\pi}_{\rm UV}  + C^{D}_0({\cal R}_a) \left( \bar{\Pi}_{\rm UV} - \bar{\pi}_{\rm UV} \right) \bar{\Pi}_{\rm UV} = 0, \label{eq:UVCPeq1} \\ 
&& \frac{4}{3} C^{\bar{\beta}_\pi}_0({\cal R}_a) 
- \frac{38}{21} \bar{\pi}_{\rm UV} + \frac{4}{5} \bar{\Pi}_{\rm UV} + C^{D}_0({\cal R}_a) \left(\bar{\Pi}_{\rm UV} - \bar{\pi}_{\rm UV} \right) \bar{\pi}_{\rm UV} =0, \label{eq:UVCPeq2} 
\ee
where $(\bar{\Pi}_{\rm UV},\bar{\pi}_{\rm UV}):= \lim_{s \rightarrow +\infty} (\bar{\Pi}(s),\bar{\pi}(s))$.
Notice that the possibility that $\bar{\Pi}_{\rm UV} = \bar{\pi}_{\rm UV}=0$ is rejected because in this case $\lim_{s \rightarrow +\infty} \alpha(s) = \sigma_{\alpha}$ and Eq.(\ref{eq:UVCPeq2}) does not give $\bar{\Pi}_{\rm UV} = \bar{\pi}_{\rm UV}=0$.
In addition, the case that $\alpha \rightarrow -\infty$ is also rejected because $|{\cal R}_a| \rightarrow +\infty$ and $(\bar{\beta}_{\pi}, D)$ becomes complex value in this domain.
Thus, the allowed situation is only $\alpha \rightarrow +\infty$ and ${\cal R}_a \rightarrow 0$.
The real solution in this situation is obtained as
\be
(\bar{\Pi}_{\rm UV}, \bar{\pi}_{\rm UV}) \approx (1.0518,-1.5367).
\ee
From Eqs.(\ref{eq:beta_UV_app})(\ref{eq:alp_UV_app}), the leading order of $(\beta, \alpha, {\cal R})$ is obtained as
\be
\beta \sim \sigma_\beta s^{-2.9218}, \qquad \alpha \sim 7.7655 \cdot \log s + \sigma_\alpha, \qquad {\cal R} \sim \sigma_{\cal R} s^{-7.7655}.
\ee
This result implies that $1/(\beta s^2) \gg 1/s$ as $s \rightarrow \infty$, so that it is contradiction to the assumption that the non-collision part is dominant.
Therefore, this possibility is rejected.
When constructing $w=1/(s \beta)$, then $w \sim \sigma_\beta^{-1} s^{1.9218}$ and $\lim_{s \rightarrow +\infty} w(s) = +\infty$, which contradicts to our assumption.

\subsubsection*{Case (II): the collision part $\approx$ the non-collision part}
In this case, $\beta \sim \sigma_\beta s^{-1}$, and thus $\bar{\Pi}_{\rm UV} - \bar{\pi}_{\rm UV} = \frac{2}{3}$.
The CP in this case is given by solving
\be
&& \frac{\bar{\Pi}_{\rm UV}}{\sigma_\beta \theta_0} - \left[  \frac{\bar{\Pi}_{\rm UV}}{3} + 2 \bar{\pi}_{\rm UV} + \frac{2}{3} C^{D}_0({\cal R}_a)  \bar{\Pi}_{\rm UV} \right] = 0, \nl \\
&& \frac{\bar{\pi}_{\rm UV}}{\sigma_\beta \theta_0} - \left[ \frac{4}{3} C^{\bar{\beta}_\pi}_0({\cal R}_a)
  - \frac{38}{21} \bar{\pi}_{\rm UV} + \frac{4}{5} \bar{\Pi}_{\rm UV} + \frac{2}{3} C^{D}_0({\cal R}_a) \bar{\pi}_{\rm UV} \right] = 0,
\ee
with $\bar{\Pi}_{\rm UV} - \bar{\pi}_{\rm UV} = \frac{2}{3}$.
The solution is given by
\be
(\bar{\Pi}_{\rm UV}, \sigma_\beta) = 
\left( 1.0868, \frac{0.5640}{\theta _0} \right) \ \mbox{and} \ \left(  0.24468 , -\frac{0.4083}{\theta _0} \right),
\ee
and the integration constant, $\sigma_\beta$, can not be taken freely.
The second solution is unphysical because of the negative $\sigma_\beta$.
The leading orders of $\alpha$ and ${\cal R}$ are obtained as
\be
\alpha \sim 2 \log s + \sigma_\alpha, \qquad {\cal R} \sim \sigma_{\cal R} s^{-2},
\ee
in both of the solutions.
Then, we consider the stability around the CPs.
It can be obtained from the Jacobian defined as
\be
\widehat{J}_{\rm UV} := \left.
\begin{pmatrix} 
  \frac{\pd \widehat{F}_{\bar{\Pi}}}{\pd \bar{\Pi}} & \frac{\pd \widehat{F}_{\bar{\Pi}}}{\pd \bar{\pi}} \\
  \frac{\pd \widehat{F}_{\bar{\pi}}}{\pd \bar{\Pi}} & \frac{\pd \widehat{F}_{\bar{\pi}}}{\pd \bar{\pi}} 
\end{pmatrix} \right|_{(\bar{\Pi}, \bar{\pi}) \rightarrow (\bar{\Pi}_{\rm UV}, \bar{\pi}_{\rm UV})} \cdot s,
\ee
where $s$ in the r.h.s. is the regularization.
The eigenvalues are obtained as
\be
   {\rm Eigen}(\widehat{J}_{\rm UV}) = (- 0.3136,3.3361) \quad \mbox{and} \quad (-3.6940, -1.7284).
\ee
Thus, on the $(\bar{\Pi},\bar{\pi})$-plane, the CPs are saddle and source, respectively.

When using $w=1/(s \beta)$ for the flow time, these CPs do not appear because $1-(\chi_\beta + \bar{\Pi}-\bar{\pi}) \rightarrow 0$ and all of Eqs.(\ref{eq:ODE_b_w})-(\ref{eq:ODE_bp_w}) are singular in the limit.
More precisely, it comes from the fact that $\tau$ and $w$ are not one-to-one correspondence.

\bibliographystyle{ieeetr}
\bibliography{massive_kinetic.bib}

\end{document}